\documentclass[a4paper,twocolumn,11pt,accepted=2023-12-03]{quantumarticle}
\pdfoutput=1
\usepackage[utf8]{inputenc}
\usepackage[english]{babel}
\usepackage[T1]{fontenc}
\usepackage{amsmath,amssymb,amsfonts}

\usepackage{multirow,makecell,booktabs}
\usepackage{graphicx}
\usepackage{hyperref}
\usepackage[table,xcdraw,dvipsnames]{xcolor}
\usepackage{physics}
\usepackage{bm}
\usepackage{bbm}
\usepackage{slashed}
\usepackage[normalem]{ulem}
\usepackage{tcolorbox}
\usepackage[caption=false]{subfig}
\usepackage{tikz}
\usepackage[shortlabels]{enumitem}

\usepackage{dcolumn}
\usepackage{mathrsfs}
\usepackage{tikz-feynman}

\usepackage{pifont}
\newcommand{\cmark}{\ding{51}}%
\newcommand{\xmark}{\ding{55}}%

\definecolor{fancy_red}{RGB}{144,58,42}
\definecolor{fancy_green}{RGB}{178,224,71}
\definecolor{fancy_purple}{RGB}{184,165,202}
\definecolor{darkgreen}{RGB}{50,205,50}

\renewcommand{\i}{\text{i}}
\newcolumntype{P}[1]{>{\centering\arraybackslash}p{#1}}

\newcommand{\numberbox}[2][]{
    \if\relax\detokenize{#1}\relax
        \tcbox[on line,boxsep=0pt,left=2pt,right=2pt,top=2pt,bottom=2pt,colback=fancy_purple]{~#2 ~}
    \else
        \href{#1}{\tcbox[on line,boxsep=0pt,left=2pt,right=2pt,top=2pt,bottom=2pt,colback=fancy_purple]{~#2 ~}}
    \fi
}
\newcommand{\numbering}[2][]{
    \stepcounter{experimentcounter}%
    \numberbox[#1]{\arabic{experimentcounter}}
}

\usepackage{xspace}
\newcommand{\melvin}{\textsc{Melvin} }
\newcounter{experimentcounter}

\newcommand{\pytheus}{\textsc{PyTheus}\xspace}
\newcommand{\theseus}{\textsc{Theseus}\xspace}
\newcommand{\mpl}{Max Planck Institute for the Science of Light, Erlangen, Germany.}
 
\usepackage[numbers,compress]{natbib}
\begin{document}

\title{Digital Discovery of 100 diverse Quantum Experiments with PyTheus}

\author{Carlos Ruiz-Gonzalez$^\S$}%
\email{cruizgo@proton.me}%
\affiliation{\mpl}%
\author{Sören Arlt$^\S$}%
\email{soeren.arlt@mpl.mpg.de}%
\affiliation{\mpl}%

\author{Jan Petermann}%
\affiliation{\mpl}%

\author{Sharareh Sayyad}%
\affiliation{\mpl}%

\author{Tareq Jaouni}%
\affiliation{Nexus for Quantum Technologies, University of Ottawa, K1N 6N5, ON, Ottawa, Canada.}%

\author{Ebrahim Karimi}%
\affiliation{\mpl}%
\affiliation{Nexus for Quantum Technologies, University of Ottawa, K1N 6N5, ON, Ottawa, Canada.}%

\author{Nora Tischler}%
\affiliation{Centre for Quantum Computation and Communication Technology (Australian Research Council),
Centre for Quantum Dynamics, Griffith University, Brisbane, Australia.}%

\author{Xuemei Gu}%
\affiliation{\mpl}%

\author{Mario Krenn}%
\email{mario.krenn@mpl.mpg.de}
\affiliation{\mpl}%

\begin{abstract}
Photons are the physical system of choice for performing experimental tests of the foundations of quantum mechanics. Furthermore, photonic quantum technology is a main player in the second quantum revolution, promising the development of better sensors, secure communications, and quantum-enhanced computation. These endeavors require generating specific quantum states or efficiently performing quantum tasks. The design of the corresponding optical experiments was historically powered by human creativity but is recently being automated with advanced computer algorithms and artificial intelligence. While several computer-designed experiments have been experimentally realized, this approach has not yet been widely adopted by the broader photonic quantum optics community. The main roadblocks consist of most systems being closed-source, inefficient, or targeted to very specific use-cases that are difficult to generalize. Here, we overcome these problems with a highly-efficient, open-source digital discovery framework \pytheus, which can employ a wide range of experimental devices from modern quantum labs to solve various tasks. This includes the discovery of highly entangled quantum states, quantum measurement schemes, quantum communication protocols, multi-particle quantum gates, as well as the optimization of continuous and discrete properties of quantum experiments or quantum states. \pytheus produces interpretable designs for complex experimental problems which human researchers can often readily conceptualize. \pytheus is an example of a powerful framework that can lead to scientific discoveries -- one of the core goals of artificial intelligence in science. We hope it will help accelerate the development of quantum optics and provide new ideas in quantum hardware and technology.
\end{abstract}

\def\thefootnote{$\S$}\footnotetext{These authors contributed equally to this work.\\ The order was decided by a coin flip.}\def\thefootnote{\arabic{footnote}}
\maketitle

\tableofcontents

\section{Introduction}

Photons, the individual particles of light, have long been used as the core player for fundamental experiments and applications in quantum information science \cite{pan2012multiphoton}. Photons do not easily interact with their environments; therefore, they can be distributed over large distances -- which makes them a key resource for long-distance quantum communication \cite{liao2017satellite,liao2018satellite} and experiments that require strict Einstein locality conditions \cite{hensen2015loophole,shalm2015strong,giustina2015significant}. Using advanced measurement-based quantum computing schemes, photons are among the most promising candidates for future quantum computers \cite{bartolucci2021fusion}. Entanglement between two or more photons can be produced without a vacuum or cooling, and therefore many advanced experimental results can be achieved directly with table-top setups. Furthermore, the bosonic nature of photons allows for the generation of complex entangled quantum states of indistinguishable photons that are a key resource for quantum-enhanced measurements \cite{polino2020photonic}. These potential applications have lead to enormous technological advances in integrated chips for fast and precise control of photonic quantum states \cite{SchaeffIntegrated,wang2018integrated,wang2020integrated,pelucchi2022potential}, high-quality single-photon sources \cite{wang2019towards,arakawa2020progress,tomm2021bright,uppu2021quantum}, novel photon-pair sources \cite{santiago2022resonant}, photon number resolving detectors \cite{singlephotons2011,slussarenko2019photonic}, and advanced high-quality multi-photon interference  \cite{bouchard2020two,menssen2017distinguishability,feng2021observation,qian2023multiphoton}.

One question now is how to utilize these technologies to build up exciting new experiments for the foundations of quantum physics and practical quantum hardware.

Historically, the design of quantum experiments strongly relied on the intuition and creativity of human experts who leverage their experience and come up with blueprints of experiments. However, due to the unintuitive phenomena and enormous combinatorial space of the potential designs, it becomes extremely difficult for human researchers to discover more complex quantum setups. It might be possible that there are high-quality solutions to experimental design questions far outside of the region where humans' intuition fails. How could we possibly find such extraordinary solutions?

This question has sparked a strong interest in the automated discovery of quantum experiments with computers, overviewed in \cite{krenn2020computer}. The invention of these tools for quantum optics experiments \cite{Melvin2016} have indeed overcome experimental limitations and allowed for new avenues in laboratories for entanglement research \cite{babazadeh2017high,srv332experimentmalik,erhard3dghz,kysela2020path}. One crucial question is whether we can also learn something about physics from these tools. And indeed, several new concepts have been published that were purely discovered through automated design \cite{Melvin2016}, such as a new general idea of entanglement structure \cite{krenn2017entanglement}, and generalized constructions of photonic quantum gates \cite{gao2020computer}. Those concepts were discovered by tedious analysis of the computer's solutions, which was time-consuming. The problem was that the algorithms were powerful enough to find unknown solutions but had no incentive to present a simple, human-understandable form of it.

This was solved by the invention of \theseus \cite{theseus}, an efficient algorithm for the discovery of new quantum experiments that can readily be interpreted by humans. The key insight was a shift in the representation. Rather than describing quantum experiments as quantum optical components on an optical table, experiments are described as graphs of correlations between photons. This representation, which has been a derivative of a computer-discovered concept itself, was developed in \cite{graphs1,graphs2,graphs3} -- and allows working with a much more natural representation, which can be translated back at any point to an experiment consisting of optical elements. (It should be noted that the representation is independent of photonic graph states for measurement-based quantum computing \cite{raussendorf2001one,raussendorf2003measurement,briegel2009measurement}, and it is so far unknown how to translate among them.)

In this paper, we introduce \pytheus\footnote{GitHub:\\ \url{https://github.com/artificial-scientist-lab/PyTheus}}, a highly-efficient, open-source, automated design and discovery framework for quantum optics experiments. At the core, \pytheus uses a much extended graph-based representation of quantum optics, which allows us not only to represent entanglement and quantum gates, but lets us design quantum measurements, quantum communication protocols, optimize experimental properties, and discover quantum systems that involve single-photon sources, mixed states, and states entangled in the photon-number basis. Besides the advances of the scientific scope, we note that \pytheus is written in \textsc{Python}, and therefore can readily be combined with machine learning frameworks such as \textsc{TensorFlow} and \textsc{PyTorch}, and allows for immediate parallelization in computer clusters.

To showcase the applicability of \pytheus, we demonstrate the discovery of 100 previously unknown or advanced implementations of quantum optics experiments, ranging from exciting new systems for entanglement research to quantum states from condensed matter physics that are interesting for quantum simulation purposes, new ways of performing quantum communication tasks such as entanglement swapping, new quantum state measurements, and quantum gates. The experiments can involve both probabilistic photon sources and deterministic single-photon sources, and many of them are readily implementable in today's modern quantum optics labs. In the GitHub repository, we present the instructions for \pytheus that discover each of the examples. We hope that \pytheus's efficiency, generality, and low entry barrier kick-starts the application of computer-discovered quantum setups in experimental laboratories worldwide, and inspires new exciting computer-inspired ideas and directions for fundamentals and applications of photonic quantum physics research.

While the goal of this paper was to demonstrate the discovery capability of \pytheus, in several cases, it was impossible \textit{not} to \textit{see} clear generalizations and reasons why the solutions work. We show this in some cases below. One of the exceptionally interesting concepts we discovered was a new quantum multiphoton interference effect that can simulate probabilistic multi-pair sources just with photon pairs. We describe this new physics concept and its application in a parallel paper \cite{halopaper}.

The article is structured in the following way: In section \ref{section:graphs}, we introduce the graph-based representation of quantum optics, which lies at the heart of \pytheus. In section \ref{sec:pytheus}, we introduce the idea of the computational \pytheus framework, which we then apply to the discovery of 100 new quantum experiments in section \ref{100exp}. In section \ref{sec:outlook}, we explain some future(istic) ideas that might lie ahead of us.

\subsection{Related Work}
The first automated and artificial-intelligence-driven design methods for new quantum experiments were introduced in 2016 (for a more detailed review on the topic see \cite{krenn2020computer,krenn2023artificial}). One of them, \melvin, was focused on specific photonic quantum information tasks such as quantum state generation and quantum transformations, using discrete learning techniques \cite{Melvin2016}. The second one, \textit{Tachikoma}, focused on the discovery of new experimental setups for quantum metrology tasks and used genetic algorithms for discrete optimization \cite{knott2016search}. \textit{Tachikoma} has been expanded to incorporate neural network surrogate models to speed up the search process for new quantum-enhanced measurements \cite{o2019hybrid,nichols2019designing}. At the same time, the ideas of \melvin have led to numerous implementations of experiments in various laboratories \cite{babazadeh2017high,srv332experimentmalik,erhard3dghz,kysela2020path} and the extraction of new ideas and concepts in quantum physics \cite{krenn2017entanglement, gao2020computer}. Automated design tools have helped to build new ways to perform quantum information tasks such as quantum cloning \cite{zhan2020experimental}. Compared to these tools, \pytheus does not work on the discrete search space. Discrete spaces are very challenging to navigate as gradients cannot be used. Rather, \pytheus uses domain knowledge in the form of a new physics-inspired representation that is entirely continuous.

These ideas have later been expanded by using reinforcement-learning \cite{melnikov2018active, melnikov2020setting}, for quantum communication \cite{wallnofer2020machine,valcarce2022automated}, recurrent neural networks \cite{adler2021quantum}, and deep generative models such as variational autoencoders \cite{flam2022learning} or logical AI \cite{cervera2021design}. Compared to these tools, \pytheus uses direct optimization on the outputs of a physical simulator and does not rely on learned simulators or strategies. That makes it significantly faster.

Various quantum physics groups and companies have also developed simulators and optimizers since. A main focus there is on the design of experimental settings for photonic quantum computing \cite{arrazola2019machine,Killoran2019strawberryfields,Quandela_Perceval, BudapestQuantumComputingGroup_Piquasso}. One remarkable simulator is \textit{Strawberryfields} \cite{Killoran2019strawberryfields}, which is focused on design and optimization tasks for continuous-variable photonic quantum computing and quantum machine learning tasks. Recent updates include auto-differentiation which significantly speeds up the rate of optimization. A related software package is \cite{walrus}, which allows for fast computation of tasks related to Gaussian boson sampling. Compared to these tools, the focus of \pytheus is different. \pytheus is built for discrete-variable quantum optics, and not targeted to photonic quantum computing (or boson sampling tasks). Furthermore, one main motivation is the interpretability of the discovered results, which is achieved via a topological optimization on the graph-based representation.

An alternative methodology that focuses not only on the design question but also on understanding the underlying physical concepts is \theseus \cite{theseus}. There, the algorithm employs a graph-based representation to describe photonic experiments, and the final results are topologically simplified graphs that can be interpreted and conceptualized in a much more straightforward way than representations that work directly on the optimization of optical elements. Compared to \theseus, \pytheus expands this idea and applies it to many new situations, inaccessible before, such as the design of quantum measurement and communication setups via the Choi–Jamiołkowski isomorphism.

Related work has shown how quantum computers could overcome the enormous computational of designing quantum optical hardware \cite{kottmann2021quantum}. 
\pytheus relies, for now, on classical computers, but recent hardware advances might enable the execution of tools like \pytheus on quantum hardware \cite{bao2023very}.

\section{Graphs and Quantum Experiments}\label{section:graphs}
\begin{table}[!b]
\renewcommand{\arraystretch}{1.1}
    \centering
    \begin{tabular}{c|c|c}
     \hline
        \multicolumn{2}{c|}{\textbf{Graph theory}} &  \textbf{Experiment} \\ \hline
        \multicolumn{2}{c|}{\!\!color weighted graph\!\!}&quantum experiment\\ \hline
        \multirow{6}{*}{\!\!vertex\!\!}& \begin{tikzpicture}[baseline={([yshift=-.5ex]current bounding box.center)}]
         \node[circle, fill=white, thick, draw=black,  scale=1.1](1) at (0,0) {};
        \end{tikzpicture} &\! path to photodetector\! \\\cline{2-3}
        
        & \begin{tikzpicture}[baseline={([yshift=-.5ex]current bounding box.center)}]
         \node[circle, fill=white, thick, draw=black, dotted, scale=1.1](1) at (0,0) {};
        \end{tikzpicture} &\! heralded optical path*\! \\\cline{2-3}
       
        &  \begin{tikzpicture}[baseline={([yshift=-.5ex]current bounding box.center)}]
         \node[regular polygon,regular polygon sides=3, fill=white, thick, draw=black, scale=0.7](1) at (0,0) {};
        \end{tikzpicture}& \!single photon source \! \\\cline{2-3}
       
        &  \begin{tikzpicture}[baseline={([yshift=-.5ex]current bounding box.center)}]
         \node[regular polygon,regular polygon sides=3, fill=white, thick, draw=black, dotted, scale=0.7](1) at (0,0) {};
        \end{tikzpicture}& incoming photon  \\\cline{2-3}
        
       & \begin{tikzpicture}[baseline={([yshift=-.5ex]current bounding box.center)}]
         \node[regular polygon,regular polygon sides=4, fill=white, thick, draw=black, scale=1](1) at (0,0) {};
        \end{tikzpicture}&  ancillary photodetector \\\cline{2-3}
        
       & \begin{tikzpicture}[baseline={([yshift=-.5ex]current bounding box.center)}]
         \node[regular polygon,regular polygon sides=6, fill=white, thick, draw=black, scale=1](1) at (0,0) {};
        \end{tikzpicture}& \!\! number resolving detector\!\! \\\cline{2-3}
       
       & \begin{tikzpicture}[baseline={([yshift=-.5ex]current bounding box.center)}]
         \node[star, star points=10, fill=white, thick, draw=black, scale=0.8](1) at (0,0) {};
        \end{tikzpicture}& \!environment interaction \!\! \\\hline
        \multirow{4}{*}{edge} & color& internal mode number\\\cline{2-3}
         &\! weight $\in\mathbb{C}$ \!& amplitude** \\\cline{2-3}
         
         &\begin{tikzpicture}[baseline={([yshift=-.5ex]current bounding box.center)}]
        \draw[-,draw=black,line width=0.4mm] (-1.1,0) -- (-0.,0);
        \node[diamond, fill=white, thick, draw=black, scale=0.6](1) at (-0.5,0) {};
        \end{tikzpicture} & negative amplitude \\\cline{2-3}
        
         &\begin{tikzpicture}[baseline={([yshift=-.5ex]current bounding box.center)}]
         \node[circle, fill=white, thick, draw=black, scale=1.1](1) at (-1,0) {};
         \node[circle, fill=white, thick, draw=black, scale=1.1](1) at (0,0) {};
        \draw[-,draw=black,line width=0.4mm] (-0.82,0) -- (-0.18,0);
        \end{tikzpicture}& correlated photon pair\\\cline{2-3}
    
         &\begin{tikzpicture}[baseline={([yshift=-.5ex]current bounding box.center)}]
         \node[regular polygon,regular polygon sides=3, fill=white, thick, draw=black, scale=0.7](1) at (-1,0) {};
         \node[circle, fill=white, thick, draw=black, scale=1.1](1) at (0,0) {};
         \draw[-,draw=black,line width=0.4mm] (-0.82,0) -- (-0.15,0);
        \end{tikzpicture}& single photon path\\\hline
       
    \end{tabular}
    \caption{The correspondence between graph theory and quantum experiments. *For some experiments one must known the total amount of photons crossing a group of optical paths (see section \ref{sec:noon_states}). This is clarified in the graph figures with a gray envelop. **Unless the contrary is specified, all weights are real values.}
    \label{tab:graph-experiment-link}
\end{table}
The connection between quantum optical experiments and graph theory was discovered a few years ago \cite{graphs1,graphs2,graphs3} and has been further developed as a design algorithm \theseus for new quantum experiments \cite{theseus}. In the graph-experiment representation, each colored weighted graph corresponds to a quantum experimental setup, and \textit{vice versa}. Each edge and each vertex of the graph represent a correlated photon pair and a photon path, respectively. Its complex weight denotes the amplitude of the photon pair, and the edge color represents a photon's internal mode number for a given path, which corresponds to the photon's degree of freedom such as polarization \cite{polarizationERP1995}, path \cite{wang2018integrated,lu2020three,pelucchi2022potential}, transverse spatial modes \cite{rubinsztein2016roadmap,padgett2017orbital,bouchard2017high,bavaresco2018measurements}, time-bin \cite{timebin1989} or frequency \cite{frequencybin2010,santiago2022resonant}. This abstract graph representation allows us to have the full information of quantum optical experiments and has been used for discovering quantum states and transformations \cite{theseus}. At first glance, it might seem that there is fundamentally no difference between the path degree of freedom (which is encoded as vertices) and the internal degrees of freedom of photons (which are encoded as colors of edges). However, this is only true information theoretically. Physically, the path degree of freedom is exceptional, because it allows to add spatially separation between photons and thereby perform non-locality experiments.

Here we significantly extend the bridge between graphs and experiments, which allows us to perform design and discovery tasks for quantum state generation (for pure and mixed states and on the photon number basis), quantum measurements, quantum communication protocols, and gates for quantum computing. In Table.~\ref{tab:graph-experiment-link}, we show the correspondence between graph theory and quantum experiments. The graph representation can be directly translated into different experimental implementations. In the remaining part of this section, we explain how these graphs encode quantum states, and how to translate them to experimental setups.

\subsection{Quantum State Generation} \label{sec:stategeneration}
\subsubsection{Probabilistic Photon-Pair Sources} \label{sec:p2ps}

\begin{figure}[!t]
	\centering
	\includegraphics[width=0.48\textwidth]{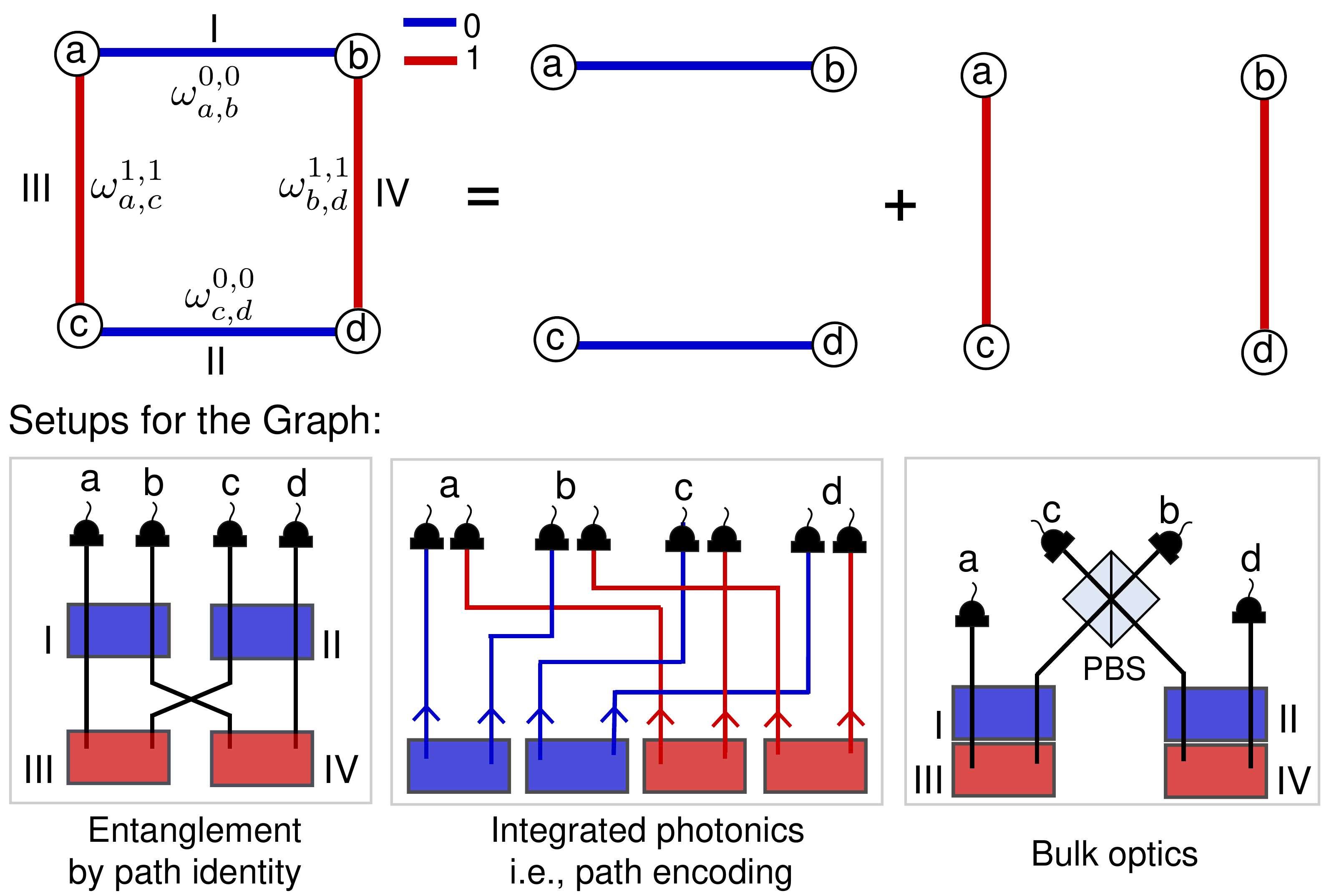} 
 \caption {An abstract graph representation of quantum experiments producing GHZ states. There, edges and vertices represent correlated photon-pair and optical paths, respectively. The edge colors and weights $\boldsymbol\omega$ are mode numbers and complex coefficients. The blue and red colors refer to the mode numbers 0 and 1, respectively. There are two perfect matchings (i.e., a subgraph that covers all vertices only once), and their coherent superposition leads to a quantum state $\ket{\psi}=\omega_{a,b}^{0,0} \omega_{c,d}^{0,0} \ket{0000} + \omega_{a,c}^{1,1} \omega_{b,d}^{1,1}\ket{1111}$ (without normalization). We can get a four-qubit GHZ state by setting all weights equal. The graph can be translated into several quantum setups with different technologies, such as entanglement by path identity or path encoding used in integrated photonics or standard bulk optics (e.g., using polarization encoding). The squares I-IV are probabilistic sources that create photon pairs. For clearer presentation, we draw the source in the color corresponding to the mode they contribute.}
 \label{fig:Twodim_graph_pm_example}
\end{figure}

Probabilistic photon-pair sources, which are typically based on nonlinear processes such as spontaneous parametric down conversion (SPDC) and four-wave mixing (FWM) \cite{boyd2020nonlinear}, are one of the most widespread resources to generate entangled and correlated pairs of photons. A range of photonic quantum experiments using probabilistic sources can be interpreted as a weighted colored graph \cite{theseus,graphs1,graphs2,graphs3}. There, each vertex represents an optical path to a detector and each edge refers to the correlated photon pair produced by a probabilistic photon-pair source. The edge weight is the amplitude associated with the photons, and the edge color describes the photon's internal mode number (i.e., the degree of freedom of a photon). The connection between the graph and the corresponding quantum state is given by the weight function \cite{theseus}
\begin{align}
    \Phi(\boldsymbol{\omega})=\sum_m \frac{1}{m!} \left( \sum_{e \in E(G)} \omega(e) x^\dagger(e) y^\dagger(e) + \text{h.c.}\right)^m,
    \label{eq:weightfunction}
\end{align}
where $E(G)$ is the set of edges of the graph. The quantum state is obtained by applying the weight function to the vacuum, i.e. $\ket{\psi}=\Phi(\boldsymbol{\omega})\ket{\text{vac}}$. The term h.c. stands for hermitian conjugate, which includes annihilation operators. As an example of states using four path (i.e., $a$, $b$, $c$, and $d$) with two-dimensional internal modes (i.e., $0$ and $1$) in Fig.~\ref{fig:Twodim_graph_pm_example}, the $\Phi(\boldsymbol{\omega)}$ is given as
\begin{align}
    \Phi(\boldsymbol{\omega})\approx\sum_{N}&\frac{1}{N!}( \omega_{a,b}^{0,0}a^\dagger_0 b^\dagger_0 + \omega_{b,d}^{1,1}b^\dagger_1 d^\dagger_1 \notag \\
    &+ \omega_{c,d}^{0,0}c^\dagger_0 d^\dagger_0 + \omega_{a,c}^{1,1}c^\dagger_1 a^\dagger_1 + \text{h.c.})^N,
    \label{eq:pairsource}
\end{align}
where $\boldsymbol{\omega}=(\omega_{a,b}^{0,0},\omega_{b,d}^{1,1},\omega_{c,d}^{0,0},\omega_{a,c}^{1,1})$ is a list of edge weights $\omega_{x,y}^{i,j}\in\mathbb{C}$ and $|\omega_{x,y}^{i,j}|^2<1$, the superscript and subscript represent the mode number and the optical path, respectively. $x^\dagger_{k}$ is the creation  operator of a photon in path $x$ with mode $k$. The pair-emission process is up to the $\text{N}$-order, and the probability of occurrence for lower-order events is higher than that of higher-order ones. In principle, the hermitian conjugate terms influence the final state. However, for the low pump regime, the effect of the annihilation terms is negligible in many cases. For example this is the case when the final state is conditioned on having a photon in every detector. In all of the examples that we present here, it is safe to neglect the annihilation operators. 

An alternative method to compute these systems in a non-approximate way is to follow the Takagi decomposition (see \cite{PhysRevA.100.032326}, specifically Eq.27) , which leads to reliable solutions also in the strong-pump regime. Naturally, this method is more expensive to compute. In our manuscript, we do not need to use this more involved method, as the low-pump approximation holds with high accuracy. 
Experimentally, a common way to obtain a quantum state is to condition the experimental results on a simultaneous detection event in each detector, which is also called the $n$-fold coincidence detection. In our graph representation, this only happens when a subset of the edges contains each of the $n$ vertices exactly once (see the subset of blue edges or red edges in Fig.~\ref{fig:Twodim_graph_pm_example}), which is the so-called perfect matching of a graph. For the example in Fig.~\ref{fig:Twodim_graph_pm_example}, we neglect the empty mode and higher-order terms $N>2$ to post-select the quantum state. There are two perfect matchings (two blue edges and two red edges) in Fig.~\ref{fig:Twodim_graph_pm_example}, which contribute to two quantum terms $\ket{0000}$ and $\ket{1111}$. The weight of a perfect matching is the product of all its edge weights. For each term in a quantum state, the weight is given by the sum of all weights of the perfect matchings that contribute to it. Therefore, the weights for quantum terms $\ket{0000}$ and $\ket{1111}$ are $\omega_{a,b}^{0,0} \omega_{c,d}^{0,0}$ and $\omega_{a,c}^{1,1} \omega_{b,d}^{1,1}$, respectively. In the end, a coherent superposition of the two perfect matchings in the graph leads to the post-selected quantum state, which is
\begin{equation}
    \ket{\psi}\approx\omega_{a,b}^{0,0} \omega_{c,d}^{0,0} \ket{0000} + \omega_{a,c}^{1,1} \omega_{b,d}^{1,1}\ket{1111}.
    \label{eq:2dGHZ}
\end{equation} 
If we set all weights the same and normalize the state, we can then reach a four-particle Greenberger-Horne-Zeilinger (GHZ) state. 

\paragraph{Translation to Experiments --}\label{sec:translation} A colored weighted graph can directly be translated into several quantum optical experiments using different technologies \cite{theseus}, as shown in Fig.~\ref{fig:Twodim_graph_pm_example}. The most straightforward way is applying the concept of entanglement by path identity \cite{krenn2017entanglement, kysela2020path,RevModPhysPI2022}, for which the graph representation has initially been designed \cite{graphs1}. In this scheme shown on the left side of Fig.~\ref{fig:Twodim_graph_pm_example}, paths of indistinguishable photons produced from multiple crystals are overlapped, leading to a coherent superposition of possible origins for a photon. Equipped with the graph-experimental correspondence in Table \ref{tab:graph-experiment-link}, we now translate four edges and vertices into four photon-pair sources and optical paths. Here, we encode photon's internal mode number in polarization such that blue and red edge colors in the graph depict horizontal and vertical polarization, respectively. The four sources are set up in such a way that two sources (blue edges) produce photons with states $\ket{00}$ while the other two (red edges) produce photons with states $\ket{11}$. We then pump all sources coherently and consider that two of the four sources produce a photon pair in this example. Conditionally on perfect matching or post-selection, we have only two cases (two-photon pairs come either from the sources I and II or from the sources III and IV) that contribute to the final quantum state in Eq.~\eqref{eq:2dGHZ}. All other combinations of sources do not lead to only one photon in each of the detectors. For example, a photon pair in crystals I and III leads to two photons in $a$, one photon in $b$, and $c$, but no photon in $d$; therefore, this case is disregarded. If the mode numbers are indistinguishable, one can observe a new form of multi-photon interference, which has first been experimentally observed in \cite{feng2021observation,qian2023multiphoton} and which forms the basis of many of the proposed experiments below.

Another promising technology is using integrated photonic, which usually uses path encoding \cite{SchaeffIntegrated,wang2018integrated,wang2020integrated,pelucchi2022potential}, where each photon's one mode goes to one detector, as shown in the bottom middle-part of Fig.~\ref{fig:Twodim_graph_pm_example}. In this example, each photon has two path modes depicted in blue and red colors, respectively. Therefore, one can consider that each vertex in Fig.~\ref{fig:Twodim_graph_pm_example} actually contains two sub-detectors that directly connect to the paths for the photon's two-mode numbers (i.e., blue and red paths). The vertex is activated when only one of the two sub-detectors clicks. In this way, perfect matching means the situations where four sub-detectors click in either blue paths or red paths, leading to the desired quantum state. Moreover, one can also use the standard bulk optics for generating photonic entanglement \cite{tenghzprl,tenghzoptica}. When the edges are in the same color, which means the corresponding photons have the same mode number, one can use either probabilistic beam splitters or directly path-identified photon-pair sources to form the edge. If the edges have different colors (for example, two-mode numbers in Fig.~\ref{fig:Twodim_graph_pm_example}), one can employ mode-dependent beam splitters such as polarizing beam splitters (PBSs) and transverse spatial mode sorters \cite{crosscrystal1999,pan2012multiphoton} to create an edge. With polarization encoding in the example, we can insert a PBS between two cross-crystal sources \cite{crosscrystal1999}, for the state generation. Yet, such a translation is not unique, and an expert may further simplify the proposed setup. 

\begin{figure}[!t]
	\centering
\includegraphics[width=0.45\textwidth]{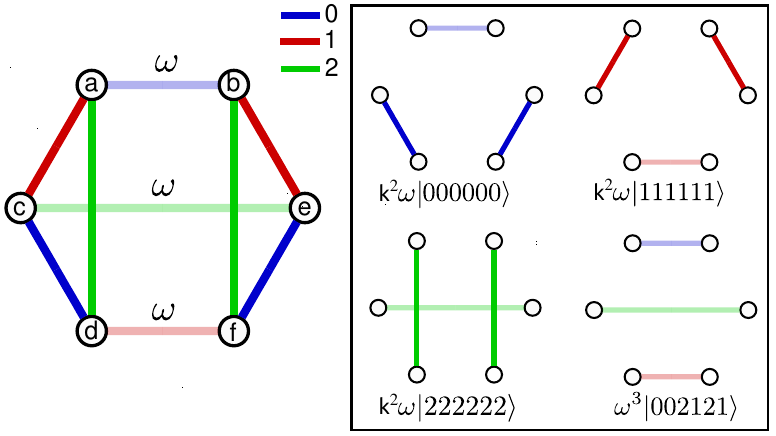} 
 \caption {Asymptotic solution for quantum experiments producing a three-dimensional six-particle GHZ state. All edges have weights $k$ but those labelled as $\omega$. The four perfect matchings of the graph lead to the quantum state $\ket{\psi}=k^2\omega(\ket{000000}+\ket{111111}+\ket{222222})+\omega^3\ket{002121}$. To minimize the amplitude of the undesired term the weights must fulfill $k^2\omega>\omega^3$. Moreover, $k^2\omega>k^4$ to guarantee that (N$>\!\!3$)-order terms from Eq.\eqref{eq:pairsource} are also negligible.} 
 \label{fig:epsilon_ghz_graph}
\end{figure}
\paragraph{Asymptotic Solutions --} Until now, we know that each perfect matching can contribute to a term of the quantum state, and their coherent superposition gives the desired state. With the above graph-experimental link, one now expects to produce arbitrary multi-particle entanglement in high dimensions, e.g., a six-particle three-dimension GHZ state. A GHZ state appears when all perfect matchings are disjoint, i.e., every edge is used only in one perfect matching \cite{graphs1,graphs2}. However, it has been shown that it is impossible to construct a graph of more than four vertices with only three disjoint perfect matchings \cite{graphs1,graphs2}. Fig.~\ref{fig:epsilon_ghz_graph} exemplifies this point. Three disjoint perfect matchings contribute to the desired quantum terms. However, there is another perfect matching that leads to an additional term that is not part of the GHZ state (the so-called Maverick term \cite{graphs1}). While we cannot fully erase the term, and therefore achieve a GHZ state with perfect fidelity, we can reduce the weight of the undesired term. Yet, one must be cautious with (N$>\!\!3$)-order terms from Eq.\eqref{eq:pairsource}, since they also need to be smaller than the GHZ state terms we are targeting. These two constraints boil down to the inequalities $k^2>\omega^2$ and $\omega>k^2$, respectively, which are both experimentally feasible. We call this and similar graphs \textit{asymptotic solutions}(Fig.~\ref{fig:epsilon_ghz_graph}) . 

\begin{figure}[!t]
	\centering
	\includegraphics[width=0.46\textwidth]{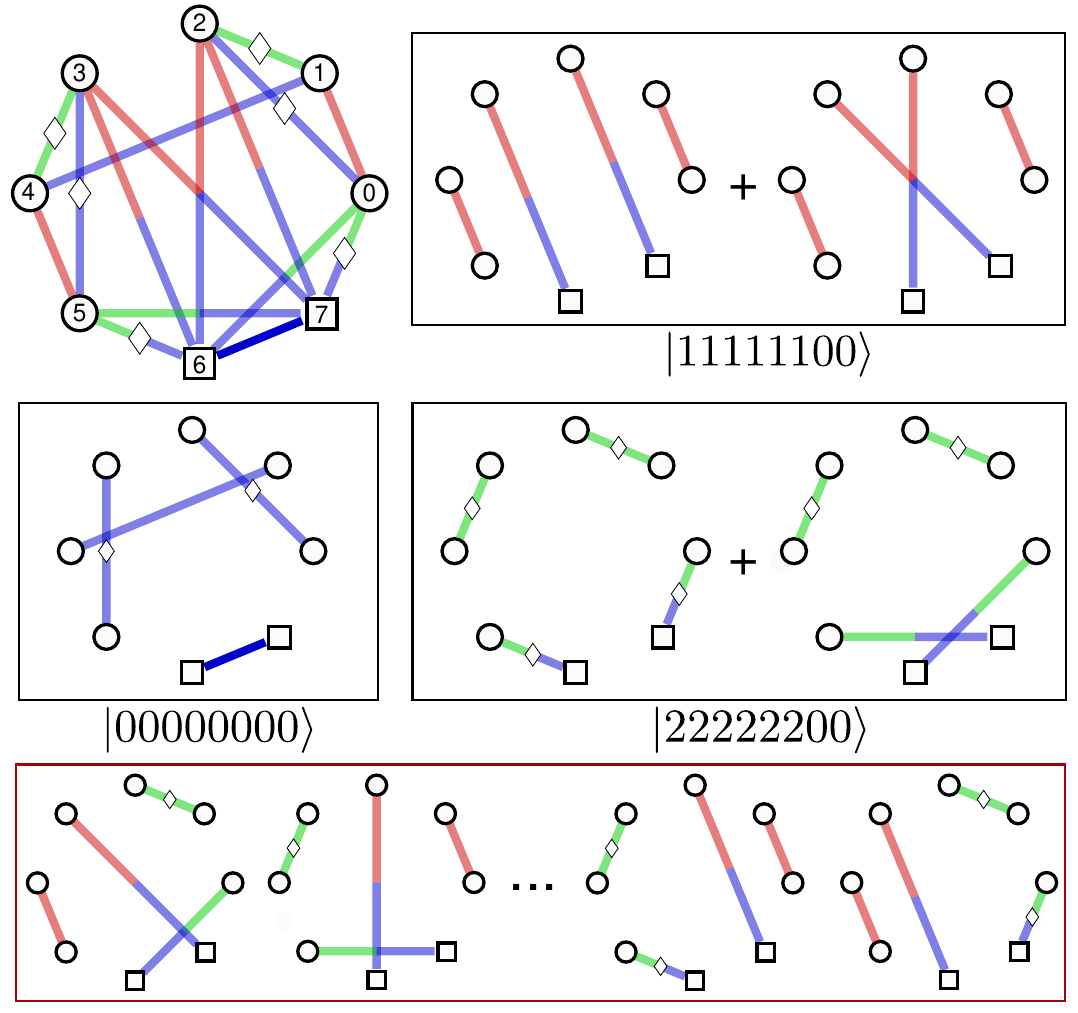} 
 \caption{An abstract graph for producing a three-dimensional six-particle GHZ state with two ancillary particles. The diamonds along the edges represent that the edge weights are negative. Vertices 6 and 7 depicted by squares are for ancillas, which are in the same mode number (e.g., blue color for mode number 0). Five perfect matchings in black boxes contribute to the three terms in the desired GHZ state, while the rest of the perfect matchings are destructively canceled out.} 
 \label{fig:ancilla_ghz_graph}
\end{figure}
\paragraph{Ancillary Particles --} Although there is a limitation to the state generation with standard linear optics and post-selection in the graph representation \cite{graphs2, calsamiglia2002generalized}, one can further employ ancillary particles to completely get rid of Maverick terms \cite{halopaper, paesani2021scheme}. For example, assisted with two ancillary photons, we can obtain the six-particle three-dimension GHZ state (details see Fig.~\ref{fig:ancilla_ghz_graph}). Additionally, to create quantum states with an odd number of particles, one can just use some odd numbers of photons produced from photon-pair sources as ancillary particles \cite{graphs3}. We present further examples in section \ref{100exp}.

\subsubsection{Deterministic Single-Photon Sources}\label{sec:single_photon}
\begin{figure}[!t] 
	\centering
	\includegraphics[width=0.48\textwidth]{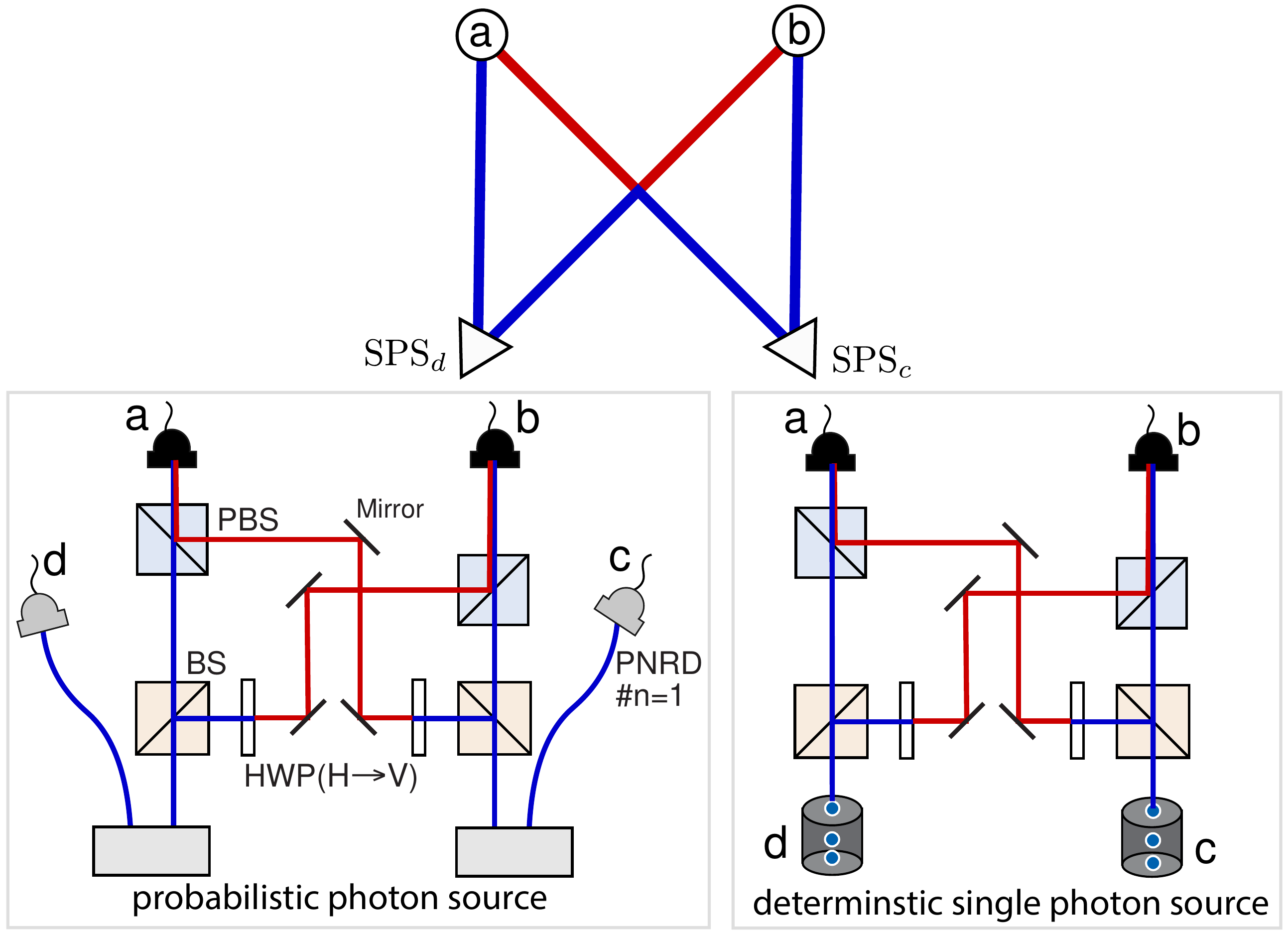} 
 \caption{An abstract graph for quantum experiments producing a Bell state with single-photon sources. The vertices $\text{SPS}_{d}$ and $\text{SPS}_{c}$ described as triangles stand for the incoming path for single-photon sources while vertices $a$ and $b$ stand for the optical paths. No edges are allowed between input vertices $\text{SPS}_{d}$ and $\text{SPS}_{c}$. The perfect matchings in this graph lead to a two-dimensional Bell state. The corresponding experiments are listed below. Experimentally, a single-photon source can be either a heralded single-photon source implemented with a photon-pair source and a photon-number-resolving detector (PNRD) or a deterministic single-photon source which can be an ideal quantum-dot-based source. In this experiment, the two photons arrive at the same time (and have the same wavelength, polarization, and spatial mode) such that they are indistinguishable.} 
 \label{fig:single_photon}
\end{figure}
In the above section \ref{sec:p2ps}, we have treated an edge in a graph as a probabilistic photon-pair source (in general, as the correlated photon pair) and a vertex for an optical path connecting to a single-photon (non)sensitive detector. Experimentally, one can construct a single-photon source that relies on a probabilistic photon-pair source, where one photon is detected heralds the presence of the other \cite{singlephotons2011}. For example, the single-photon source in path $\text{SPS}_{d}$ ($\text{SPS}_{c}$) is conditionally on the detection of only one photon in the path $d$ ($c$) using a probabilistic photon-pair source and a photon-number-resolving detector, as shown in Fig.~\ref{fig:single_photon}. In the graph representation, one can consider that there is an edge between the vertex $\text{SPS}_{d}$ ($\text{SPS}_{c}$) and a virtual vertex which always needs to have exactly one incoming edge. For simplicity, we can just use a input vertex (e.g., $\text{SPS}_{d}$ or $\text{SPS}_{c}$, without drawing its adjacent virtual vertex) for such a single-photon source based on a probabilistic photon-pair source, see Fig.~\ref{fig:single_photon}. There are no connecting edges between two input vertices. The edge connecting to an input vertex is reinterpreted as photons propagate towards a detector \cite{theseus,chin2021graph}. 

Interestingly, this can also be explained in Klyshko's advanced-wave picture, described as ``an intuitive treatment of two-photon correlation with the help of the concept of an effective field acting upon one of the two detectors and formed by parametric conversion of the advanced wave emitted by the second detector'' \cite{belinskii1994two,arruda2018klyshko}. In other words, we can treat an edge as a quantum information transfer of one single system; thus, one can describe many abstract quantum information flows instead of just probabilistic photon-pair sources. In general, our input vertex representation is directly applicable to deterministic single-photon sources such as the ideal on-demand sources based on semiconductor quantum dots \cite{wang2019towards,arakawa2020progress,tomm2021bright,uppu2021quantum} or multiplexed single-photon sources \cite{meyer2020single}. Therefore, \pytheus can design experiments that use single-photon sources.

\subsubsection{Mixed States}\label{sec:mixed}
So far, we have considered coherent creation processes, resulting in pure states. However, experimentally it is possible to introduce incoherence. Whenever distinguishing information about the quantum state escapes from the experiment to the environment, a partial trace of the density matrix is performed, producing a mixed state. This fact can be used systematically in the graph representation. To do so, we dedicate one of the vertices as the environment (environment vertex). We produce the full quantum state
\begin{equation}
    \ket{\psi} = \ket{\psi_0}\ket{0}_{\text{env}}+\ket{\psi_1}\ket{1}_{\text{env}}+\ket{\psi_2}\ket{2}_{\text{env}}+\ldots.
\end{equation}
Then, tracing out the environment contribution, we obtain
\begin{equation}
    \rho = \ket{\psi_0}\bra{\psi_0}+\ket{\psi_1}\bra{\psi_1}+\ket{\psi_2}\bra{\psi_2}+\ldots .
\end{equation} 
\begin{figure}[t]
	\centering
	\includegraphics[width=0.48\textwidth]{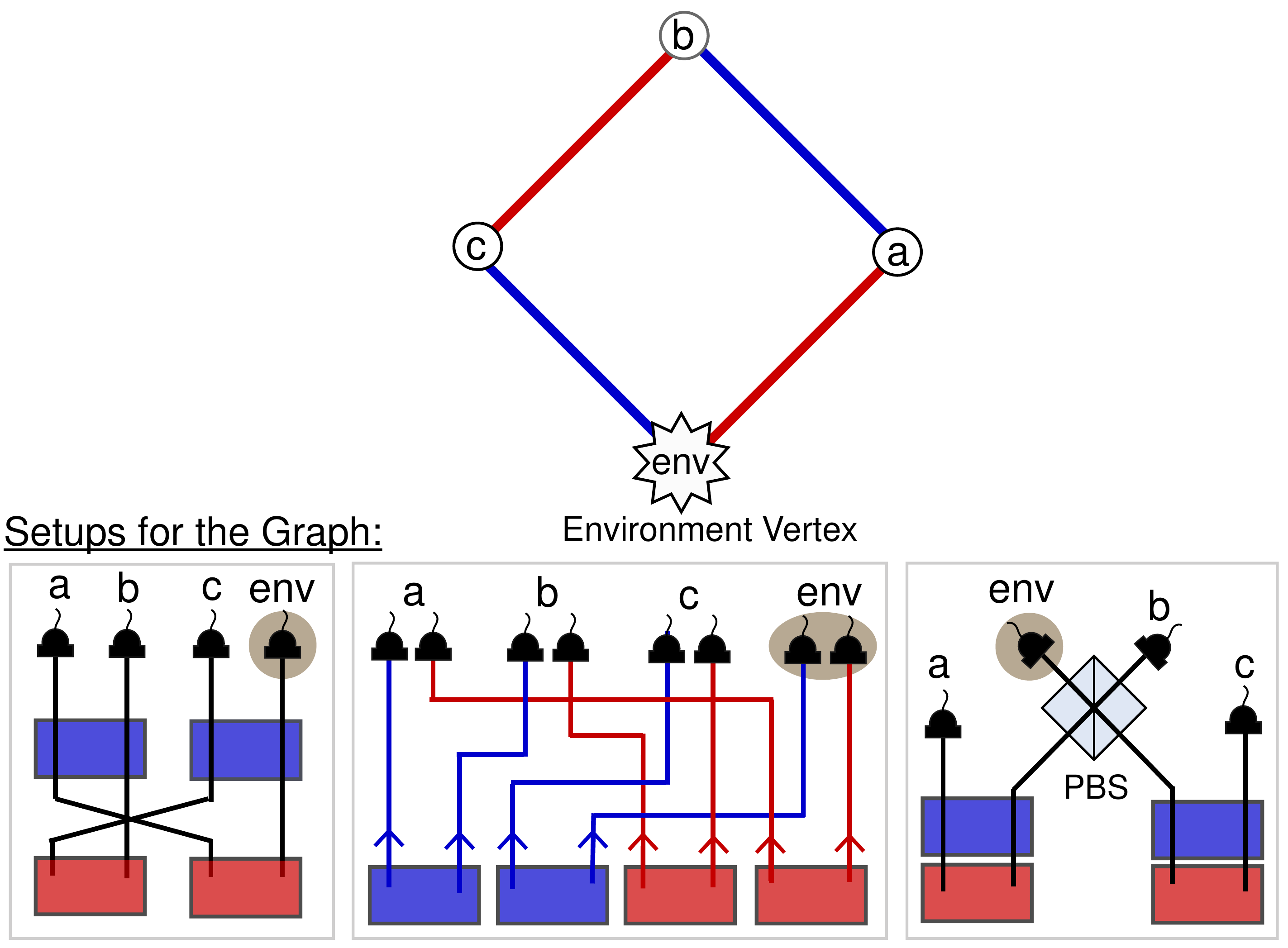} 
	\caption {An abstract graph for producing mixed states. The detection paths were labeled $a$, $b$, $c$, and $\text{env}$, and internal modes described in blue and red are mode numbers 0 and 1. A mixed state can be generated by using one of the vertices as the environment (vertex env in this example). Experimentally this could be implemented by measurements that do not discriminate between the photon modes, for instance, detectors with a large active area. The corresponding setups for the graph are listed below.}
	\label{fig:IntroMixedGHZ}
\end{figure}

We illustrate this procedure by a simple example in Fig.~\ref{fig:IntroMixedGHZ}. From a graph that produces
\begin{equation}
    \ket{\psi} =  \ket{000}_{abc}\ket{0}_\text{env}+\ket{111}_{abc}\ket{1}_\text{env},
\end{equation}
we trace out the last photon obtaining
\begin{equation}
    \rho_{000111} = \ket{000}\bra{000}+\ket{111}\bra{111}.
    \label{eqIntroMixedGHZ}
\end{equation}

This procedure is experimentally achieved by using detectors that do not distinguish between the modes in vertex $\text{env}$. For example, if spatial modes are used, one can use a wide-area photon detector or multi-mode fibers in front of the detector. If the time degree of freedom is used to encode the mode numbers, one can integrate over the entire time such that one cannot distinguish between time-bins. Doing so, we introduce mixedness and recover the state in Eq.~\eqref{eqIntroMixedGHZ}. Similarly to the translation in \ref{sec:translation}, we can also translate the graph for creating mixed states into different schemes, as shown in Fig.~\ref{fig:IntroMixedGHZ}. Therefore, \pytheus can be used to design experiments for complex and interesting mixed states.

\subsubsection{States Entangled in the Photon-Number Basis}\label{fock_basis}

Fock states containing a fixed number of particles in a given spatial mode form a complete basis for many-body Hilbert spaces. The superposition of Fock states brings, among others, the well-known N00N states \cite{Sander1989,lee2002quantum} that promise many advantages such as the best-possible quantum-enhanced precision, super-sensitivity and super-resolution \cite{giovannetti2011advances, polino2020photonic}. The $N$-particle N00N state is (up to normalization)
\begin{equation}
\ket{\text{N00N}}^{N}_{2}:=\ket{N,0}_{a,b}+\ket{0,N}_{a,b},
\label{eq:NOON}
\end{equation}
where the subscript number $2$ means there are two paths (e.g., $a$ and $b$), $\ket{0}$ indicates an unoccupied mode. The concept of the N00N state in Eq.~\eqref{eq:NOON} can be extended to a multi-mode case, where $N$ particles are distributed in one of several optical paths \cite{zhang2018scalable,hong2021quantum}.

We know the total number of photons $N$ going through a set of optical paths, but in contrast to the previous sections, we ignore how many photons occupy each of the paths. Therefore, perfect matchings are not the only contributions to the final state that we must consider. Here, we need to compute all combinations of edges that lead to a total number of $N$ photons for a given set of all optical paths. 
This includes combinations with repeated edges, which represent multiple photon pairs from the same source, and self-loops, in which a source produces a collinear photon pair \cite{PhysRevA.60.R4209,burlakov2001collinear}. Moreover, for some contributions there can be vertices with degree zero, as long as the total number of photons -- the total degree -- is $N$. The conditioning on the total number of photons must be considered when employing the N00N states in metrology experiments, as described in previous work \cite{polino2020photonic, afek2010high}. 
\begin{figure}[!t]
	\centering
	\includegraphics[width=0.35\textwidth]{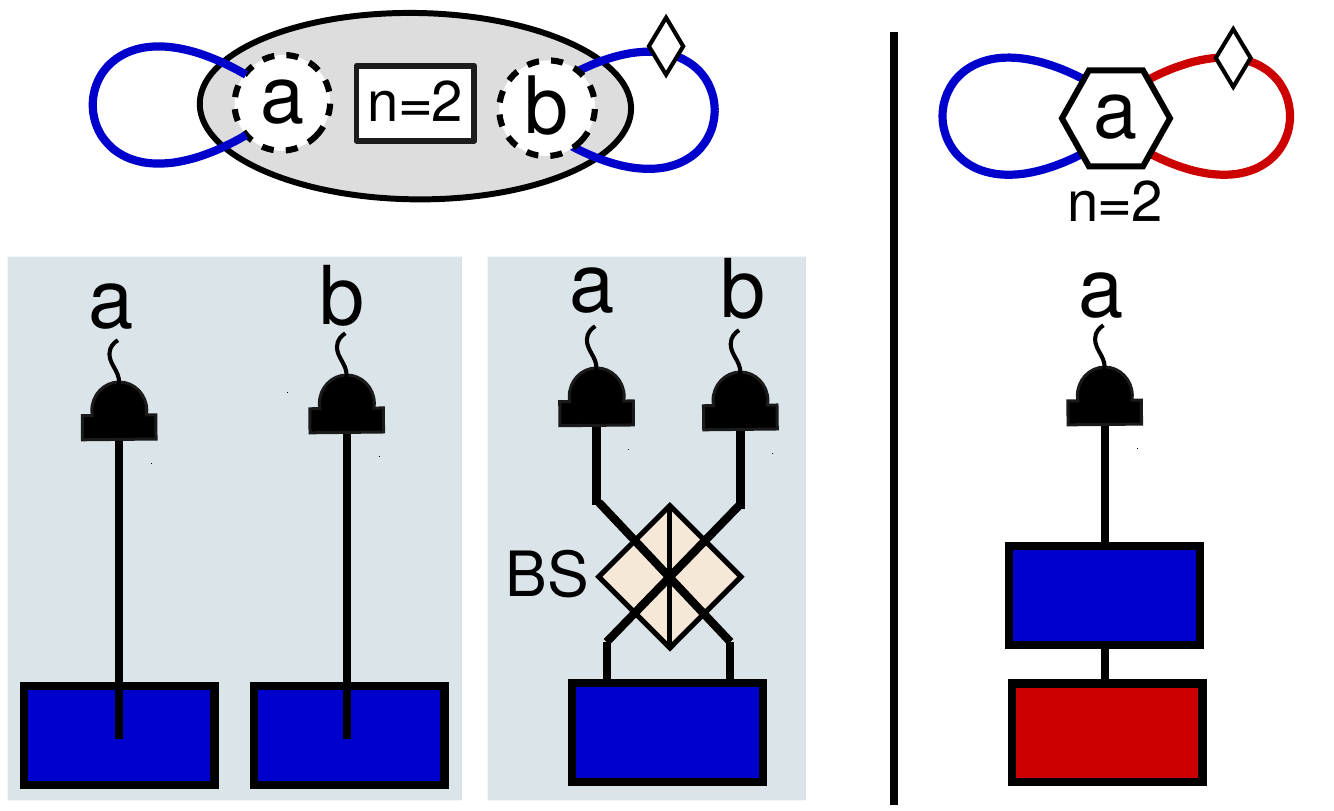}
	\caption {Graphs for producing $\ket{\text{N00N}}^{2}_{2}$ states and the related experiments. On the left side, the graph with two blue edges (i.e., two self-loops) is for a path-entangled N00N state $(\ket{2,0}_{a,b}-\ket{0,2}_{a,b})/\sqrt{2}$. We translate this graph into several setups (below the graph). The setups using path identity or path encoding are the same in this case. With bulk optics, one can achieve the state with a Hong-Ou-Mandel interferometry \cite{HOM1987,bouchard2020two}, where two identical photons enter a 50:50 beam splitter. One can also perform such a state in polarization, i.e., $(\ket{2,0}_{H,V}-\ket{0,2}_{V,H})/\sqrt{2}$ on the right side. The blue and red colors are, respectively, for horizontal and vertical polarization. The coherent superposition of blue and red self-loops gives the N00N state. Its related setup is described below the graph. There either two horizontally polarized photons (blue edge) or two vertically polarized photons (red edge) are in path $a$ with a single-photon sensitive detector.}
	\label{fig:HOMeffect}
\end{figure}

As an example, we show a graph for the $\ket{\text{N00N}}^{2}_{2}$ state in Fig.~\ref{fig:HOMeffect}. In this case, the total photon number is $N=2$. All edges are in the same color, thus corresponding to the standard path-entangled N00N states. A self-loop edge indicates that there are two photons in its connected vertex. Thus the coherent superposition of two photons being in one of the two vertices (one of the self-loops produced a photon-pair, but we ignore it) leads to a coherent superposition of $\ket{2,0}_{a,b}$ and $\ket{0,2}_{a,b}$, i.e., a two-mode two-photon N00N state. One can now translate the graph into quantum experiments, similarly, as we did in section \ref{sec:p2ps}. In this example, the setups using entanglement by path identity and path encoding are the same. With bulk optics, one can use probabilistic beam splitters that provide the mixing between the two input photons, to bunch two photons in one of the two optical paths $a$ and $b$, which is the well-known Hong-Ou-Mandel (HOM) effect \cite{bouchard2020two,HOM1987}. Moreover, we can also generate such a N00N state in polarization or transverse spatial modes instead of the path; see Fig.~\ref{fig:HOMeffect} for details.  

\begin{figure}[!t]
\centering
	\includegraphics[width=0.45\textwidth]{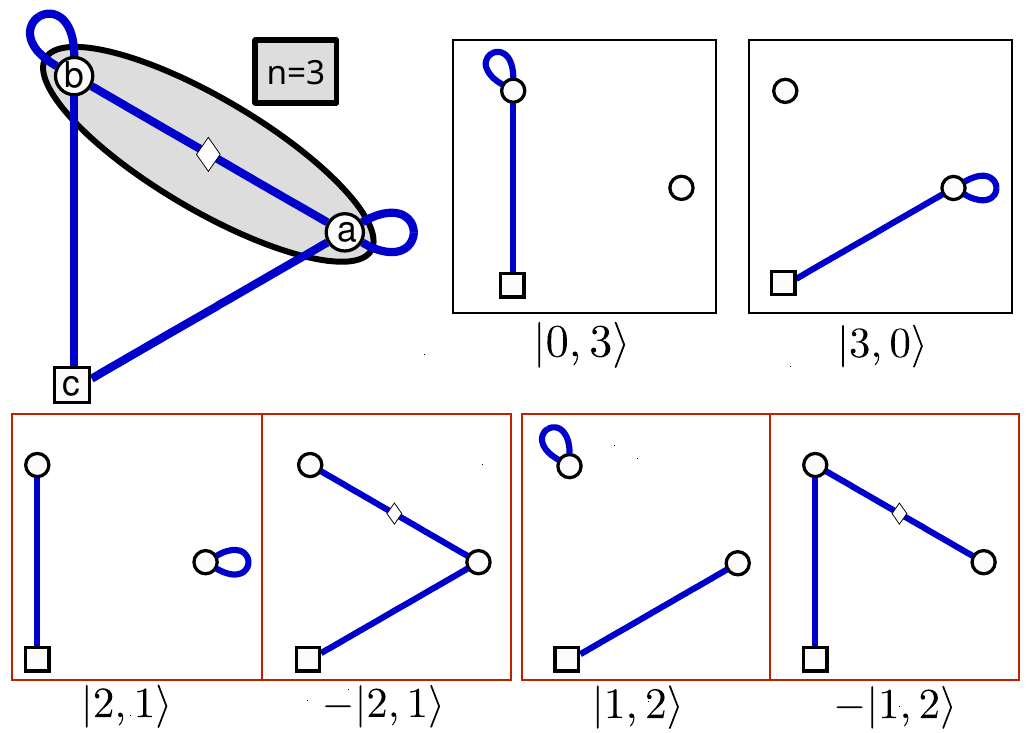} 
 \caption{Graph for producing $\ket{\text{N00N}}^{3}_{2}$ state with an additional ancillary particle. The total photon number (including ancillary one) for each component in the state is four. There are six combinations of two edges that cover the ancillary vertex only once. Two of them (black box) contribute to the required terms $\ket{3,0}$ and $\ket{0,3}$, and the others ($\ket{2,1}$ and $\ket{1,2}$) inside the red box cancel out.}
 \label{fig:noon_2mode_3ph}
\end{figure}

As for the previous states, ancillary paths can assist in the creation of states on the photon-number basis. Here we condition the final state on the existence of a total number of photons in the non-ancilla detectors but, again, the ancilla paths are reached by single photons. Fig.~\ref{fig:noon_2mode_3ph} shows an example in which the ancilla (vertex 3) receives only one photon, and the other two detectors get a total of three photons.

\subsection{Quantum Communication}\label{sec:qcommunication}
\begin{figure}[!t]
	\centering
	\includegraphics[width=0.4\textwidth]{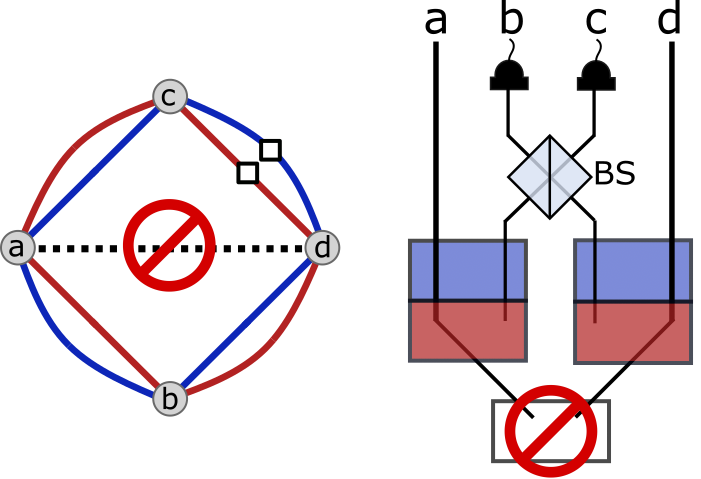} 
	\caption {An abstract graph for entanglement swapping. On the right, an experimental setup for entanglement swapping is shown. The two particles a and d are entangled when two ancillary detectors click. A common source of particles a and d (as shown crossed out below the setup) would circumvent the point of entanglement swapping. The two particles are to be entangled without having interacted with each other. This restriction manifests in the graph corresponding to the experiment (shown on the left). An edge between the vertices a and d (dashed line) is not permitted.}
	\label{fig:qcomm}
\end{figure}
In future quantum networks that connect individual users with quantum resources and quantum computers, the distribution of quantum entanglement is essential. One key concept used in quantum communication in quantum networks is entanglement swapping \cite{zukowski1993event,pan1998experimental}, where two qubits that never interacted can be entangled. 
This provides an important route for interesting tests of quantum foundations and plays an indispensable role in quantum technology such as quantum repeaters \cite{sangouard2011quantum}.  \pytheus allows us to explore new directions that could lead to solutions that require fewer resources than current techniques or are implemented in surprising ways. Many quantum network communication tasks have been experimentally implemented \cite{pan1998experimental,basset2019entanglement,llewellyn2020chip,samara2021entanglement}, thus the new solutions by \pytheus can readily be implemented in laboratories.

We are interested in finding experimental setups for creating entanglement between particles that have not interacted or originated from a common source. With \pytheus this can be done for scenarios involving higher dimensions and multiple particles. Finding the corresponding graph works analogously to the state generation task shown in section \ref{sec:stategeneration}, but comes with extra constraints on the graph. We set an entangled state (e.g. two particles in a Bell state) as a target for the optimization. The experiment corresponding to the resulting graph should create this state. To ensure that the two particles have no direct interaction, the graph must fulfill the following additional constraint. It should have no edge connecting the two corresponding vertices. Such an edge translated to a photon pair source would mean that the two photons could come out of the same source. This is shown in Fig.~\ref{fig:qcomm}. Similarly, the connection of two vertices to the same single-photon source vertex implies that they can not be space-like separated. Such constellations would not be valid for entanglement swapping, either. In the same way, as for incoming photons (single photon sources or input photons), the constraints are enforced by removing the edges from the starting graph.

Approaching these tasks with \pytheus makes it possible to explore new ways of distributing entanglement. We show the examples found by \pytheus in section \ref{100exp}.

\subsection{Quantum Measurements}\label{sec:measurements}
A graph may also be interpreted as a quantum measurement on an input quantum state. Such measurements are of utmost importance in many quantum communication tasks, for example, a Bell state measurement for quantum teleportation. The goal of measurements here is to distinguish different orthogonal incoming states, e.g., in the case when all detectors click simultaneously. Photons entering the experiment through input paths are represented analogously to single-photon sources introduced in section \ref{sec:single_photon}. An input photon corresponds to one vertex in the graph. An edge connecting an input vertex to a detector vertex corresponds to the photon traveling to that detector, following Klyshko's advanced-wave picture \cite{belinskii1994two} as explained in section \ref{sec:single_photon}. The same constraints apply, excluding edges connecting two vertices belonging to input photons.
\begin{figure}[t]
	\centering
	\includegraphics[width=0.35\textwidth]{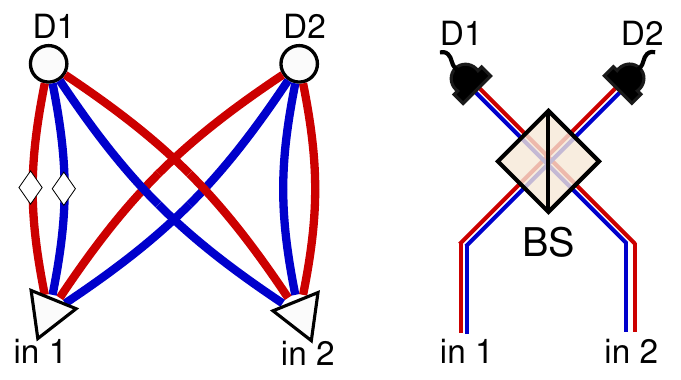} 
	\caption {An abstract graph for Bell state measurement. A measurement event happens if the photons \textbf{in1} and \textbf{in2} are in the Bell state $\ket{\Psi^-}$. Simultaneous clicks in the detectors $D1$ and $D2$ signify a successful Bell measurement. The corresponding setup is given for the graph, which is a typical Bell state measurement setup.}
	\label{fig:GraMeasureBellMeas2D}
\end{figure}

As an example, we show a graph in Fig.~\ref{fig:GraMeasureBellMeas2D} for the well-known Bell state measurement (BSM) \cite{weinfurter1994experimental, michler1996interferometric}, which plays an important role in many quantum information tasks. 
When the two detectors click, the incoming photons are projected to one Bell state $\ket{\Psi^-}$, which can be seen as the generation of a state given by a superposition of the two perfect matchings in the graph. To search for an experimental setup for measuring a particular state, we let \pytheus search for a graph that would produce the state under consideration of the topological constraints on the graph. The graph in Fig.~\ref{fig:GraMeasureBellMeas2D} interpreted as state generation would produce the state $\ket{\Psi^-}$ under the condition that there is no connection between vertices in1 and in2. 

With a graph that `produces' the state $ \ket{\varphi}_{in}$, we can effectively perform a projective measurement ($P_0=\ketbra{\varphi}$ and $P_1=\mathbb{I} - \ketbra{\varphi}$) on the input photons. A coincidence of all detectors corresponds to the output 0 and any other pattern is interpreted as output 1.

The representation of measurements in terms of graphs constitutes an extension of the previous interpretation given in \cite{theseus}. This allows us to use \pytheus for the discovery of measurement setups for any state and various constraints and conditions.

\subsection{Quantum Computation}\label{sec:gates}
Quantum gates are a crucial building block in quantum computation \cite{nielsen2010quantum,knill2001scheme}. A quantum gate performs a unitary transformation on an input state. These transformations can also be interpreted in terms of graphs. In these graphs, similar to measurements (described in subsection \ref{sec:measurements}), the input photons are described by designated vertices. Likewise, we impose restrictions on which types of vertices can be connected by edges. Photons exiting the setup (output) are also represented by vertices, as in state generation. Additional ancillary photons can stem from probabilistic photon-pair sources and single-photon sources can also be involved. With \pytheus we can search for arbitrary quantum gates under a wide range of experimental conditions.

\begin{figure}[!t]
	\centering
	\includegraphics[width=0.5\textwidth]{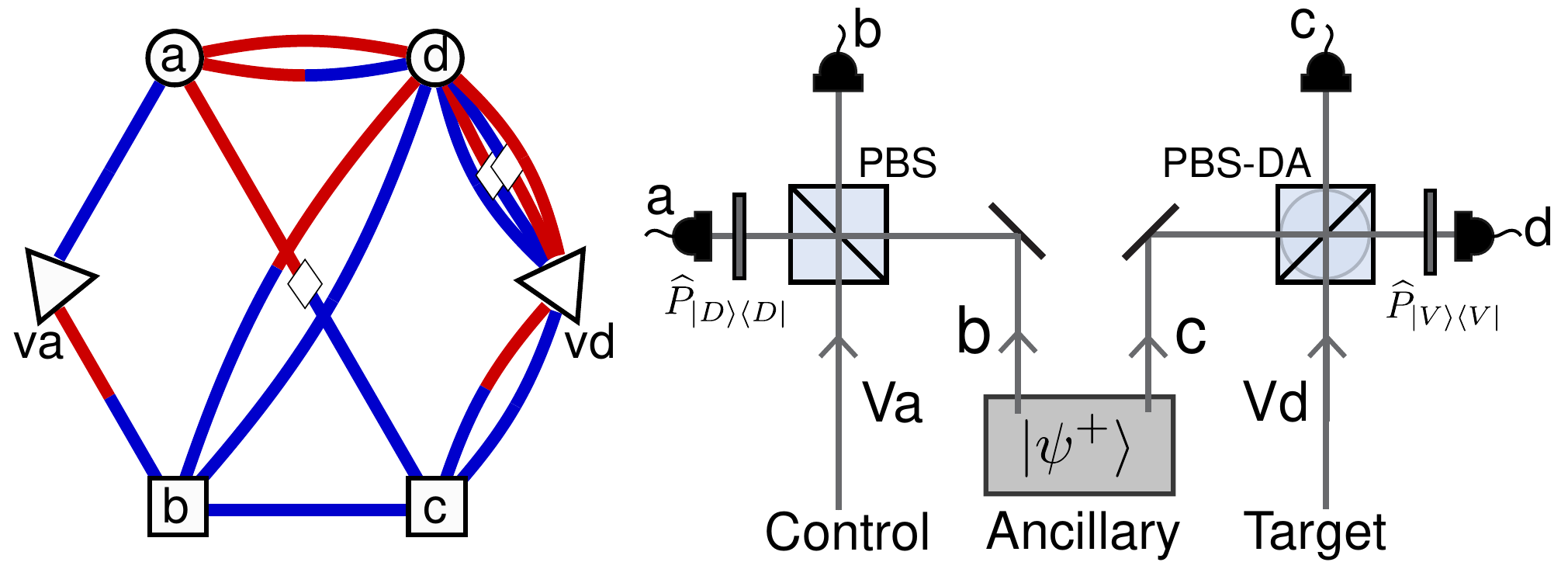} 
	\caption {An abstract graph for a quantum CNOT gate shown on the left side. The right panels show the corresponding setup for a CNOT gate, which was realized by \cite{gasparoni2004realization}.}
	\label{fig:qcomp}
\end{figure}

A target unitary quantum gate acting on an $N$-dimensional Hilbert space is specified by how each element of an orthonormal basis transforms. One canonical example in quantum gates is the controlled-NOT (CNOT) gate, which acts on two qubits and is described by the mapping:
\begin{align*}
    \ket{00} \rightarrow \ket{00}, \\
    \ket{01} \rightarrow \ket{01}, \\
    \ket{10} \rightarrow \ket{11}, \\
    \ket{11} \rightarrow \ket{10}.
\end{align*}
To find this gate, we let \pytheus search for a graph that would produce the state 
\begin{equation}
    \ket{00}\ket{00} + \ket{01}\ket{01} + \ket{10}\ket{11} + \ket{11}\ket{10};
\end{equation}
under consideration of the topological constraints on the graph (no connections between input photons). In Fig. \ref{fig:qcomp}, we show an experimental setup realized \cite{gasparoni2004realization} together with its graph representation.

Photonic quantum gates fall into two main categories \cite{RevModPhys.79.135}. In a post-selected gate, the outgoing photons are detected directly after the gate. This ensures that there is exactly one photon in each path after the gate \cite{postselectedgates,Knill2002,ralphcnot,CNOTfirst2003,Langford2005,q_mem_advantage,fredkin16,MetrologyHeraldedGate}. Because possibilities where two outgoing photons enter the same path are excluded, fewer experimental resources are required. However, this procedure destroys the outgoing state, and the particles can not be used further. The second category is heralded gates \cite{gasparoni2004realization,PhysRevLett.94.030501,bao2007optical,gao2010teleportation,okamoto2011realization,spscnot,Zeuner2018}, where only ancillary photons are detected, and the outgoing photons of the gate remain undetected. These gates are ultimately more useful since the output state can be used further. A graph for a heralded quantum gate is harder to discover and needs more experimental resources since the looser selection rules can lead to more unwanted terms. Both types of quantum gates (post-selected and heralded) can be represented with graphs and thereby be designed using \pytheus. Terms produced in post-selected experiments are represented by perfect matching, and terms produced in heralded experiments are represented by collections of edges that cover the ancillary vertices.

The ability to perform design for quantum measurements and quantum gates relies on an idea that state generation and state propagation are closely related \cite{theseus}. In the experimental quantum optics community, this idea is known as Klyshko’s advanced-wave picture \cite{belinskii1994two, aspden2014experimental}, while in the quantum information science community it is often referred as Choi–Jamiołkowski isomorphism \cite{jiang2013channel}.

\begin{figure*}[!t]
	\centering
	\includegraphics[width=.92\textwidth]{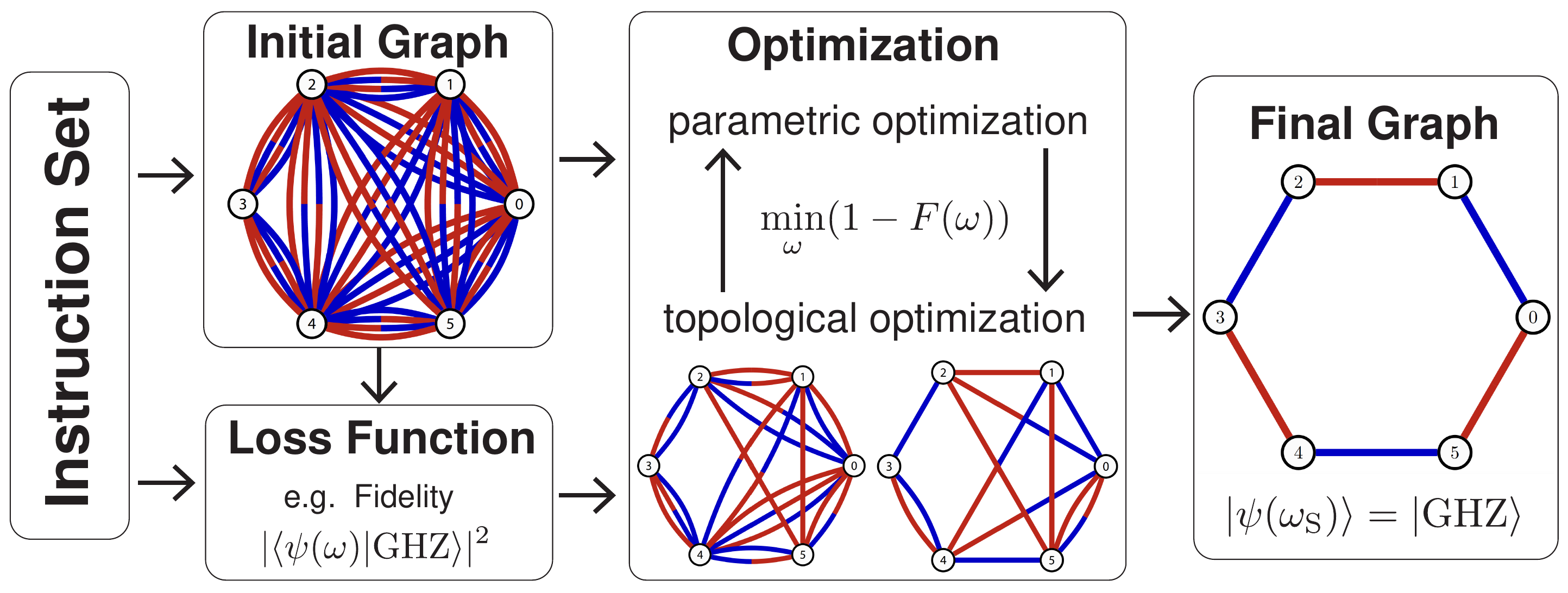}
 \caption {The \pytheus workflow. From the number of vertices and available dimensions, \pytheus builds the initial graph, which is fully connected except for vertex 3, which is only connected to its nearest neighbors. We impose this based on previous knowledge, but other constraints may arise depending on the available tools. Computing all the graph's perfect matchings, we find the more general state that the initial graph can produce: $\ket{\psi(\omega)}$. As the loss function, we choose the \textit{Fidelity} of the state $\ket{\psi}$ with respect the target: $\ket{\text{GHZ}}$. Aiming for an interpretable graph with perfect fidelity, we iterate the weight optimization with the edge removal until we find the simplest graph that produces $\ket{\text{GHZ}}=\ket{\psi(\omega_S)}$. The available resources, the loss function and further optimization details are specified in the \textit{Instruction Set.}}
    \label{fig:workflow}
\end{figure*}
\section{The PyTheus Library}\label{sec:pytheus}
Starting from a dense or fully connected graph, \pytheus uses gradient descent combined with topological optimization to find minimal graphs corresponding to some target quantum experiment. The \pytheus library greatly expands the range of applications of its predecessor, and it is significantly faster \footnote{For an eight-dimensional three-particle highly-entangled state ($\text{SRV}(8,5,3)$) \pytheus was almost five times faster than the current state-of-the-art method \cite{theseus}. Examples of higher particle numbers were not readily available for the previous method, but we expect increasingly higher advantages for \pytheus.}. The package is written in Python and is available on GitHub for further applications and development of the source code. The graph encoding is compatible with other tools such as \textit{The Walrus}\cite{walrus}, a library to (among other applications) compute perfect matchings from the graphs' adjacency matrix. \footnote{The loops' weights in the adjacency matrix must be multiplied by 2.}.

\pytheus applications range from the creation of quantum states to the design of quantum communication protocols. While diverse, all these tasks are performed following the same steps, illustrated in Fig.~\ref{fig:workflow}. In this section, we explain the software workflow, which kind of loss functions we employ, and how to start using PyTheus.

\textbf{Workflow Overview} -- The \textit{Instruction Set} file details \textit{what} we want (e.g., a quantum state or a communication protocol) and some instructions about \textit{how} to get it. The latter includes the entire allowed topology of the final solution, such as the type of photon sources/detectors and ancillary particles, how many of each we have, or which polarization modes can be used. It also contains the \textit{Loss Function} to minimize (fidelity, count rate, or other metrics) as well as further optimization settings. In the example of Fig.~\ref{fig:workflow}, we want to obtain a post-selected six-particle, two-dimensional GHZ state, $\ket{\text{GHZ}}^2_6=\ket{000000}+\ket{111111}$. We employ 6 standard photodetectors and the photon pair sources described in section \ref{sec:p2ps}. In this example, we also specify further topological constraints of the final solution (thus of the initial graph) -- here, vertex 3 must not have connections to vertex 0, 1 and 5. 

The experimentally available topology specified in the instruction set lead to the \textit{Initial Graph}. This weighted graph represents all possible states/experiments that can be produced with the available resources. The number of vertices, the type of edges, or which vertices can be directly connected are some of the \textit{Topological Constraints} that one can impose on the graph or that may follow from the available tools or from the task we want to implement -- the sources described in section \ref{sec:single_photon} and the protocols from section \ref{sec:qcommunication} are representing examples. On the initial graph of Fig.~\ref{fig:workflow} the six vertices are connected with bicolored edges and based on previous knowledge, we only connect vertex 3 with its nearest neighbors. 

From the initial graph, $\mathcal{G}(\omega)$, we can compute the more general state produced by the weight function $\Phi(\omega)$ acting on the vacuum (see Eq.~\eqref{eq:pairsource}). However, we do not consider the infinite terms produced by $\Phi(\omega)$, we only compute the creation events which fulfill certain \textit{Conditional rules}. For example, in Fig.~\ref{fig:workflow} the post-selection rule results in terms for which all detectors click. These creation events are represented by the graph's perfect matchings, and produce a coherent superposition of kets leading to the state $\ket{\psi(\omega)}$. We can specify more complex conditioning rules, which will be relevant, for example, for N00N states. There, we condition on states with a total final photon number in the output nodes. The conditioning rules are imposed by experimental circumstances \ref{section:graphs}. 

Once we extract a general state $\ket{\psi(\omega)}$ from the initial graph, we can optimize it according to some \textit{Loss Function}. As in  Fig.~\ref{fig:workflow} we seek a particular state, $\ket{\text{GHZ}}$, we maximize the \textit{Fidelity} of our state accordingly: $F=|\braket{\text{GHZ}}{\psi(\omega)}|^2$. To find a specific states we can also use the \textit{Count Rate}, a metric that approximates how often our experiment produces such a state. Alternatively, we can maximize physical properties like entanglement rather than looking for a specific state. To perform quantum information tasks like measurements or communication protocols, we must translate each task into a state creation process -- see sections \ref{sec:measurements}, \ref{sec:qcommunication}, and \ref{sec:gates} for further details.

If the available experimental resources suffice, by optimizing the weights, we will find a graph that satisfies our needs. However, the optimal solution is not unique, and even if many of the weights vanish, we will likely obtain a very dense final graph, which will be hard to interpret. Therefore, to extract useful insights from our setups, we must simplify the graphs as much as possible, alternating the optimization of the weights' values with the removal of edges. This \textit{Topological Optimization} will lead us to a simple graph, for which the removal of any additional edge would unacceptably raise the loss function's value.

For some graphs, like the one shown in Fig.~\ref{fig:workflow}, no smaller graph can produce the target state \cite{graphs1}. However, in most cases, it is challenging to ascertain whether there exists a simpler solution, especially when we choose to optimize for real weights instead of complex ones. With respect to these subtleties, following the workflow, we first generate graphs with abundant ancillary particles and reassess the results with the goal of reducing the number of resources. Using this procedure, we propose one hundred experiments to be realized in quantum optics.



\begin{table*}[!t]
    \centering\resizebox{\textwidth}{!}{%
\begin{tabular}{|cllcclclccll|}
\hline
\multicolumn{4}{|c}{\hspace{1cm}\subfloat[4-dimensional 4-photon GHZ state (asymptotic and exact solutions).]{\includegraphics[width=0.24\textwidth]{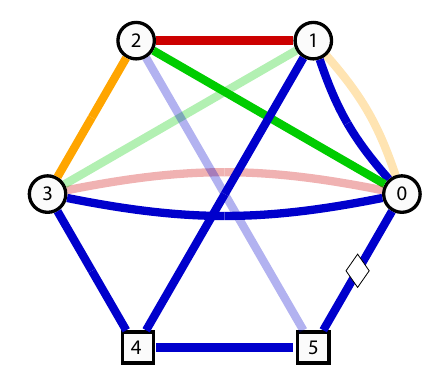}}}         
& \multicolumn{4}{c}{\hspace{1cm}\subfloat[2-mode 5-photon N00N state $\ket{50}+\ket{05}$ (symmetric graph with an inscribed pentagram).]{\includegraphics[width=0.21\textwidth]{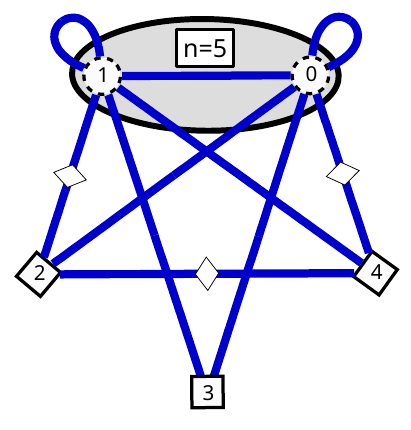}}}        
& \multicolumn{4}{c|}{\subfloat[Post-selected CNOT quantum gate acting on a qutrit target with two ancilla photons]{\includegraphics[width=0.24\textwidth]{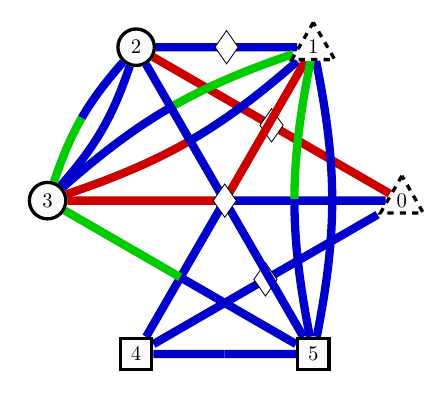}}}         \\
\multicolumn{3}{|c}{\subfloat[A 4-qubit entangled state which requires weights with the golden ratio (or complex-valued) to be generated.]{\includegraphics[width=0.23\textwidth]{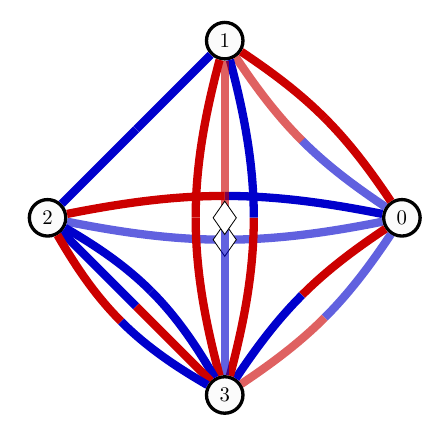}}} 
& \multicolumn{3}{c}{\subfloat[Quantum measurement for a quantum communication task with quantum advantage (Mean King's Problem).]{\includegraphics[width=0.23\textwidth]{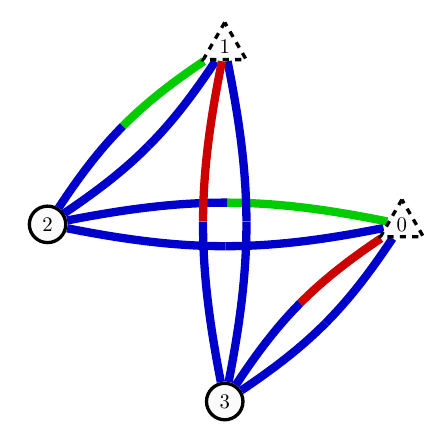}}} & \multicolumn{3}{c}{\subfloat[Entangling photons that never interacted (related to entanglement swapping) without initial Bell states.]{\includegraphics[width=0.23\textwidth]{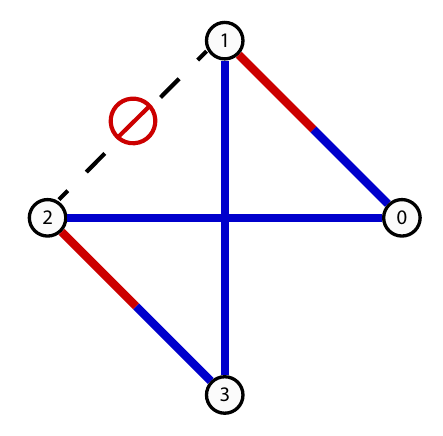}}} & \multicolumn{3}{c|}{\subfloat[Mixed state with bound entanglement that violates Bell inequality (counterexample to the Peres conjecture from 1999, solved 2014).]{\includegraphics[width=0.23\textwidth]{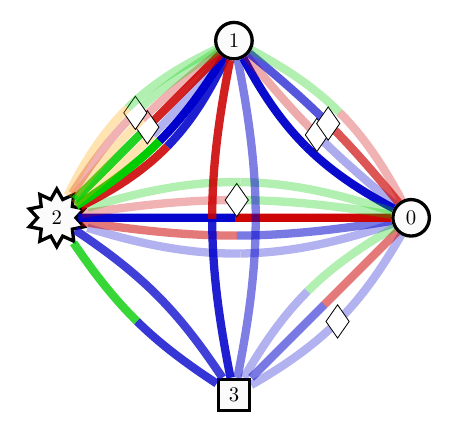}}} \\  \hline
\end{tabular}%
}
    \caption{A collection of seven diverse highlights, each of which is interesting in its own right.}
    \label{fig:highlights}
\end{table*}

\section{Hundred Experiments}\label{100exp}

In the following, we introduce one hundred experiments conceived by \pytheus. A broad catalog of designs that, hopefully, will contribute to multiple branches of quantum optics, including quantum computing, communications, and quantum sensing. For each proposed experimental setup which is new and has not been described in any theoretical or experimental paper, we mark the corresponding graph with a \numberbox[https://github.com/artificial-scientist-lab/PyTheus/]{number in a box} counting from 1 to 100. The weights of each graph can be found in the GitHub repository, together with the instruction sets used to search them.  

We start by showing seven highlights from our discoveries in Table.~\ref{fig:highlights}.
\begin{enumerate}[(a)]
  \item A multi-photon entangled state that goes beyond the barrier of 3-dimensions \cite{graphs2}. It requires only two ancillas and has a small number of edges which makes it a very promising proposal for practical application, and observation of new properties at the foundation of quantum mechanics \cite{lawrence2014rotational}. 
  \item The generation of a two-mode N00N state with 5 photons. The associated graph, very symmetric, makes use of 3 ancillas.
  \item A post-selected 3-photon control gate (or Toffoli) that does not require any ancilla photons.
  \item A 4-photon qubit quantum state with equal coefficients $c=\frac{1}{\sqrt{7}}$, that requires the ratio between certain weights to be the golden ratio. Alternatively, a lower number of complex-weighted edges can also lead to such state.
  \item A previously unknown quantum measurement scheme that allows the experimental implementation of a quantum communication protocol, the Mean King problem, proposed in 1987 \cite{vaidman1987}.
  \item A very surprising form of quantum entanglement swapping, which does not rely on the generation of two Bell pairs and a Bell state measurement.
  \item Experimental setup of a 2-photon mixed entangled state that falsified the Peres conjecture \cite{peres1999all}. The state is bound entangled (its entanglement cannot be distilled), however, it can be used to violate Bell's inequality \cite{moroder2014steering,vertesi2014disproving}. This could lead to an experimental falsification of the Peres conjecture.
\end{enumerate}

\subsection{Generation of Entangled States}
In this part, we propose ways to generate entangled states, which play an important role not only in our understanding of entanglement and the non-local nature of quantum mechanics but also in many quantum information applications.

\tocless\subsubsection{GHZ States}
\begin{figure}[!t]
	\centering
	\subfloat[$\ket{\text{GHZ}}^{4}_{3}$]{\includegraphics[width=0.24\textwidth]{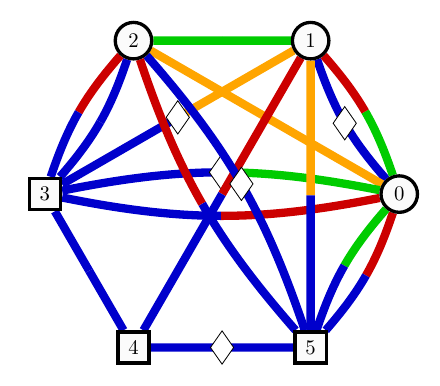}}\vspace{-5mm}\\
	\subfloat[$\ket{\text{GHZ}}^{5}_{3}$]{\includegraphics[width=0.24\textwidth]{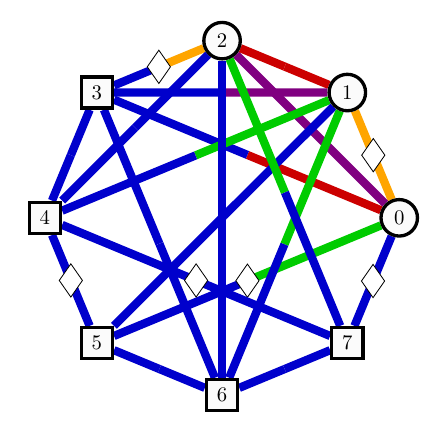}}\!\!
	\subfloat[$\ket{\text{GHZ}}^{6}_{3}$]{\includegraphics[width=0.24\textwidth]{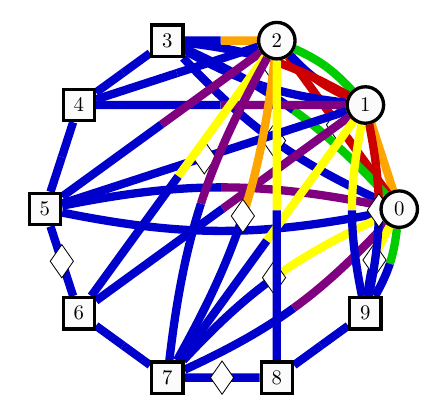}}\\
	\subfloat[Asymptotic $\ket{\text{GHZ}}^{4}_{4}$]{\includegraphics[width=0.24\textwidth]{graph_ghz_f446_boosted.pdf}}\!\!
	\subfloat[Exact $\ket{\text{GHZ}}^{4}_{4}$]{\includegraphics[width=0.24\textwidth]{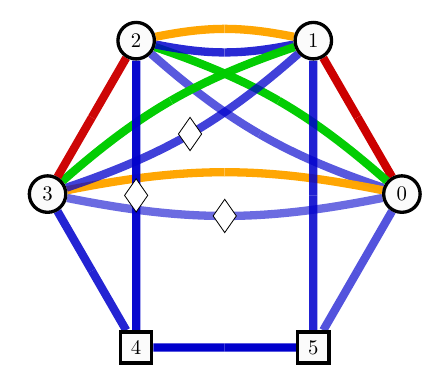}}\\
	\subfloat[$\ket{\text{GHZ}}^{3}_{5}$]{\includegraphics[width=0.24\textwidth]{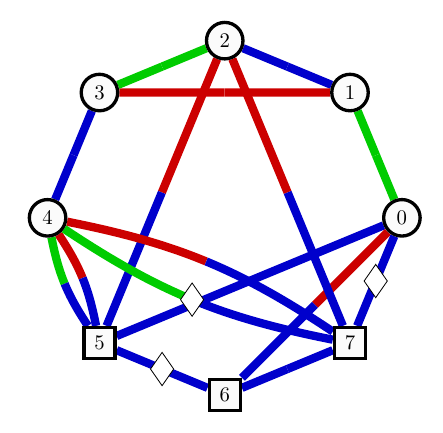}}\!\!
	\subfloat[$\ket{\text{GHZ}}^{4}_{5}$]{\includegraphics[width=0.24\textwidth]{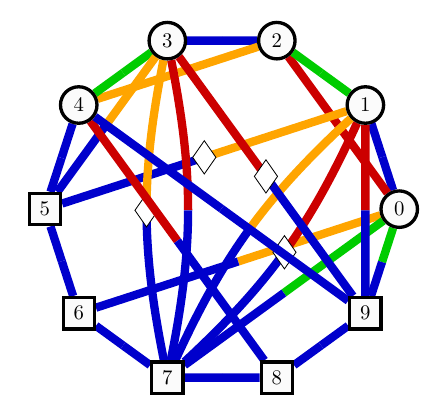}}
	\caption {High-dimensional multiparticle GHZ states with ancillas. The weights here in these graphs are real numbers and the diamonds inserted in the edges represent the negative sign. The vertices in square refer to the detection of an ancillary photon. For the state $\ket{\text{GHZ}}^{4}_{4}$ we have two possible solutions.}
	\label{fig:GHZGraph}
\end{figure}
Bell's theorem shows that Einstein-Podolsky-Rosen's propositions about local realism \cite{epr1935} are inconsistent when we apply them to quantum systems of two particles, revealed by the violation of the Bell inequality with quantum mechanical statistical correlations \cite{Belltheorem}. In the late 1980s, Greenberger, Horne, and Zeilinger took a step further from two to three particles. Interestingly, with a tripartite quantum system, local realism can be violated by quantum mechanics with perfect correlations rather than statistical correlation, and one can completely dispense with inequalities \cite{ghz, greenberger1990bell}. It enables the performance of experiments where the \textit{quantum physical predictions are mutually contradictory with expectations based on local realism} \cite{pan2000experimental}. Such a GHZ state forms an important class of entanglement and its generalization is given as
\begin{equation}
\label{eq:ghz}
	\ket{\text{GHZ}}_{n}^{d}=\frac{1}{\sqrt{d}}\sum_{i=0}^{d-1}\ket{i}^{\otimes n}, 
\end{equation}
where $n$ denotes the number of particles and $d$ is the dimension for each particle. 

Increasing the number of particles and dimensions in the GHZ states not only enlarges the Hilbert space  but also leads to many new exciting classically paradoxical phenomena \cite{ryu2014multisetting, lawrence2014rotational,lawrence2019many}. Besides their fundamental interest, high-dimensional multiparticle GHZ states have also served as an important resource for many quantum information applications \cite{erhard2020review}.

With the size and dimension increasing, it is very challenging to know how to experimentally create GHZ states. Experimental progress has been made in this direction to push the size and dimension, especially in the linear optics regime \cite{tenghzprl, tenghzoptica, 18qubitghz,erhard3dghz}. Until now, only the three-particle three-dimensional GHZ state has been demonstrated with linear optics \cite{erhard3dghz} and superconducting qutrits \cite{cervera2022experimental}. Going beyond the $\ket{\text{GHZ}}_{3}^{3}$ state with arbitrary particle numbers and higher dimensions will give rise to exciting new possible quantum applications and new insights on the foundations of quantum mechanics.
This remains to be explored, in any experimental platform. 

\pytheus enables the discovery of many new multi-particle high-dimensional GHZ states for which no experimental implementations are known yet. That includes the 3-particle GHZ state for 4, 5 and 6 dimensions \numbering[https://github.com/artificial-scientist-lab/PyTheus/tree/main/pytheus/graphs/HighlyEntangledStates/ghz_346]{experimentcounter}\numbering[https://github.com/artificial-scientist-lab/PyTheus/tree/main/pytheus/graphs/HighlyEntangledStates/ghz_358]{experimentcounter}\numbering[https://github.com/artificial-scientist-lab/PyTheus/tree/main/pytheus/graphs/HighlyEntangledStates/ghz_3610]{experimentcounter}, the 2 solutions (asymptotic and exact) for the 4-particle 4-dimension GHZ state \numbering[https://github.com/artificial-scientist-lab/PyTheus/tree/main/pytheus/graphs/HighlyEntangledStates/ghz_asym446]{experimentcounter}\numbering[https://github.com/artificial-scientist-lab/PyTheus/tree/main/pytheus/graphs/HighlyEntangledStates/ghz_446]{experimentcounter}, and the 5-particle GHZ state for 3 and 4 dimensions \numbering[https://github.com/artificial-scientist-lab/PyTheus/tree/main/pytheus/graphs/HighlyEntangledStates/ghz_538]{experimentcounter}\numbering[https://github.com/artificial-scientist-lab/PyTheus/tree/main/pytheus/graphs/HighlyEntangledStates/ghz_5410]{experimentcounter}, as shown in Fig.~\ref{fig:GHZGraph}.

\tocless\subsubsection{Bell Gems}
Bell states offered the first proof of the non-locality of quantum mechanics. Later on, they became the cornerstone of many quantum communication schemes and are nowadays widely used in quantum computing. These maximally entangled bipartite states, described in the Hilbert space $\mathcal{H}_2\otimes\mathcal{H}_2$, can be generalized for more particles and dimensions in multiple ways while keeping their main properties \cite{generalbellstates}. 

\begin{figure}[!h]
	\centering
	\includegraphics[width=0.24\textwidth]{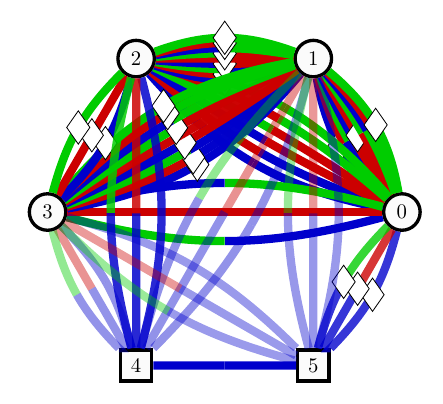}
 \caption {Graph corresponding to the creation of a 3 qutrit Bell gem.}
	\label{fig:gem4}
\end{figure}

In 2004, Gregg Jaeger suggested a generalization for sets of $2^N$ qubits, the \textit{Bell Gems}~\cite{bellgems}. To construct these states we need to pick a pair of orthogonal states $\ket{\alpha}$ and $\ket{\beta}$ from the Bell basis: $\ket{\Phi^\pm}=\ket{00}\pm \ket{11}$ and $\ket{\Psi^\pm}=\ket{01}\pm \ket{10}$. Then, we iterate the following mappings in all possible ways:
\begin{gather}
    \{\ket{\alpha},\ket{\beta}\} \rightarrow \left(\ket{\alpha}\ket{\alpha} \pm \ket{\beta}\ket{\beta}\right) \\
    \{\ket{\alpha},\ket{\beta}\} \rightarrow \left(\ket{\alpha}\ket{\beta} \pm \ket{\beta}\ket{\alpha}\right)
\end{gather}
Therefore, for $2^N$ qubits, we get a basis of $2^{2^N}$ states. For the case of a pair of 3 dimensional Bell states 84 states exist. Fig.~\ref{fig:gem4} shows how to generate one of these high-dimensional Bell gems which, up to normalization, reads \numbering[https://github.com/artificial-scientist-lab/PyTheus/tree/main/pytheus/graphs/HighlyEntangledStates/BellGem3D]{experimentcounter}
\begin{equation}
    \ket{\text{GEM}} = \ket{\psi_a}\ket{\psi_a} + \ket{\psi_b}\ket{\psi_b} + \ket{\psi_c}\ket{\psi_c},
\end{equation}
where $\ket{\psi_a} = \ket{00} + \ket{11} + \ket{22}$, $\ket{\psi_b} = \ket{01} + \ket{12} + \ket{20}$, and $\ket{\psi_c} = \ket{02} + \ket{10} + \ket{21}$.\newline


\tocless\subsubsection{Nine Ways to Entangle Four Qubits}
In 2001, F. Verstraete et.\ al.\ \cite{4qubits9entanglements} introduced a classification of any form of entanglement between 4 qubits into nine categories. Their classification establishes an equivalence relation between states generated by reversible stochastic local quantum operations assisted by classical communication (SLOCC) operations. States inside the same category can perform the same quantum information tasks, albeit with different probabilities.

\begin{figure}[!h]
	\centering
	\subfloat[$\ket{0}^{\otimes4}$]{\includegraphics[width=0.165\textwidth]{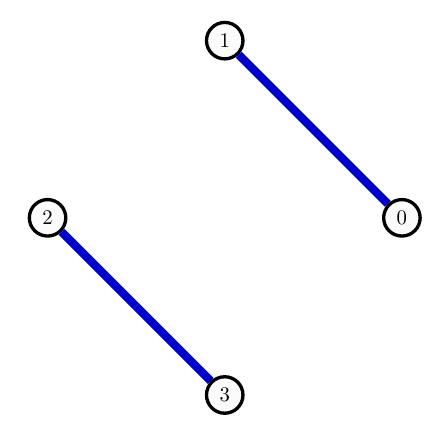}}\hspace{-3mm}
	\subfloat[$\ket{\text{W}}_{3}$]{\includegraphics[width=0.165\textwidth]{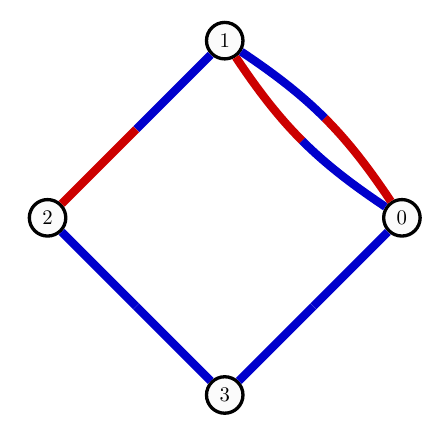}}\hspace{-3mm}
	\subfloat[$\ket{\text{W}}_{4}$]{\includegraphics[width=0.165\textwidth]{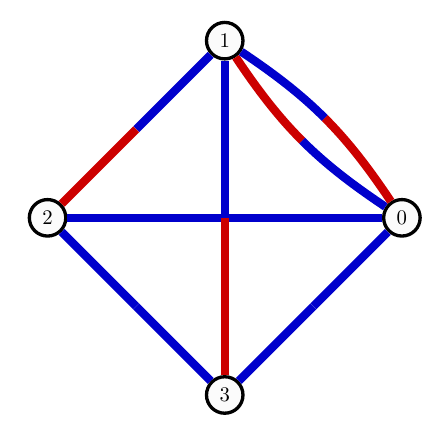}}
 \\
	\subfloat[$\ket{\Phi^+}$]{\includegraphics[width=0.165\textwidth]{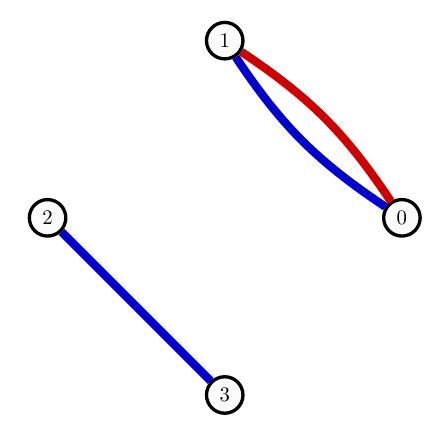}}\hspace{-3mm}
	\subfloat[$\ket{\text{GHZ}}^{2}_{3}$]{\includegraphics[width=0.165\textwidth]{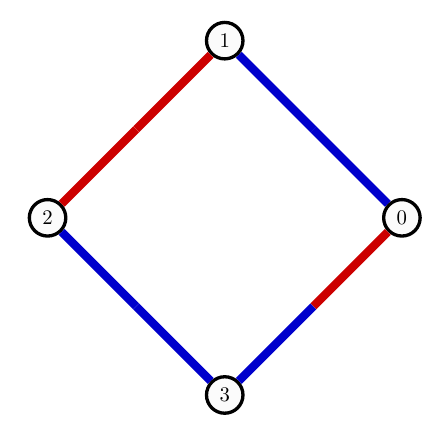}}\hspace{-3mm}
	\subfloat[$\ket{\text{GHZ}}^{2}_{4}$]{\includegraphics[width=0.165\textwidth]{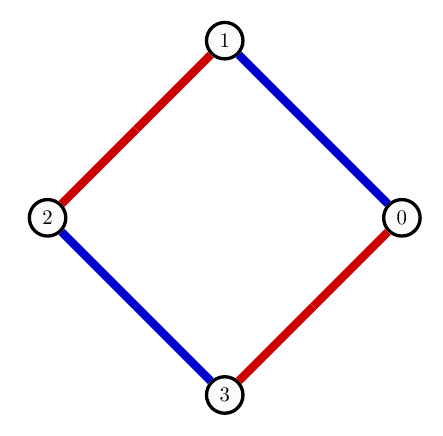}}	
	\caption {Six well-known examples from different entanglement categories described by Verstraete et.\ al. \cite{4qubits9entanglements}}
	\label{fig:Ent9Graphs6}
\end{figure}

Out of the nine categories, six of them include well-known quantum states: a separable state of 4 qubits, the W states of 3 and 4 qubits, the product of 2 Bell states, and the GHZ states of 3 and 4 qubits. For the 3 qubit states, the fourth particle is separable (like an ancilla). Fig.~\ref{fig:Ent9Graphs6} shows the graphs to produce these states. 

\begin{figure}[!t]
	\centering
	\subfloat[$\text{L}_{a_{4}}$]{\includegraphics[width=0.165\textwidth]{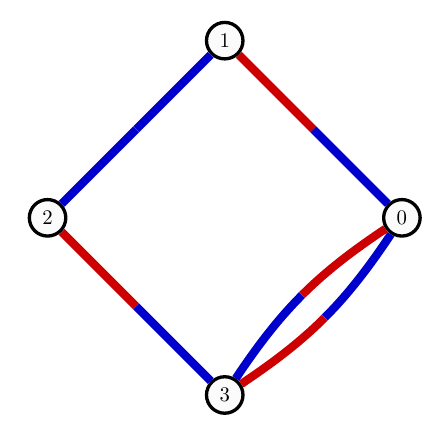}}\hspace{-3mm}
	\subfloat[$\text{L}_{0_{5\otimes\overline{3}}}$]{\includegraphics[width=0.165\textwidth]{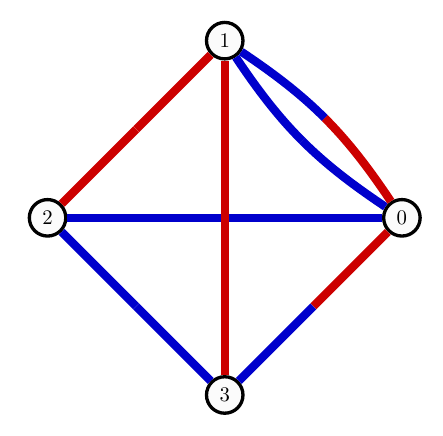}}\hspace{-3mm}
	\subfloat[$\text{L}_{0_{7\otimes\overline{1}}}$]{\includegraphics[width=0.165\textwidth]{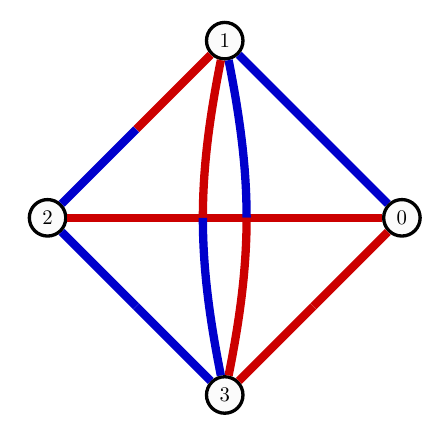}}
	\caption {Graphs for $L_{a_4}$, $L_{0_{5\oplus \overline{3}}}$, and $L_{0_{7\oplus \overline{1}}}$.}
	\label{fig:Ent9Graphs3}
\end{figure}

The remaining three categories, out of the initial nine ones, refer to less-known ways of entanglement. Verstraete's work refers to them as $L_{a_4}$, $L_{0_{5\oplus \overline{3}}}$, and $L_{0_{7\oplus \overline{1}}}$. In Fig.~\ref{fig:Ent9Graphs3} we see a graph for each category, producing the states \numbering[https://github.com/artificial-scientist-lab/PyTheus/tree/main/pytheus/graphs/HighlyEntangledStates/ent9_la4real]{experimentcounter}\numbering[https://github.com/artificial-scientist-lab/PyTheus/tree/main/pytheus/graphs/HighlyEntangledStates/ent9_053]{experimentcounter}\numbering[https://github.com/artificial-scientist-lab/PyTheus/tree/main/pytheus/graphs/HighlyEntangledStates/ent9_071]{experimentcounter}
\begin{align}
\ket{\text{L}_{a_4}} &= \ket{0001} + \ket{0110} + \ket{1000}, \\
\ket{\text{L}_{0_{5\oplus \overline{3}}}} &= \ket{0000} + \ket{0101} + \ket{1000} + \ket{1110}, \\
\ket{\text{L}_{7\oplus \overline{1}}} &= \ket{0000} + \ket{1011} + \ket{1101} + \ket{1110}. 
\end{align}

\tocless\subsubsection{Quantum Error Correction}
The path to the quantum computer requires qubits with low error rates independent of their realization platform. However, dealing with noise on any platform is inevitable. Thus, we need error-correcting codes to protect the information and achieve fault-tolerant quantum computation. This is where \textit{Logical Qubits} play a crucial role.

Logical qubits are sets of $N$ physical qubits that define two orthogonal states $\ket{0_L}$ and $\ket{1_L}$. These states are defined in such a way that errors in the physical qubits can be detected by applying a set of global measurements. These measurements will produce a different combination of outcomes for each potential error, allowing its correction. The graphs from this section produce 3 well-known logical qubits to detect and correct arbitrary errors on single qubits. 

The first one, the Shor code \cite{shor1995scheme}, employs 9 physical qubits to define \numbering[https://github.com/artificial-scientist-lab/PyTheus/tree/main/pytheus/graphs/HighlyEntangledStates/Shor]{experimentcounter}
\begin{align}
    \ket{0_L^{(9)}} = (\ket{000} + \ket{111})^{\otimes 3},\\
    \ket{1_L^{(9)}} = (\ket{000} - \ket{111})^{\otimes 3}.
\end{align}

The second example, the Steane code \cite{steane1996multiple}, employs 7 physical qubits \numbering[https://github.com/artificial-scientist-lab/PyTheus/tree/main/pytheus/graphs/HighlyEntangledStates/Steane]{experimentcounter}
\begin{align}
\ket{0_L^{(7)}} &=  \ket{0000000} + \ket{1010101} + \ket{0110011} \notag\\
 & + \ket{1100110} + \ket{0001111} + \ket{1011010} \notag\\
 & + \ket{0111100} + \ket{1101001}, \\
\ket{1_L^{(7)}} &=  X^{\otimes 7} \ket{0_L^{(7)}} = X_L \ket{0_L^{(7)}}. 
\end{align}
We go from the logical qubit to the other by applying a logical gate $X_L$, that is, by applying a Pauli $X$ gate on each physical qubit.

The third and last one, with only five physical qubits \cite{Laflamme_code}, is the Laflamme code \numbering[https://github.com/artificial-scientist-lab/PyTheus/tree/main/pytheus/graphs/HighlyEntangledStates/Laflamme]{experimentcounter}
\begin{align}
\ket{0_L^{(5)}} & =  \ket{00000} + \ket{11000} + \ket{01100} + \ket{00110} \notag\\
 & + \ket{00011} + \ket{10001} - \ket{10100} - \ket{01010} \notag\\
 & - \ket{00101} - \ket{10010} - \ket{01001} - \ket{11110} \notag\\
 & - \ket{01111} - \ket{10111} - \ket{11011} - \ket{11101},\\
\ket{1_L^{(5)}} & = X^{\otimes 5} \ket{0_L^{(5)}} = X_L \ket{0_L^{(5)}}. 
\end{align}
These two logical qubits constitute the smallest error-correcting code resilient to arbitrary single qubit errors \cite{min_error_correct}.

\begin{figure}[!t]
	\centering
	\subfloat[$|0_L^{(5)}\rangle$]{\includegraphics[width=0.24\textwidth]{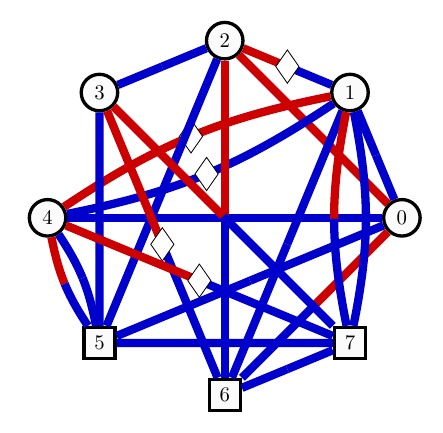}}\\
 \vspace{-5mm}
	\subfloat[$|0_L^{(7)}\rangle$]{\includegraphics[width=0.24\textwidth]{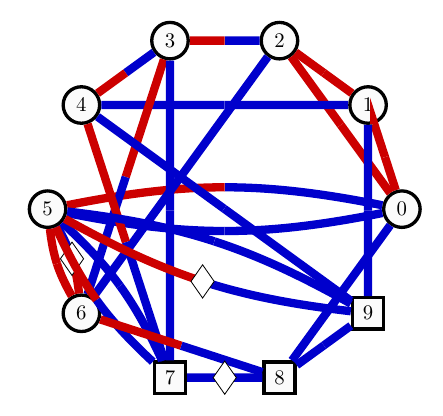}}\!\!
	\subfloat[$|0_L^{(9)}\rangle$]{\includegraphics[width=0.24\textwidth]{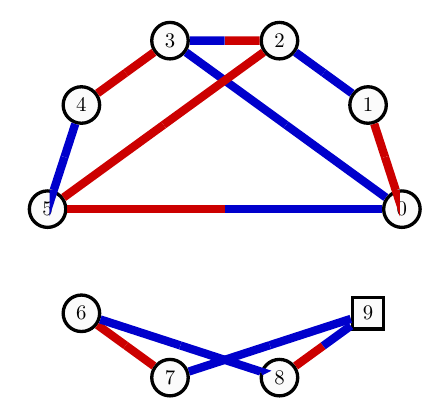}}
	\caption {Graphs for generating the logical 0 state using 5, 7, and 9 physical qubits (+ ancillas, which are in a fixed state).}
	\label{fig:error_correction}
\end{figure}
Fig.~\ref{fig:error_correction} shows the graphs for producing each of the logical 0 states for 5, 7, and 9 physical qubits. For the Shor code, the second logical qubit state can be obtained by multiplying the weights of every bicolored edge by -1 (or alternatively, taking the 3rd, 6th, and 9th qubits and applying a minus at each red incoming edge). For the other two codes, one can obtain the graphs associated with the logical 1 state by switching the colors blue and red. Such transformation is equivalent to applying the gate $X$ on each qubit, the logical $X_L$.\newline

\tocless\subsubsection{Products of W States}
\begin{figure}[!t]
	\centering
	\includegraphics[width=0.25\textwidth]{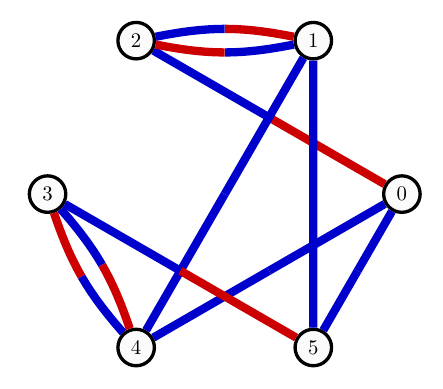}\hspace{-3mm}
	\includegraphics[width=0.24\textwidth]{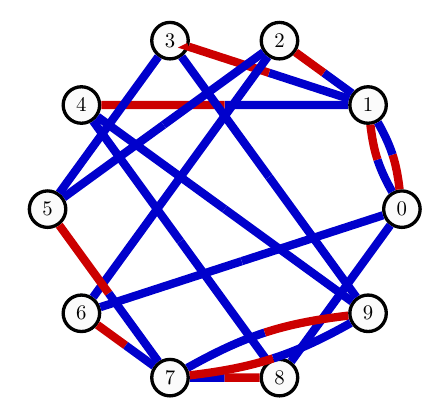}
	\caption {Graphs for generating the states $\ket{W_3}^{\otimes2}$ and $\ket{W_5}^{\otimes2}$ without ancillas.}
	\label{fig:w3w3}	
\end{figure}

Previous work with the graph representation showed how to create W states for an arbitrary even number of qubits without using any ancilla \cite{graphs3}. On the other hand, to generate a W state with an odd number of qubits, we always needed at least one ancillary detector. Accordingly, if we want to duplicate such odd-qubit states, we can simply duplicate the graph together with the ancillas. However, this straightforward method is not the most efficient, at least for the states $\ket{W_3}^{\otimes2}$ and $\ket{W_5}^{\otimes2}$. As shown in Fig.~\ref{fig:w3w3}, these two states can be obtained without using ancillas \numbering[https://github.com/artificial-scientist-lab/PyTheus/tree/main/pytheus/graphs/HighlyEntangledStates/W3W3]{experimentcounter}\numbering[https://github.com/artificial-scientist-lab/PyTheus/tree/main/pytheus/graphs/HighlyEntangledStates/W5W5]{experimentcounter}.\newline

\tocless\subsubsection{Dicke States}
\begin{figure}[!h]
	\centering
    \subfloat[$\ket{D^2_5}$]{\includegraphics[width=0.24\textwidth]{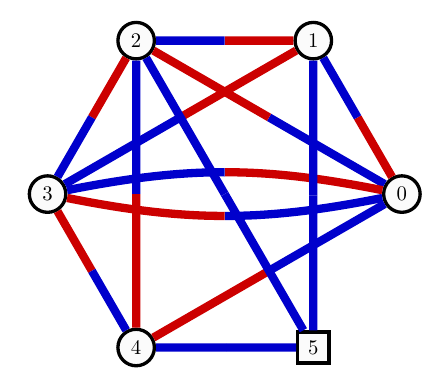}}\!\!
	\subfloat[$\ket{D^2_7}$]{\includegraphics[width=0.24\textwidth]{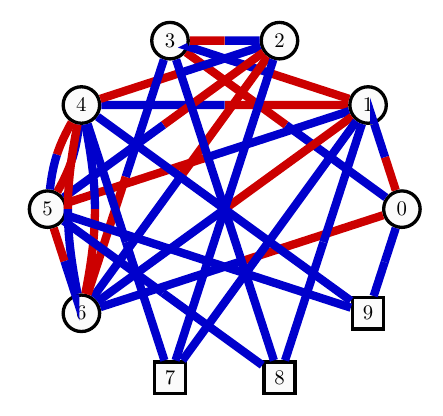}}
    \caption{Dicke states with two $\ket{1}$ for five and seven qubits.}
	\label{fig:odddicke}	
\end{figure}

The equal superposition of every $N$-qubit state containing a single $\ket{1}$ gives us the W states. These states can be generalized, superposing all permutations of $N$ qubits with exactly $k$ $\ket{1}$s. They are the \textit{Dicke states}
\begin{equation}
    \ket{D^k_N}=\sqrt{\frac{k!(N-k)!}{N!}}\sum_{i} P_i( \ket{0}^{\otimes N - k} \ket{1}^{\otimes k}),
\end{equation}
where $\sum_{i} P_i( \cdot )$ stands for all possible permutations of $k$ $\ket{1}$ among $N$ qubits. These states, which have applications in multiparty quantum communication and quantum metrology \cite{dicke_persistent,dicke_telecloning,dicke_networking,dicke_metrology}, can be produced for an even number of qubits with a general graph introduced in previous works \cite{graphs3}. Here we produce a pair of Dicke states with an odd number of qubits and two excitations \numbering[https://github.com/artificial-scientist-lab/PyTheus/tree/main/pytheus/graphs/HighlyEntangledStates/dicke52]{experimentcounter}\numbering[https://github.com/artificial-scientist-lab/PyTheus/tree/main/pytheus/graphs/HighlyEntangledStates/dicke72]{experimentcounter} (see Fig.~\ref{fig:odddicke}).

\begin{figure}[!h]
	\centering
    \subfloat[$\ket{D(3,(1,1,1)}$]{\includegraphics[width=0.24\textwidth]{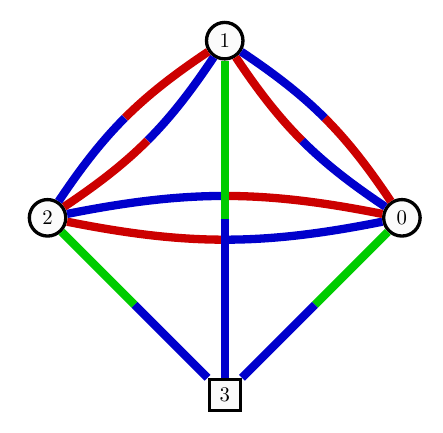}}\!\!
	\subfloat[$\ket{D(4,(2,1,1))}$]{\includegraphics[width=0.24\textwidth]{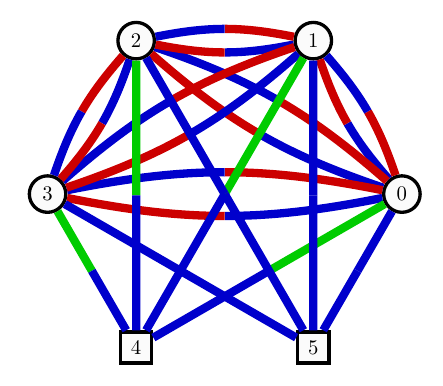}}
 \\
    \subfloat[$\ket{D(5,(2,2,1)}$]{\includegraphics[width=0.24\textwidth]{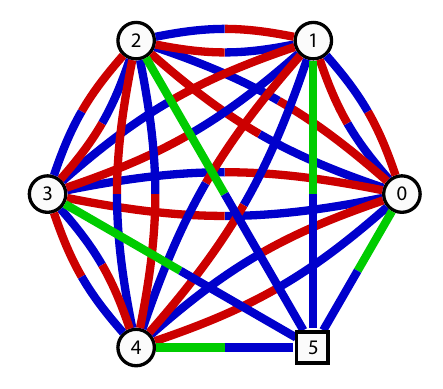}}\!\!
	\subfloat[$\ket{D(4,(1,1,1,1))}$]{\includegraphics[width=0.24\textwidth]{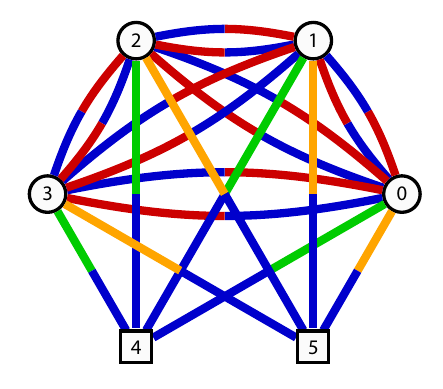}}
    \caption{The three-dimensional Dicke states for three and four particles (top) were found by \pytheus. The 2 graphs in the bottom row were manually produced following the observed pattern.}
	\label{fig:symm}	
\end{figure}

Moreover, there exists a generalization of Dicke states for higher dimensions \cite{dicke_qudit}. Given a set of integers $\{k_i\}$, such that $\sum_i k_i=N$, we describe its corresponding d-dimensional Dicke state as
\begin{align}
    \ket{D(N,\{k_i\})} = \sqrt{\frac{\prod_i k_i!}{N!}}\sum_i P_i &(\ket{0}^{\otimes k_0}\ket{1}^{\otimes k_1} \notag \\
    &\dots\ket{d-1}^{\otimes k_{d-1}} )
\end{align}
where $\sum_{i} P_i( \cdot )$ stands for all possible permutations of the $N$ qudits. In Fig.~\ref{fig:symm} we plot the graphs that produce the 3-dimensional Dicke state with three particles \numbering[https://github.com/artificial-scientist-lab/PyTheus/tree/main/pytheus/graphs/HighlyEntangledStates/dicke33]{experimentcounter}
\begin{align}
    \ket{D(3,(1,1,1)} &= \ket{012} + \ket{021} + \ket{102} \notag\\ 
    & +\ket{120} + \ket{201} + \ket{210},
\end{align}
and with 4 particles (with uneven coefficients $\{k_i\}$) \numbering[https://github.com/artificial-scientist-lab/PyTheus/tree/main/pytheus/graphs/HighlyEntangledStates/dicke43]{experimentcounter}
\begin{align}
    \ket{D(4,(2,1,1))} &= \ket{0012}+\ket{1200}+\ket{1020} \notag\\
    & + \ket{0102}+\ket{1002}+\ket{0120} \notag\\
    & + \ket{0021}+\ket{2100}+\ket{2010} \notag\\
    & + \ket{0201}+\ket{2001}+\ket{0210}.
\end{align}

After plotting the results found by \pytheus to generate the last two states, we found a pattern on the graphs that led us to states like $\ket{D(5,(2,2,1)}$ and $\ket{D(4,(1,1,1,1))}$: We start taking a fully connected graph of $N$ particles, in which the vertices are connected by 2 bicolored edges, blue-red and red-blue, indicating two modes (the same as in Ref.\ \cite{graphs3}). Then one can add ancillas that are connected to the $N$ first particles by bicolored edges, one for each of the $N$ particles. These new edges introduce more modes, leading to the following series of states for an even number of particles, $2k$:
\begin{align*}
    \text{0 ancillas, 2 modes} &\rightarrow  \ket{D(2k,(k,k))}  \\
    \text{2 anc, 3 modes} &\rightarrow  \ket{D(2k,(k,k\text{ - }1,1))}  \\
    \text{2 anc, 4 modes} &\rightarrow  \ket{D(2k,(k\text{ - }1,k\text{ - }1,1,1))}  \\
    & \cdots
\end{align*}
 \vspace{-7mm}
\begin{equation*}
    \text{2($k\text{ - }1$) anc, 2$k$ modes} \rightarrow  \ket{D(2k,(1,\cdots,1))}
\end{equation*}
For an odd number of particles, $2k\!+\!1$, starting with 1 ancilla, we obtain
\begin{align*}
    \text{1 anc, 3 modes} &\rightarrow  \ket{D(2k\!+\!1,(k,k,1))}  \\
    \text{3 anc, 4 modes} &\rightarrow  \ket{D(2k\!+\!1,(k,k\text{ - }1,1,1))}  \\
    \text{3 anc, 5 modes} &\rightarrow  \ket{D(2k\!+\!1,(k\text{ - }1,k\text{ - }1,1,1))}  \\
    & \cdots 
\end{align*}
 \vspace{-7mm}
\begin{equation*}
    \text{$2k\text{ - }1$ anc, $2k\!+\!1$ modes} \rightarrow  \ket{D(2k\!+\!1,(1,\cdots,1))}.
\end{equation*}

\tocless\subsubsection{Yeo-Chua State}
\begin{figure}[!t]
	\centering
	\includegraphics[width=0.26\textwidth]{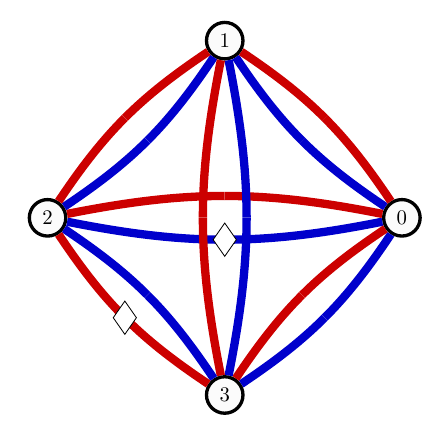}
	\caption {Graph for generating the Yeo-Chua state.}
	\label{fig:yc_state}	
\end{figure}

In 1993, Bennett et.\ al.\ showed how to use Bell states to teleport arbitrary and unknown single-qubit states \cite{quantum_teleportation_paper}. Going one step further, Yeo and Chua developed a protocol to teleport unknown states of two qubits \cite{Yeo_2006}, and instead of using Bell states, they employed the Yeo-Chua state
\begin{align}\label{eq:YC}
\numbering[https://github.com/artificial-scientist-lab/PyTheus/tree/main/pytheus/graphs/HighlyEntangledStates/YC]{experimentcounter}\ket{YC}  =& \ket{0000}-\ket{0011}-\ket{0101} \notag\\ 
    &  + \ket{0110} + \ket{1001} + \ket{1001} \notag\\
    & + \ket{1100} + \ket{1111}.
\end{align}
This highly entangled state can be produced with the graph of Fig.~\ref{fig:yc_state}.\newline

\tocless\subsubsection{Complex Weights as a Resource}
\begin{figure}[!t]
	\centering
    \subfloat[Real solution]{\includegraphics[width=0.24\textwidth]{graph_golden.pdf}}\!\!
    \subfloat[Complex solution]{\includegraphics[width=0.24\textwidth]{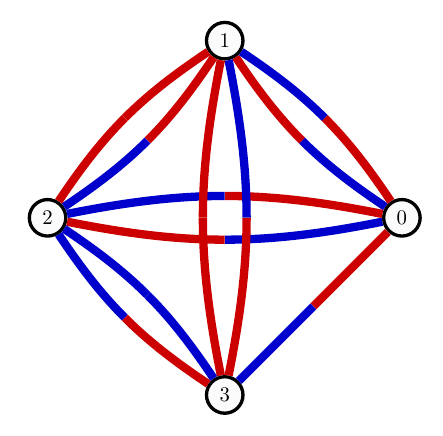}}
	\caption {Graphs for the four-qubit state defined in Eq.~\eqref{eq:4qubitcomplex}. The left solution employs the golden ratio and the right one complex weights.}
	\label{fig:complex}
\end{figure}

\begin{figure}[t]
	\centering
	\includegraphics[width=0.49\textwidth]{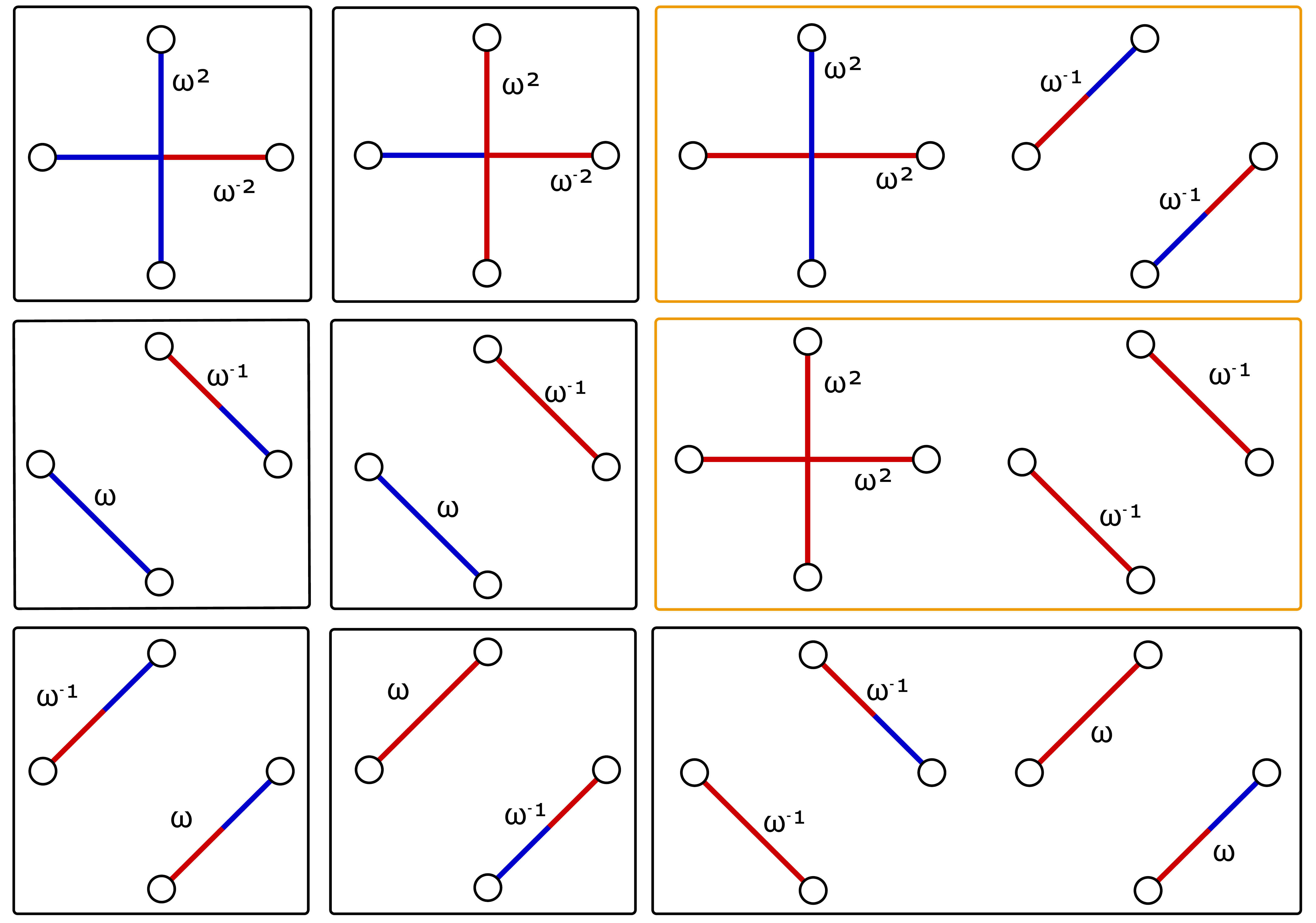}
	\caption {Perfect matchings of graph shown in Fig.~\ref{fig:complex} for the state defined in Eq.~\eqref{eq:4qubitcomplex}. $\omega=e^{i\pi/6}$. Each green box contributes one term of the state. The perfect matchings in the orange boxes interfere destructively. It is especially notable that the weights of the two $\ket{0111}$ perfect matchings (bottom right) are complex conjugates and add to one as $\omega^2 + \omega^{-2} = 1$.}
	\label{fig:compexPMs}
\end{figure}

Exploring the applicability range of \pytheus, we searched for four-qubit states of the form
\begin{align}
    \ket{\psi} = c_1 \ket{0000} + c_2 \ket{0001} + ... + c_{16}\ket{1111},
\end{align}
where $c_i \in \{ 0,1 \}$. A total of $2^{16}-1 = 65535$ states.
Graphs with $\pm1$ weights suffice to produce most of the states, many of them do not even require interference. Yet, some states, such as 
\begin{align}\label{eq:4qubitcomplex}
    \ket{\psi} &= \ket{0011} + \ket{0100} + \ket{0111} + \ket{1000} \notag\\ 
    &+ \ket{1100} + \ket{1101} + \ket{1110},
\end{align}
require not only destructive interference between creation terms, but very specific ratios between the weights. In particular, the weights to produce the state \eqref{eq:4qubitcomplex} make use of the \textit{Golden Ratio}, since they must fulfill the expression
\begin{equation}\label{eq:golden_ratio}
    x^2 + x = 1.
\end{equation}
The graph that employs the golden ratio is shown in Fig~\ref{fig:complex}, and similar graphs can produce a whole family of 64 states which are equivalent under permutations and bit-flips. However, there is an alternative complex-weighted graph, with less edges, to produce these states. As shown in Fig~\ref{fig:complex}, the absolute value is the same for all weights and the interference patterns due to the complex phases are shown in Fig~\ref{fig:compexPMs}. Even thought they are computationally more expensive to obtain, complex-weighted solutions can provide new interesting designs.\newline

\tocless\subsubsection{Single Photon Sources for State Creation}
\begin{figure}[!h]
    \centering
 \subfloat[$\ket{\text{GHZ}}^2_6$ (s.p.s)]{\includegraphics[width=0.24\textwidth]{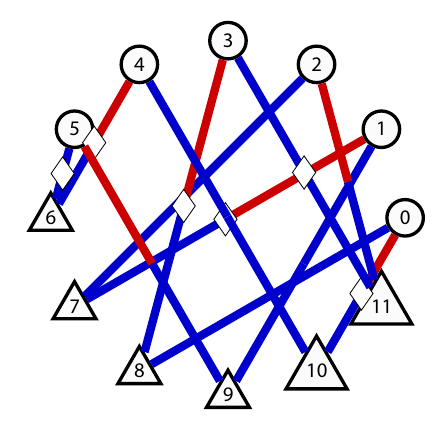}}\!\!
 \subfloat[$\ket{\text{GHZ}}^3_4$ (s.p.s)]{\includegraphics[width=0.24\textwidth]{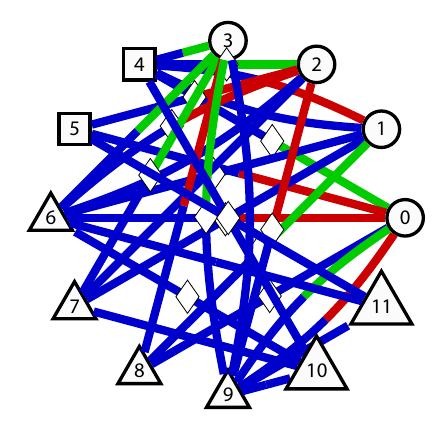}}\\
 \subfloat[$\ket{\text{GHZ}}^3_4$ (hybrid)]{\includegraphics[width=0.24\textwidth]{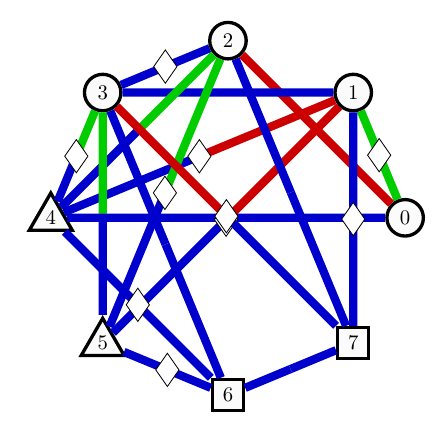}}
    \caption{Graphs corresponding to the creation of three-, four- and five-particle GHZ states using single-photon sources.}
    \label{fig:heralded_sp_ghz}
\end{figure}

Significant progress has been made toward using deterministic single-photon sources in quantum optics experiments during the last years \cite{wang2019towards,meyer2020single,singlephotons2011}. In this section, we present graphs corresponding to quantum experiments for the creation of several highly entangled states using single-photon sources. Deterministic sources do not suffer from the same trade-off between high fidelity and high count rate as probabilistic sources. When using deterministic sources, the number of photons required can be smaller than with probabilistic sources \cite{PhysRevA.67.030101,PhysRevA.102.012604}.
\paragraph{Post-Selected}
We have found graphs for creating different GHZ states with single-photon sources as a resource. One can create a six qubit GHZ from six input photons \numbering[https://github.com/artificial-scientist-lab/PyTheus/blob/main/pytheus/graphs/HighlyEntangledStates/ghz_62_sp/graph_ghz_62_sp.pdf]{experimentcounter}, a four qutrit GHZ state from six input photons \numbering[https://github.com/artificial-scientist-lab/PyTheus/tree/main/pytheus/graphs/HighlyEntangledStates/ghz_62_sp]{experimentcounter}, as well as the same four qutrit GHZ state from a combination of two single photon sources and two simultaneous SPDC events \numbering[https://github.com/artificial-scientist-lab/PyTheus/blob/main/pytheus/graphs/HighlyEntangledStates/ghz_43_2p2a/graph_ghz_43_2p2a.pdf]{experimentcounter} (see Fig.~\ref{fig:heralded_sp_ghz}).

\begin{figure}[!h]
    \centering
    \includegraphics[width=0.26\textwidth]{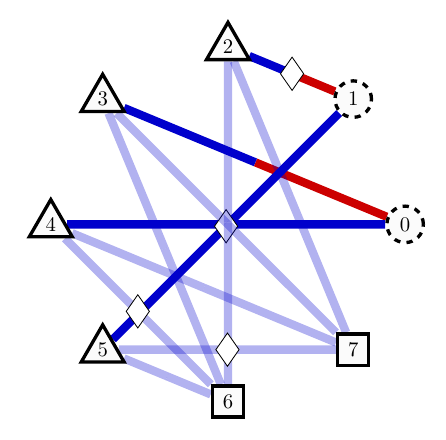}
    \caption{Graph corresponding to the creation of a heralded two-dimensional Bell state in the outgoing photons (0-1) from four single photon sources (2-5) and two heralding detectors (6-7). The heralding detectors do not require photon-number-resolution. The light edges have weights $w<1$ corresponding to loss in the photon path.
}
    \label{fig:heralded_sp_2dbell}
\end{figure}

\paragraph{Heralded --}
There have been proposals for the creation of multi-particle high-dimensional entangled states from single photon sources using discrete Fourier transform \cite{paesani2021scheme}. In Fig. \ref{fig:heralded_sp_2dbell} we show a graph \numbering[https://github.com/artificial-scientist-lab/PyTheus/tree/main/pytheus/graphs/HighlyEntangledStates/heralded_bell_sp]{experimentcounter}, which achieves a heralded Bell state using two ancillary photons without the need for photon-number-resolving detectors with almost perfect fidelity (F = 99\%) for $w=0.1$. There are cross-term that occur due to the lack of photon- number resolution. We want to acknowledge that the probability of a heralding event is very low for this example (P = 0.08\%) and it is thus unlikely that this setup is experimentally feasible. As $w$ approaches 0, fidelity approaches $100\%$ but count-rate approaches 0. The trade-off between high fidelity and high count-rate is interesting to consider.

\subsection{Maximizing Entanglement}\label{sec:max_entanglement}
When a quantum state cannot be expressed as the product of its parts, we call it entangled. However, if one asks \textit{how} entangled a state is, the answer is not unique, there are multiple measures of entanglement. Here we present some of these metrics, together with states that are known to maximize them. Finally, we choose a particular metric and optimize a graph, not to reach a particular state -- as done so far by maximizing the fidelity -- but to instead maximize the entanglement metric itself. In this way, we will recover some well-known entangled states and also find new ones.

\tocless\subsubsection{Schmidt Rank Vector}
Multiparticle entangled states in high dimensions allow for a much deeper structure underlying the potential ways in which the particles can be entangled. These structures can be characterized by the Schmidt Rank Vector (SRV) and give rise to new phenomena that only exist if both the number of particles and the number of dimensions go beyond two \cite{HuberSRV,HuberSRVentropy, CadneySRV}. Here we consider tripartite states. The rank vector is a list of the ranks of the reduced density matrices. The state $\text{SRV(A,B,C)}$ refers to a state, where
\begin{equation}
	A=\text{rank}(\rho_{A}), 	B=\text{rank}(\rho_{B}), 	C=\text{rank}(\rho_{C}),\nonumber
\end{equation}
and $\rho_{X}$ is the density matrix of the system with the particle $X$ traced out. In other words, the SRV is a vector of the dimensionalities of entanglement of every bipartition, which shows the dimensionality of entanglement between one particle and the rest of the quantum state. This master-slave-slave configuration is very useful for quantum applications such as layered quantum communication \cite{LayeredQKD}. Taking a $\text{SRV(3,3,2)}$ state as an example, we then know that the first two particles are both three-dimensionally entangled with the third particle, whereas the third one is only two-dimensionally entangled with the rest. The dimensionality for each particle cannot be increased with linear operations and classical communication (LOCC).

Searching the $\text{SRV(A,B,C)}$ states has been investigated with the computer algorithm \melvin \cite{Melvin2016} and graph theory \cite{graphs2, hypergraphs}. Interestingly, it has been shown from graph theory that without using any ancilla particles, there are several $\text{SRV(A,B,C)}$ states that cannot be created. \cite{graphs2,hypergraphs}. Here we list these states (up to normalization):
\begin{equation}
\begin{split}
\numbering[https://github.com/artificial-scientist-lab/PyTheus/tree/main/pytheus/graphs/MaxEntanglement/srv_554]{experimentcounter} &\text{SRV(5,5,4)}:\\
&\ket{000}+\ket{111}+\ket{222}+\ket{333}+\ket{443}\notag\\
\numbering[https://github.com/artificial-scientist-lab/PyTheus/tree/main/pytheus/graphs/MaxEntanglement/srv_632]{experimentcounter} &\text{SRV(6,3,2)}:\\
&\ket{000}+\ket{101}+\ket{210}+\ket{311}\notag\\+&\ket{420}+\ket{521}\notag\\
\numbering[https://github.com/artificial-scientist-lab/PyTheus/tree/main/pytheus/graphs/MaxEntanglement/srv_655]{experimentcounter}&\text{SRV(6,5,5)}:\\
&\ket{000}+\ket{111}+\ket{222}+\ket{334}\notag\\+&\ket{443}+\ket{544}\notag\\
\numbering[https://github.com/artificial-scientist-lab/PyTheus/tree/main/pytheus/graphs/MaxEntanglement/srv_733]{experimentcounter} &\text{SRV(7,3,3)}:\\
&\ket{000}+\ket{101}+\ket{210}+\ket{311}\notag\\+&\ket{422}+\ket{520}+\ket{621}\notag
\end{split}
\end{equation}
\begin{figure}[!b]
	\centering
	\subfloat[SRV(5,5,4)]{\includegraphics[width=0.24\textwidth]{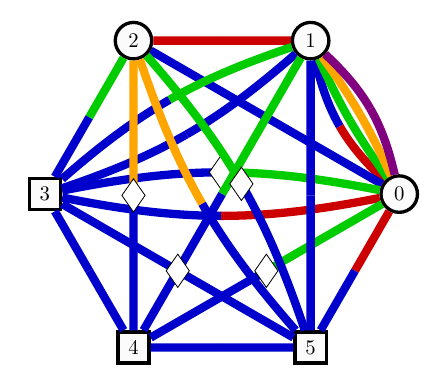}}
	\subfloat[SRV(6,3,2)]{\includegraphics[width=0.24\textwidth]{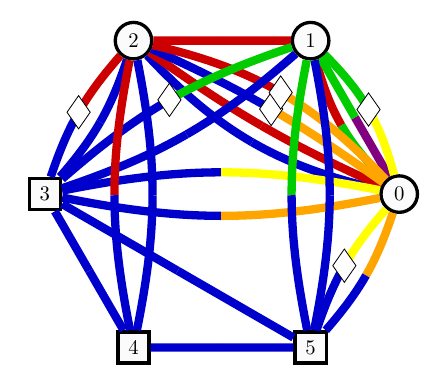}}
    \\ \vspace{-2mm}
	\subfloat[SRV(6,5,5)]{\includegraphics[width=0.24\textwidth]{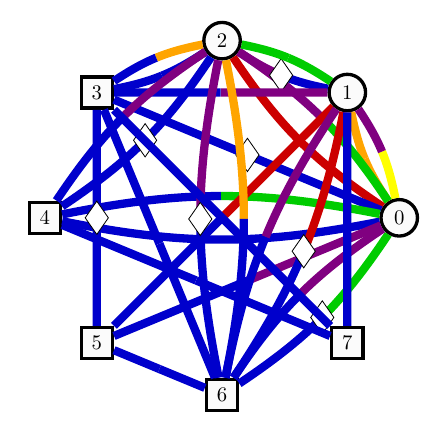}}\!\!
	\subfloat[SRV(7,3,3)]{\includegraphics[width=0.24\textwidth]{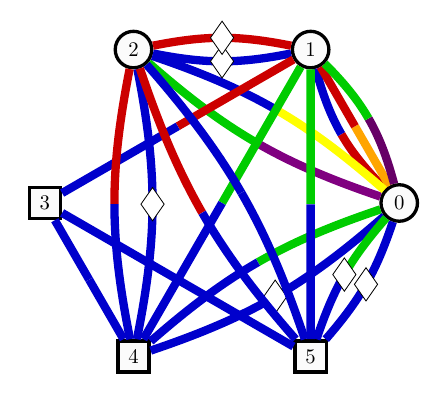}}
	\caption {Graphs for producing $\text{SRV(A,B,C)}$ states using ancillas.}
	\label{fig:SRVgraphs}	
\end{figure}
It has been unknown whether we can experimentally create these states with additional particles, and this challenging question has been open for more than three years. Also, until now, $\text{SRV(A,B,C)}$ states in high dimensional systems have only been recently demonstrated for $\text{SRV(3,3,2)}$ \cite{srv332experimentmalik}, $\text{SRV(3,3,3)}$ \cite{erhard3dghz}, and $\text{SRV(4,4,2)}$ \cite{srv442experimenthu}. By using PyTheus, one can learn how to generate much more complex SRV states of high dimensions and more particles, such as the impossible states described in \cite{graphs2,hypergraphs}. We show the results in Fig.~\ref{fig:SRVgraphs}.

\tocless\subsubsection{Hyperdeterminant}
\begin{figure}[!h]
	\centering
	\includegraphics[width=0.26\textwidth]{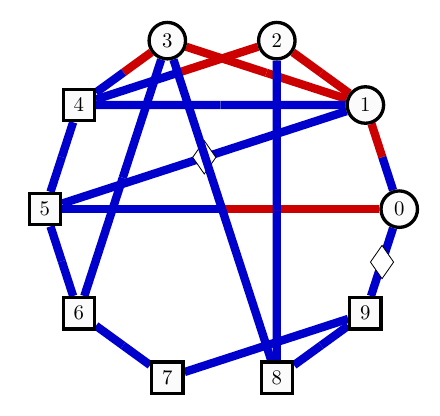}
	\caption {Graph to generate the state $\ket{\text{HD}}$, which maximizes the hyperdeterminant.}
	\label{fig:hyperdeterminant}	
\end{figure}

Expanding the definition of determinant to tensors of arbitrary dimensions, we obtain \textit{Hyperdeterminants} \cite{hyperdeterminant}, which can also be used as an entanglement metric. In particular, the hyperdeterminant of a tensor $2\times2\times2\times2$ characterizes the entanglement of a 4 qubit state, which is maximal for the state (up to normalization) \numbering[https://github.com/artificial-scientist-lab/PyTheus/tree/main/pytheus/graphs/MaxEntanglement/HD]{experimentcounter}
\begin{equation}
    \ket{\text{HD}} = \ket{\Psi^+00}+\ket{00\Psi^+} + \ket{1111},
\end{equation}
where $\sqrt{2}\ket{\Psi^+}=\ket{01}+\ket{10}$, and whose associated graph is shown in Fig.~\ref{fig:hyperdeterminant}.\newline

\tocless\subsubsection{Negative Partial Trace}
Given a state $\rho_{AB}$, its \textit{partial transpose} with respect to the subsystem $A$ is
\begin{equation}
    (\rho^{T_A})_{ab,mn} = \rho_{mb,an}.
\end{equation}
As shown by Peres, it is a necessary condition for the separability of $A$ and $B$ that none of the eigenvalues of $\rho^{T_A}$ are negative \cite{peres_separability}. Moreover, it is also a sufficient condition for bipartite states of dimensions $2\times2$ and $2\times3$ \cite{horodecki2001}. 

\begin{figure}[!h]
	\centering
    \includegraphics[width=0.22\textwidth]{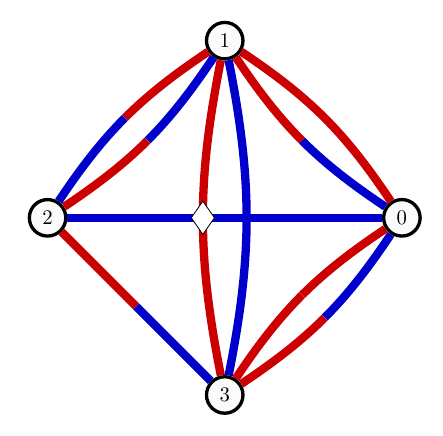}\!\!
	\includegraphics[width=0.24\textwidth]{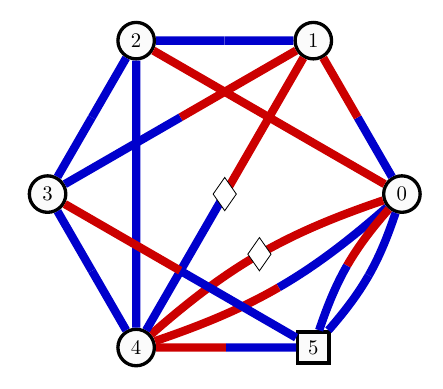}
	\caption {Graphs to generate the BSSB states of 4 and 5 qubits, respectively. The latter requires an ancilla.}
	\label{fig:bssb}	
\end{figure}

Aiming for highly entangled states, Brown et.\ al.\ looked for states of 2, 3, 4, and 5 qubits, whose partial traces for each bipartition had the lowest possible eigenvalues \cite{brown05}. They called this metric \textit{Negative Partial Trace}, and it leads to the following states of 4 and 5 qubits \numbering[https://github.com/artificial-scientist-lab/PyTheus/tree/main/pytheus/graphs/MaxEntanglement/bssb4]{experimentcounter} \numbering[https://github.com/artificial-scientist-lab/PyTheus/tree/main/pytheus/graphs/MaxEntanglement/bssb5]{experimentcounter}
\begin{align}
    \ket{\text{BSSB4}} &= \ket{0000} + \ket{+011} \notag\\ 
    &+ \ket{1101} + \ket{-110} \\
     \ket{\text{BSSB5}} &= \ket{000\Phi^+} + \ket{100\Phi^-} \notag\\
     & + \ket{010\Psi^+}  + \ket{111\Psi^-}.
\end{align}
Where $\sqrt{2}\ket{\pm}=\ket{0}\pm\ket{1}$, $\sqrt{2}\ket{\Phi^\pm}=\ket{00}\pm\ket{11}$, and $\sqrt{2}\ket{\Psi^\pm}=\ket{01}\pm\ket{10}$. These states can be produced with the graphs shown in Fig.~\ref{fig:bssb}.

\tocless\subsubsection{R\'enyi Entropy}
Among all the entanglement metrics in this document, as well as the ones that are left out, the \textit{R\'enyi Entropy} is one of the most important \cite{renyi_entropy}. Given a state with $\rho_{AB}$, the R\'enyi entropy of order $\alpha$ between the subsystems $A$ and $B$ is
\begin{equation}
\label{eq:renyi}
    \mathcal{S}_\alpha (\rho_A) = \frac{1}{1-\alpha} \log \Tr (\rho_A^\alpha) = \mathcal{S}_\alpha(\rho_B),
\end{equation}
where $\rho_A=\Tr_B(\rho_{AB})$ and $\rho_B=\Tr_A(\rho_{AB})$ are reduced density matrices. Similarly, the \textit{Tsallis entropy} also describes the entanglement between $A$ and $B$
\begin{equation}
 \label{eq:tsallis}
   \mathcal{T}_\alpha (\rho_A) = \frac{1}{1-\alpha} \left(\Tr (\rho_A^\alpha) - 1\right) = \mathcal{T}_\alpha(\rho_B).
\end{equation}
Notice that in the limit $\alpha\rightarrow1$, both expressions approach the Von Neumann entropy.

\begin{figure}[!h]
	\centering
	\includegraphics[width=0.24\textwidth]{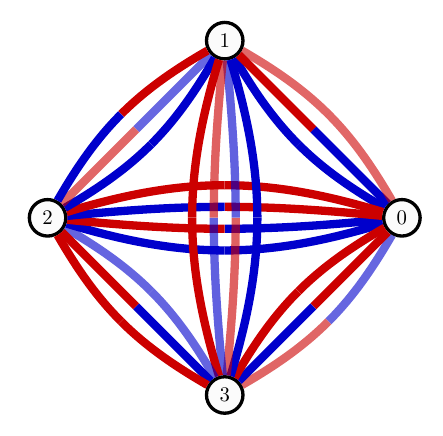}
	\caption {Graph to generate the 4-qubit L-state. All the weights in the graph are complex.}
	\label{fig:Lstate}	
\end{figure}

A relevant difference between both metrics is that, for $\alpha>1$, only the Tsallis entropy is convex while the R\'enyi one is Schur convex \cite{renyi_concavity}. Taking this difference into account, Gour and Wallach showed that the 4-qubit state $\ket{L}$ maximizes the average of Tsallis entropies with $\alpha=2$ for 2-qubit partitions \cite{Gour_2010}, a metric of entanglement also referred to as the \textit{Meyer-Wallach measure} \cite{Brennen03,Meyer2002}. This state, created by the graph shown in Fig.~\ref{fig:Lstate}, reads \numbering[https://github.com/artificial-scientist-lab/PyTheus/tree/main/pytheus/graphs/MaxEntanglement/Lstate]{experimentcounter}
\begin{align}
    \ket{\text{L}} =& (1+\omega)(\ket{0000}+\ket{1111}) \notag \\
    +& (1-\omega)(\ket{0011}+\ket{1100}) \notag \\ 
    +& \omega^2(\ket{0101} + \ket{0110} + \ket{1001} + \ket{1010}),
\end{align}
where $\omega=\exp(2i\pi/3)$.

\tocless\subsubsection{R\'enyi–Ingarden–Urbanik Entropy}
Minimizing the R\'enyi entropy over all local unitary transformations $U_{\text{LOC}}$, Enriquez et.\ al.\ described the minimal \textit{R\'enyi–Ingarden–Urbanik} entropy \cite{Enriquez2015}
\begin{equation}
  \label{eq: RIU-entropy}
  \mathcal{S}_\alpha^{\text{RIU}}(\phi) = \min_{U_{\text{LOC}}} \mathcal{S}_\alpha( p(U_{\text{LOC}}\ket{\phi})),
\end{equation}
where $\mathcal{S}_\alpha$ is the R\'enyi entropy of order $\alpha\geq0$, and $p$ is a normalized probability vector resulting from expanding the state to a product basis.

\begin{figure}[!h]
	\centering
    \includegraphics[width=0.24\textwidth]{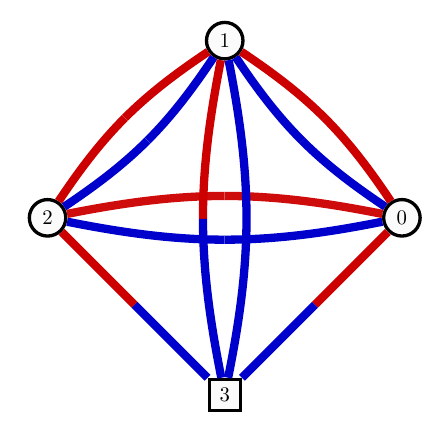}\!\!
	\includegraphics[width=0.24\textwidth]{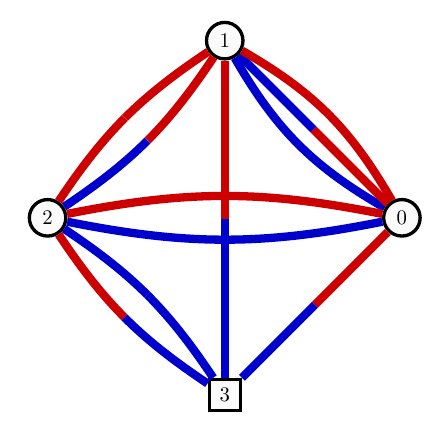}
	\caption {Graphs to generate the states that optimize the RIU entropy for $\alpha=1$ and $\alpha=2$.}
	\label{fig:maxRIU}	
\end{figure}

Together with the definition, the authors numerically find two states of three qubits that minimize the entropy for $\alpha=1$ and $\alpha=2$ \numbering[https://github.com/artificial-scientist-lab/PyTheus/tree/main/pytheus/graphs/MaxEntanglement/randmax1]{experimentcounter} \numbering[https://github.com/artificial-scientist-lab/PyTheus/tree/main/pytheus/graphs/MaxEntanglement/randmax2]{experimentcounter}
\begin{align}
    \ket{\Phi}_{\alpha=1} &=  0.27 \ket{000} + 0.377 \ket{100} + 0.326 \ket{010}  \notag\\
        &+ 0.363 \ket{001} + 0.74 e^{-0.79 \pi i} \ket{111}, \\
    \ket{\Phi}_{\alpha=2} &= 0.438 \ket{000} + 0.29 \ket{100} + 0.371 \ket{010} \notag\\
    & +  0.316 \ket{001} +  0.698 e^{ -0.826 \pi i}\ket{111}. 
\end{align}
These states correspond to the graphs of Fig.~\ref{fig:maxRIU}.

\tocless\subsubsection{Maximizing Entanglement for Each Partition}\label{sec:concurrence}
The previous states were maximally entangled according to different metrics. Indeed, many of them were found numerically when maximizing such metrics. Similarly, we can use \pytheus to maximize arbitrary physical properties of the states produced by a graph, which can be immediately translated into experiments. This is a great advantage over previous work, in which an analytically described quantum state may not be experimentally feasible.

We optimize the graph according to the Tsallis entropy for $\alpha=2$ (see Eq.~\eqref{eq:tsallis}) using as loss function
\begin{equation}
\label{eq:concurrence}
    \min_{\omega} \mathcal{L}(\rho(\omega)) = \sum_{A}\Tr\rho_A^2,
\end{equation}
where $\rho=\ketbra{\psi(\omega)}$ is the state defined by a graph with weights $\{\omega\}$, $\rho_A$ is the reduced density matrix with respect to a subsystem $A$. A subsystem and its complement $A^c$ are a bipartition of the full system. The loss function is minimized for the sum of all bipartitions. 

Given a bipartition, the entanglement between them is maximal if
 \begin{equation}
     \rho_A = \Tr_{A^c}(\rho)= \frac{1}{D}\mathbb{I},
 \end{equation}
with $D$ being the dimension of the Hilbert space in which $\rho_A$ is defined, and $\mathbb{I}$ the identity matrix. Accordingly, multipartite states with maximal entanglement for every bipartition are called \textit{Absolutely Maximally Entangled} (AME); such kinds of states have applications in quantum protocols like threshold secret sharing or open-destination teleportation \cite{AME_definitions_and_application}. On the other hand, given a state of $N$ qudits, $\rho\in\mathcal{H}_d^{\otimes N}$, we call it \textit{k-uniform} when the entanglement is maximal for all bipartitions of $k$ qudits \cite{k-uniform-paper}. GHZ states of N-qudits (see Eq.~\eqref{eq:ghz}) are examples of 1-uniform states. 

With \pytheus we can pick types of bipartitions the entanglement is to be maximized. Choosing all bipartitions targets the production of an AME state, bipartitions of size $|A| = k$ target k-uniform states. A mathematical result of entanglement theory is that AMEs only exist for certain combinations of $N$ and $d$ (see table \ref{tab: AME_overview}). In such situations we choose to optimize for k-uniform states.

\begin{table}
\centering
\resizebox{0.45\textwidth}{!}{%
\begin{tabular}{c|ccccccccc}
\hline
\multirow{2}{*}{\:\textbf{Dim}\:} &
  \multicolumn{9}{c}{\textbf{Number of particles}} \\ \cline{2-10} 
 &
  \multicolumn{1}{c|}{\:2\:} &
  \multicolumn{1}{c|}{\:3\:} &
  \multicolumn{1}{c|}{\:4\:} &
  \multicolumn{1}{c|}{\:5\:} &
  \multicolumn{1}{c|}{\:6\:} &
  \multicolumn{1}{c|}{\:7\:} &
  \multicolumn{1}{c|}{\:8\:} &
  \multicolumn{1}{c|}{\:9\:} &
  10 \\ \hline
2 &
  \multicolumn{1}{c|}{\textcolor{Green}{\cmark}} &
  \multicolumn{1}{c|}{\textcolor{Green}{\cmark}} &
  \multicolumn{1}{c|}{\textcolor{red}{\xmark}} &
  \multicolumn{1}{c|}{\textcolor{Green}{\cmark}} &
  \multicolumn{1}{c|}{\textcolor{Green}{\cmark}} &
  \multicolumn{1}{c|}{\textcolor{red}{\xmark}} &
  \multicolumn{1}{c|}{\textcolor{red}{\xmark}(3)} &
  \multicolumn{1}{c|}{\textcolor{Green}{\cmark}(3)} &
  \textcolor{red}{\xmark}(3) \\ \hline
3 &
  \multicolumn{1}{c|}{\textcolor{Green}{\cmark}} &
  \multicolumn{1}{c|}{\textcolor{Green}{\cmark}} &
  \multicolumn{1}{c|}{\textcolor{Green}{\cmark}} &
  \multicolumn{1}{c|}{\textcolor{Green}{\cmark}} &
  \multicolumn{1}{c|}{\textcolor{Green}{\cmark}} &
  \multicolumn{1}{c|}{\textcolor{Green}{\cmark}} &
  \multicolumn{1}{c|}{\textcolor{red}{\xmark}(3)} &
  \multicolumn{1}{c|}{\textcolor{Green}{\cmark}} &
  \textcolor{Green}{\cmark} \\ \hline
4 &
  \multicolumn{1}{c|}{\textcolor{Green}{\cmark}} &
  \multicolumn{1}{c|}{\textcolor{Green}{\cmark}} &
  \multicolumn{1}{c|}{\textcolor{Green}{\cmark}} &
  \multicolumn{1}{c|}{\textcolor{Green}{\cmark}} &
  \multicolumn{1}{c|}{\textcolor{Green}{\cmark}} &
  \multicolumn{1}{c|}{\textcolor{Green}{\cmark}} &
  \multicolumn{1}{c|}{\textbf{¿?}} &
  \multicolumn{1}{c|}{\textcolor{Green}{\cmark}} &
  \textcolor{Green}{\cmark} \\ \hline
5 &
  \multicolumn{1}{c|}{\textcolor{Green}{\cmark}} &
  \multicolumn{1}{c|}{\textcolor{Green}{\cmark}} &
  \multicolumn{1}{c|}{\textcolor{Green}{\cmark}} &
  \multicolumn{1}{c|}{\textcolor{Green}{\cmark}} &
  \multicolumn{1}{c|}{\textcolor{Green}{\cmark}} &
  \multicolumn{1}{c|}{\textcolor{Green}{\cmark}} &
  \multicolumn{1}{c|}{\textcolor{Green}{\cmark}} &
  \multicolumn{1}{c|}{\textcolor{Green}{\cmark}} &
  \textcolor{Green}{\cmark} \\ \hline
\end{tabular}%
}
\caption{\textbf{Known AME states up to 10 particles and 5 dimensions.} \textcolor{red}{\xmark}: no AME exist for the 2-dimensional systems of 4, 7, 8, and 10 particles, nor for the 3-dimensional system of 8 particles. \textbf{¿?}: it is unknown whether there is an AME for the 4-dimensional system of 8 particles. \textcolor{Green}{\cmark}: there exist an AME state for the rest of systems. For the 2-dimensional systems of 8, 9, and 10 particles, as well as for the 3-dimensional system of 8 particles, there exist a state with maximally entangled partitions of 3 particles \cite{k-uniform-paper,shi2020constructions}. The existence of some of these state does not imply that they have been (or can be) experimentally realized.}
\label{tab: AME_overview}
\end{table}

\begin{figure}[!t]
	\centering
	\subfloat[$\ket{\text{AME}(5,2)}_a$]{\includegraphics[width=0.24\textwidth]{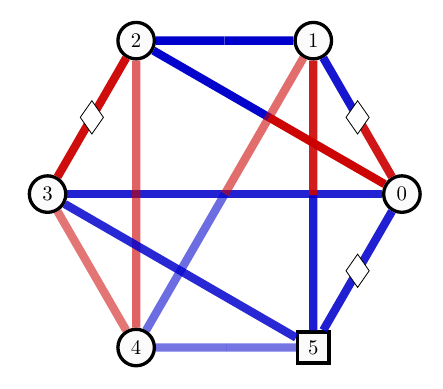}}\!\!
	\subfloat[$\ket{\text{AME}(5,2)}_b$]{\includegraphics[width=0.24\textwidth]{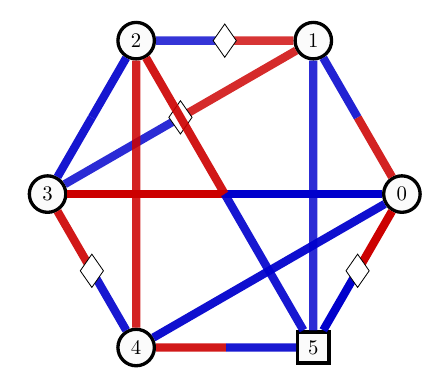}}
  \vspace{-3mm}
  \vspace{-3mm}
	\subfloat[$\ket{\text{AME}(6,2)}$]{\includegraphics[width=0.24\textwidth]{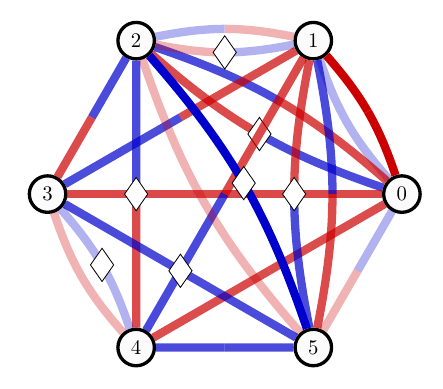}}\\
	\caption {Absolutely Maximally Entangled states with 5 and 6 qubits. The 6-qubit graph produces an asymptotic state.}
	\label{fig:AMEs}	
\end{figure}

\paragraph{AME States --} 
Since no AME exist for a system of four qubits (see table \ref{tab: AME_overview}), we start optimizing the entanglement for all possible partitions for five qubits, obtaining two different AME states \numbering[https://github.com/artificial-scientist-lab/PyTheus/tree/main/pytheus/graphs/MaxEntanglement/ame_5qubit_a]{experimentcounter} \numbering[https://github.com/artificial-scientist-lab/PyTheus/tree/main/pytheus/graphs/MaxEntanglement/ame_5qubit_b]{experimentcounter}
\begin{align}
    \ket{\text{AME}(5,2)}_a &= \ket{00000} + \ket{01101} + \ket{01110} \notag \\
    &+ \ket{10110} + \ket{11000} + \ket{11011} \notag \\
    &- \ket{00011} - \ket{10101}, \\
      \ket{\text{AME}(5,2)}_b  &=    \ket{01011} + \ket{01100} + \ket{10110} \notag \\ 
      &+ \ket{11010} - \ket{00000} - \ket{00111} \notag \\
      &- \ket{10001} - \ket{11101}. 
\label{eq: AME5Qubits}
\end{align}
For six qubits we found an asymptotic state \numbering[https://github.com/artificial-scientist-lab/PyTheus/tree/main/pytheus/graphs/MaxEntanglement/AMEepsilon6qubits]{experimentcounter}
\begin{align}
    \ket{\text{AME}(6,2)} & \approx \ket{000010} + \ket{000100} + \ket{001001} \notag\\ 
  & + \ket{010101}  +  \ket{011000} + \ket{011110} \notag\\ 
  & + \ket{101010} + \ket{110000} +  \ket{111011} \notag\\ 
  & + \ket{111101} - \ket{001111} - \ket{010011} \notag\\ 
  & - \ket{100001} - \ket{100111} - \ket{101100} \notag\\ 
  & -  \ket{110110} + \varepsilon (\ket{000000} + \ket{000110} \notag\\ 
  & +  \ket{010001} + \ket{010111}  +  \ket{110010} \notag\\ 
  & + \ket{110100} + \ket{111001} + \ket{111111} ).
\end{align}
This is not exactly the AME state of eight qubits, it has eight terms which vanish asymptotically as $\varepsilon$ goes to zero. However, by doing so, we reduce the creation rate of the whole state. The graphs to produce this and the 5-qubits AME states are shown in Fig.~\ref{fig:AMEs}.

\paragraph{k-Uniform States --}
While the AME solution for six qubits was asymptotic, we can generate the exact 2-uniform state of such a system \numbering[https://github.com/artificial-scientist-lab/PyTheus/tree/main/pytheus/graphs/MaxEntanglement/k2uniform6qubits]{experimentcounter}
\begin{align}
      \ket{\text{U}2}_6^2 &= \ket{100011} + \ket{101110} - \ket{000000}\notag \\
 &  - \ket{001101} -  \ket{010111} - \ket{011010} \notag \\
 & - \ket{110100} - \ket{111001}. 
\end{align}
Twelve out of twenty three-qubit bipartitions of the above state are maximally entangled. 

\begin{figure}[!h]
	\centering
	{\includegraphics[width=0.24\textwidth]{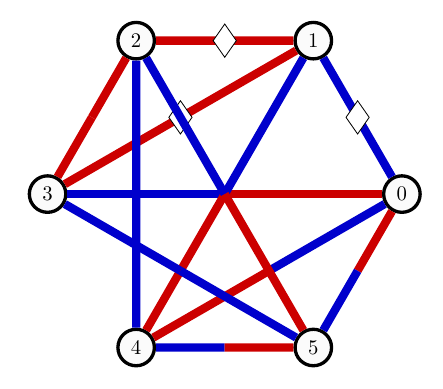}}\!\!
	{\includegraphics[width=0.24\textwidth]{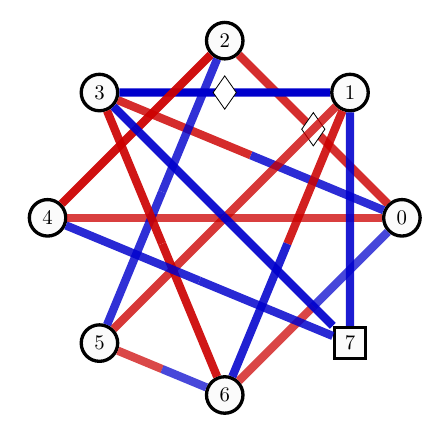}}
	\caption {$k$=2 uniform state for 6 qubits and $k$=2 (almost) uniform state for 7 qubits (+ one ancilla).}
	\label{fig:uniforms}	
\end{figure}

Similarly, we generated a 7-qubit state that maximizes the entanglement between each of the 21 possible $k=2$ partitions except one \numbering[https://github.com/artificial-scientist-lab/PyTheus/tree/main/pytheus/graphs/MaxEntanglement/k2uniform7qubits]{experimentcounter}
\begin{align}
    \ket{\sim \text{U}2}_7^2 &= \ket{0011110} + \ket{0101000} + \ket{0110111}  \notag \\
 & + \ket{1001101} + \ket{1010010} + \ket{1100100}  \notag \\
 & - \ket{0000001} - \ket{1111011}.
\end{align}
The graph to produce this `almost' uniform state and the previous one are shown in Fig.~\ref{fig:uniforms}.

\paragraph{Other States --}
As we have seen, even when an AME or k-uniform state for a given system of qudits exists, it is not always realizable with linear optics. However, even if not perfect, we have found several highly entangled states which are worth mentioning. We refer to them as
\begin{equation}
    \ket{\text{Ent(n,d,k)}},
\end{equation} 
where $n$ stands for the number of particles, $d$ for their dimensions, and $k$ for the size of the partitions. 

\begin{figure}[!h]
	\centering
	\subfloat[$\ket{\text{Ent(4,2,2)}}_{\mathbb{R}}$]{\includegraphics[width=0.24\textwidth]{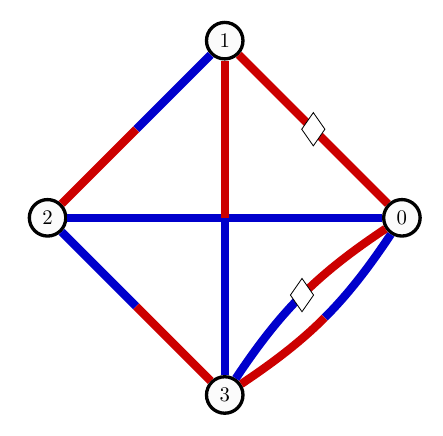}}\!\!
	\subfloat[$\ket{\text{Ent(4,2,2)}}_{\mathbb{C}}$]{\includegraphics[width=0.24\textwidth]{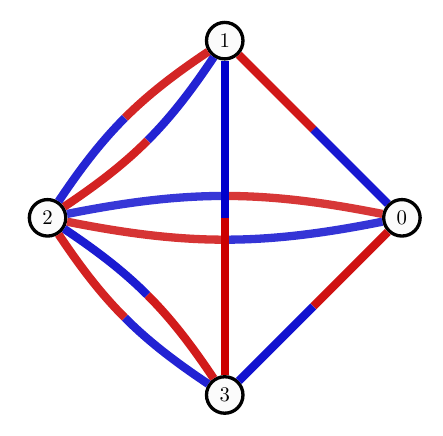}}
 \caption {The graphs produced maximally entangled states for single qubit partitions on a 4 qubit system, they also present high entanglement between 2 qubit partitions (even if not maximal). The weights on the left graph are real, and the weights on the right graph are complex (the phase of every edge can be found in the repository).}
	\label{fig:max4qubits}	
\end{figure}

For a system of four qubits, we find two interesting states when maximizing the entanglement (see Eq.~\eqref{eq:concurrence}) for every 2-qubit partition (see Fig.~\ref{fig:max4qubits}). We first optimize using real weights \numbering[https://github.com/artificial-scientist-lab/PyTheus/tree/main/pytheus/graphs/MaxEntanglement/k2maximal4qubitsREAL]{experimentcounter}
\begin{equation}
    \ket{\text{Ent(4,2,2)}}_{\mathbb{R}} =   \ket{1010} + \ket{1101} -   \ket{0011} - \ket{0100},
\end{equation}
obtaining a state which maximizes all possible partitions except one. For the second solution, we extend the weights to the complex domain, finding a local optimal \numbering[https://github.com/artificial-scientist-lab/PyTheus/tree/main/pytheus/graphs/MaxEntanglement/k2maximal4qubitsCOMPLEX]{experimentcounter}
\begin{align}\label{eq:complex4qubitk2} 
    \ket{\text{Ent(4,2,2)}}_{\mathbb{C}} &= \ket{0011} +   e^{ - i 0.54 \pi } \ket{0101}  \notag \\
    & + e^{ - i 0.93\pi }\ket{0110} +  e^{ i 0.45 \pi}\ket{1001} \notag \\ 
    & + e^{ i 0.71 \pi}\ket{1010}  +  e^{ i 0.86 \pi}\ket{1100}. 
\end{align}
This state minimizes the sum of partial traces for $k=2$ partitions, following Eq.~\eqref{eq:concurrence}. However, while the sum its (locally) minimal, and all qubit partitions are equally entangled, they are not maximally entangled for $k=2$. and all are equally  for all $k=1$ partitions and for none of the $k=2$. However, all $k=2$ partitions lead to the same value for the partial trace defined in Eq.~\eqref{eq:concurrence}. The entanglement seems to be `equally distributed'.

The complex state in Eq.~\eqref{eq:complex4qubitk2} gives the same entanglement structure as the Higuchi-Sudbery state presented in \cite{Higuchi2000}
\begin{align}\label{eq:HSstate}
    \ket{\text{HS}} = \ket{0011} + \ket{1100} &+ \omega(\ket{1010} + \ket{0101}) \notag \\
    &+ \omega^2(\ket{1001} + \ket{0110}),
\end{align}
where $\omega = e^{i2\pi/3}$.

Finally, for a system of 8 qubits, \pytheus found a state for which 48 out of 56 total three-qubit bipartitions are maximally entangled. \numbering[https://github.com/artificial-scientist-lab/PyTheus/tree/main/pytheus/graphs/MaxEntanglement/k3maximal8qubits]{experimentcounter}
\begin{align}
        \ket{\text{Ent(8,2,3)}} &= \ket{00010000} + \ket{00010111} \notag \\
 & + \ket{01101011} + \ket{01101100} \notag \\
 & + \ket{01110001} + \ket{10100010} \notag \\
 & + \ket{10100101} + \ket{10111111} \notag \\
 & + \ket{11000100} + \ket{11011001} \notag \\
 & - \ket{00001010} - \ket{00001101} \notag \\
 & - \ket{01110110} - \ket{10111000} \notag \\
 & - \ket{11000011} - \ket{11011110}.
\end{align}

\begin{figure}[!t]
\centering
	\subfloat[$\ket{\text{Ent(8,2,3)}}$]{\includegraphics[width=0.26\textwidth]{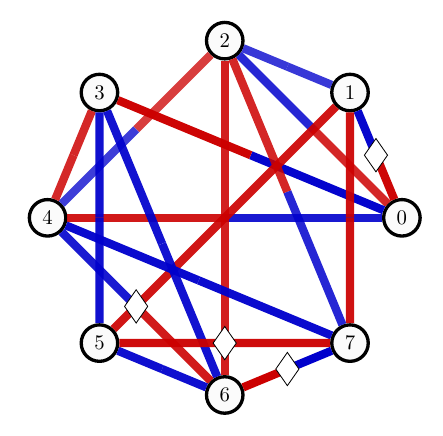}}
	\caption {The state generated by this graph is maximally entangled for 48 of the 56 possible 3-qubit partitions.}
	\label{fig:max8qubits}	
\end{figure}

The graph to produce this state is shown in Fig.~\ref{fig:max8qubits}.

It would be interesting to extend the maximization of other entanglement measures, such as experimental feasible, strong forms of all-vs-nothing violations \cite{lawrence2014rotational}, Hardy's version of local realism experiments which goes beyond the violations in Ref. \cite{hardy1993nonlocality,chen2017experimental} (for example, by involving more particles), or other measures of quantum correlations. Such extensions would contribute to the study of artificial intelligence for the foundations of quantum mechanics \cite{bharti2020machine}.

\subsection{Generation of Mixed States}
\tocless\subsubsection{Werner State}
\begin{figure}[!h]
	\centering
	\includegraphics[width=0.25\textwidth]{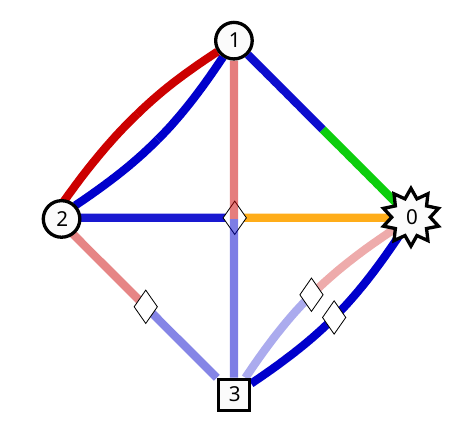}
	\caption {Graph for the Werner state with $\alpha = 0.69$. Information at vertex 0 is to be discarded. Vertex 3 is an ancillary detector}
	\label{fig:werner}
\end{figure}

The Werner state with the density matrix
\begin{align}
    \rho_\alpha = \alpha \ket{\Phi^+}\bra{\Phi^+} + (1-\alpha)\frac{\mathbb{I}}{4},
\end{align}
has different properties for different values of $\alpha$. For $\alpha < 0.683$, it is known to be Bell-local. For $\alpha > 0.697$, it is known to be Bell nonlocal \cite{Bowles2021singlecopy}. This leaves the gap between $0.683$ and $0.697$ open without a mathematical proof of its properties. By creating this state, it could be possible to test the Bell-locality of a state in the middle of the gap, $\alpha = 0.69$. The graph for the creation of this mixed state is shown in Fig.~\ref{fig:werner} \numbering[https://github.com/artificial-scientist-lab/PyTheus/tree/main/pytheus/graphs/MixedStates/werner]{experimentcounter}
.

\tocless\subsubsection{Counter-Example to the Peres Conjecture}
Relations between different notions of quantum correlations are studied in quantum information theory. These insights are important to understand the underlying resources of quantum correlations and their applicability in quantum protocols. In 1999, Asher Peres conjectured that the notion of Bell nonlocality and distillability is equivalent \cite{peres1999all}. Bell nonlocality, as witnessed by a violation of Bell's inequality, says that no local hidden-variable model can explain the correlations observed in an experiment. A state is called distillable if a maximally entangled Bell pair can be extracted from multiple copies. The conjecture was open for 15 years until a counter-example was discovered in 2014, first for a strong version of the conjecture \cite{moroder2014steering} and shortly after in its full form \cite{vertesi2014disproving}. These violations are based on the uncovering of a two-particle bound state (a state that cannot be distilled) which can be used for quantum steering and violation of Bell's inequality. The Moroder-Gittsovich-Huber-G\"uhne-Vertesi-Brunner (MGHG-VB) state can be written as a mixed state that is an incoherent superposition of four entangled two-qutrit states,
\numbering[https://github.com/artificial-scientist-lab/PyTheus/tree/main/pytheus/graphs/MixedStates/peres]{experimentcounter}
\begin{align}
    \rho = \sum_{i=1}^4 \lambda_i \ket{\psi_i}\bra{\psi_i}
\end{align}
with $\lambda=\left(\frac{3257}{6884},\frac{450}{1721},\frac{450}{1721},\frac{27}{6884}\right)$ and
\begin{align*}
    \ket{\psi_1}&=\frac{1}{\sqrt{2}}\left(\ket{00}+\ket{11} \right) \\
    \ket{\psi_2}&=\frac{\sqrt{131}}{12\sqrt{2}}\left(\ket{01}+\ket{10} \right)+\frac{1}{60}\ket{02}-\frac{3}{10}\ket{21}\\
    \ket{\psi_3}&=\frac{\sqrt{131}}{12\sqrt{2}}\left(\ket{00}-\ket{11} \right)+\frac{1}{60}\ket{12}-\frac{3}{10}\ket{20}\\
    \ket{\psi_4}&=\frac{1}{\sqrt{3}}\left(-\ket{01}+\ket{10}+\ket{22} \right).
\end{align*}
\begin{figure}[t]
	\centering
	\includegraphics[width=0.25\textwidth]{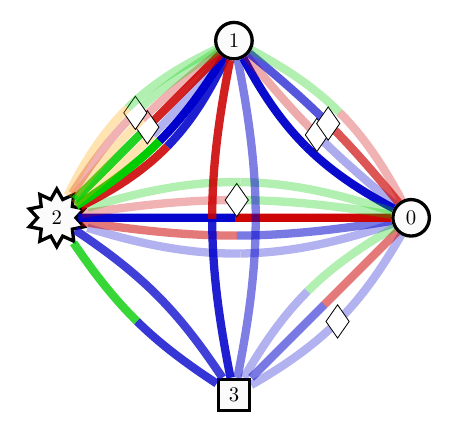}
	\caption {Graph for the Moroder-Gittsovich-Huber-G\"uhne-Vertesi-Brunner (MGHG-VB) state, which is a counter-example to the Peres conjecture. Here, vertex 3 is an ancilla photon and vertex 2 is the environment vertex. Due to the structure of the MGHG-VB state, the edge weights are not simple fractions.}
	\label{fig:PeresGraph}
\end{figure}

The graph that corresponds to this state is shown in Fig.~\ref{fig:PeresGraph}. It might be interesting to generate this state and show experimentally that the state can violate Bell's inequality and thereby demonstrate an experimental falsification of Peres's conjecture.

\subsection{Generation of Entanglement in the Photon-Number Basis}\label{sec:state_in_fock}
\tocless\subsubsection{N00N States}\label{sec:noon_states}

The ability to perform precise measurements is always affected by several limitations, some of them avoidable by careful design of experiments and others of a fundamental nature. Classically, one can estimate a phase parameter more precisely by increasing the number of particles $N$ in the measurement process; the precision is typically limited to $1/\sqrt{N}$, also known as the standard quantum limit (SQL). With the aid of quantum resources, one can beat the SQL, and the measurement precision can be enhanced to scale as $1/N$, approaching the Heisenberg limit that is governed by the physical law of quantum mechanics. One prominent representative of quantum states with the ability to break the SQL is the well-known N00N states \cite{giovannetti2004quantum,polino2020photonic}, as described in Eq.~\eqref{eq:NOON}. 

These N00N states can not only be used to test fundamental physics via violating Bell-type inequalities \cite{bellviolationNOON} but also play an important role in quantum-enhanced applications, quantum-enhanced microscopes and imaging systems \cite{SupersensitivePRL,ono2013entanglement,gao2022high,ndagano2022quantum}, super-resolving phase measurements \cite{mitchell2004super, walther2004broglie,fourNOON2006, sixNOON2007}, and quantum lithography \cite{dowling2000}, to name a few. Interestingly, as the two-mode N00N states are formed in a superposition of two distinct states $\ket{N,0}$ and $\ket{0,N}$, one could consider that the two terms correspond to the ``dead cat" and ``alive cat" in the ``Schr\"{o}dinger-cat'' form \cite{afek2010high}. With high-N00N states, we might step closer to a better understanding of macroscopic entanglement envisioned by Schr\"{o}dinger \cite{schrodinger1935gegenwartige}. In the optical regime, achieving an efficient photonic source of N00N states with large $N$ is very challenging. Several schemes and experiments have been toward the direction of generating photonic high-N00N states in different ways, such as by using strong optical nonlinearities \cite{DowlingPathEPR2007}, by linear optics and feed-forward \cite{feedforwardNOON}, and by mixing coherent light with SPDC photon-pair sources on a standard beam splitter \cite{PhysRevLett.100.073601,PhysRevA.76.031806,afek2010high,Afek2012}.

Additionally, interest has also been growing in the simultaneous estimation of multiple parameters using multi-mode quantum states \cite{zhang2018scalable,hong2021quantum}. Using a multimode N00N state, one can reach the Heisenberg limit with an $O(d)$ advantage over what is obtained with $d$ copies of a two-mode N00N state \cite{MultiplePhase2013,multiparameter2016}. The generalization of the two-mode N00N states in Eq.~\eqref{eq:NOON} is given as \cite{zhang2018scalable,hong2021quantum}
\begin{align}
\ket{\text{N00N}}^{N}_{d}:=&\frac{1}{\sqrt{d}}(\ket{N,0,...,0,0}\pm\ket{0,N,...,0,0}\pm...\\\nonumber
&\pm\ket{0,0,...,N,0}\pm\ket{0,0,...,0,N})_{1,...,d}.
\end{align}

Here we show how to produce perfect N00N states without using any coherent states or feed-forward. We list some examples that might give a different conceptual understanding and new insight. For more path-entangled states, one can directly employ \pytheus. For the ancillary detectors, we always assume they are reached by single photons, like in previous sections. 

\paragraph{Two Particles --} As we have introduced in section \ref{fock_basis}, the simplest N00N state is the $\ket{\text{N00N}}^{2}_{2}$ state that can be deterministically created via HOM interference when two indistinguishable photons are incident on a standard beam splitter simultaneously \cite{HOM1987}. From the graph in Fig.~\ref{fig:HOMeffect}, we can easily generalize the state to an arbitrary number of modes by adding more vertices, each one with a loop. The coherent superposition of a loop being in one of the vertices indicates the resulting multi-mode two-particle N00N states.

\begin{figure}
    \centering
    \subfloat[(a) $\ket{\text{N00N}}^3_2$]{\includegraphics[width=0.246\textwidth]{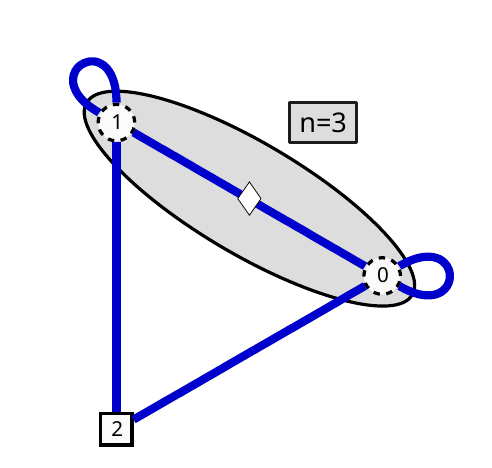}} \hspace{-4mm}
    \subfloat[(b) $\ket{\text{N00N}}^3_3$]{\includegraphics[width=0.246\textwidth]{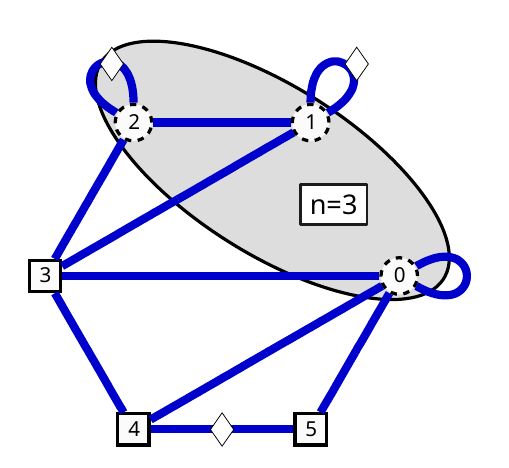}}
    \\ \vspace{-4mm}
    \subfloat[(c) $\ket{\text{N00N}}^3_4$]{\includegraphics[width=0.246\textwidth]{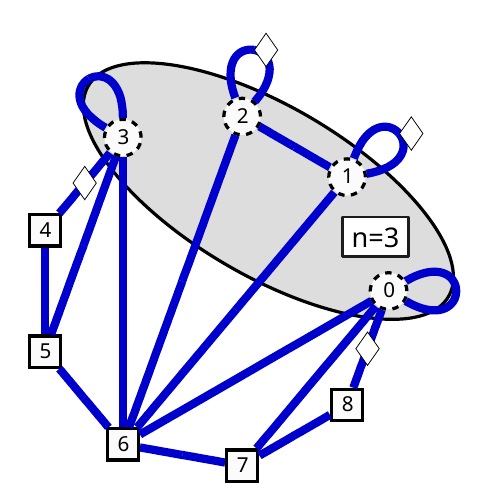}} \hspace{-4mm}
    \subfloat[(d) $\ket{\text{N00N}}^3_5$]{\includegraphics[width=0.246\textwidth]{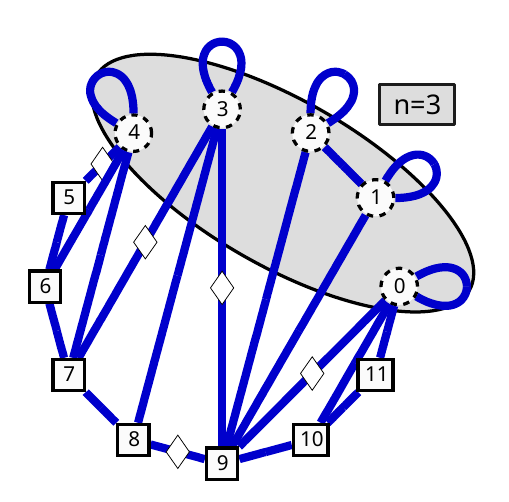}}
    \caption{Graphs for multi-mode three-particle N00N states $\ket{\text{N00N}}^{3}_{d}$ with increasing number of ancillas.}
    \label{fig:noon_multimode_all}
\end{figure}

\paragraph{Three Particles --} For producing 2-mode 3-photon N00N states $\ket{\text{N00N}}^{3}_{2}$, we additionally employ an ancilla particle in the experiment. We show the graph found by \pytheus in Fig.~\ref{fig:noon_multimode_all}, which uses only one color, thus corresponding to the standard path-entangled N00N states. The edge can also be in different colors such that we are able to construct graphs for polarization-entangled N00N states \cite{KimNOON2009,threeNOON2017,KimNOON2011}. We further explore the states in multi-mode cases, as described in Fig.~\ref{fig:noon_multimode_all} (b), (c) and (d) for 3-, 4-, 5-mode three-photon N00N states. Interestingly, the graphs found by \pytheus exhibit a nice structure, which may be amenable to generalization, to systematically produce other multi-mode NOON states without any optimization process.

\begin{figure}[!h]
    \centering
    \subfloat[(a) $\ket{40}-\ket{04}$]{\includegraphics[width=0.24\textwidth]{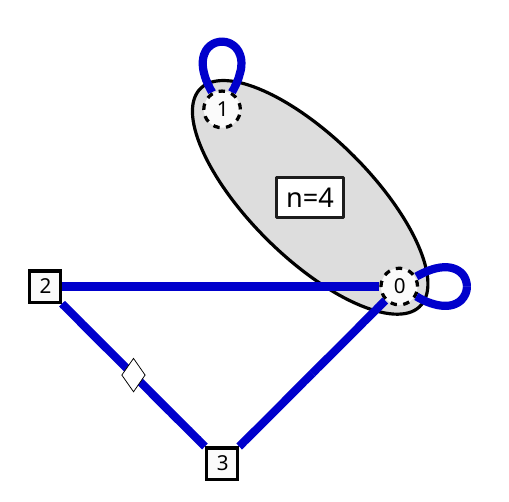}} \!\!
    \subfloat[(b) $\ket{400}-\ket{040}-\ket{004}$]{\includegraphics[width=0.24\textwidth]{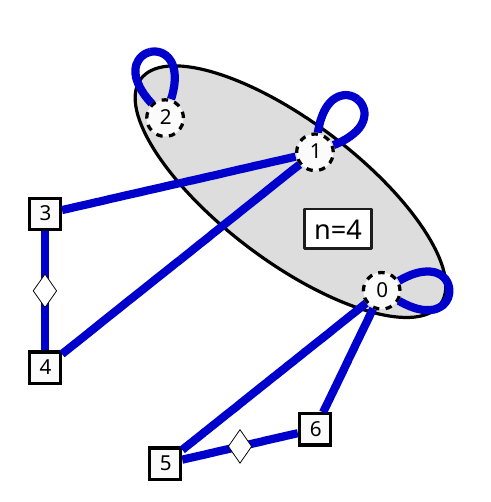}}
    \caption{Graphs with only real weights to generate 4-particle N00N states for 2 and 3 modes. Multiplying the $i$ the loops of the disconnected vertices all the amplitudes of the state's terms become positive. To increase the number of modes, we add a triangular subgraph connected to a loop. The edge connecting the two new ancillas must be negative.} 
    \label{fig:noon2m4ph}
\end{figure}

\paragraph{Four Particles --} For the previous N00N states, we optimized the graphs using real weights, finding the states $\ket{40}-\ket{04}$ and $\ket{400}-\ket{040}-\ket{004}$ \numbering[https://github.com/artificial-scientist-lab/PyTheus/blob/main/pytheus/graphs/FockStates/noon2m4ph2anc]{experimentcounter}\numbering[https://github.com/artificial-scientist-lab/PyTheus/tree/main/pytheus/graphs/FockStates/noon3m4ph4anc]{experimentcounter}. However, to generate the same N00N states with zero phase difference between the terms we need to, either add a $\pi/2$ phase on the solitary loops from Fig.~\ref{fig:noon2m4ph} or, using only real weights, add 2 more ancillas. Additionally, after generating the 4-particle N00N states for 2 and 3 modes, it became obvious how to generate the graphs for an arbitrary number of modes.

It is worth noting that given a ket $\ket{40}$, each combination of non-repeating edges contributes twice as much as one with duplicate edges (assuming equal weights). This is a consequence of the multinomial theorem applied to the creation operators described in Eq.~\eqref{eq:pairsource}. 

\begin{figure}
    \centering
    \subfloat[(a) $\ket{\text{N00N}}^5_2$]{\includegraphics[width=0.23\textwidth]{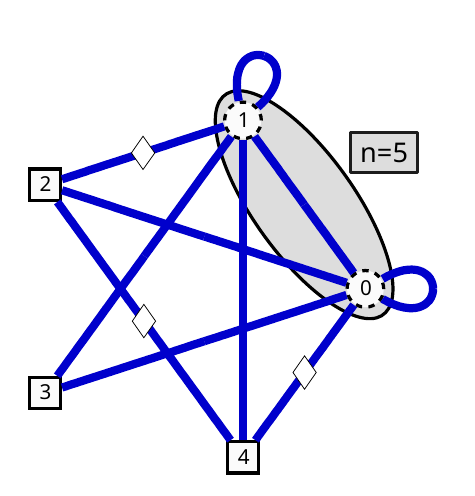}} \!\!
    \subfloat[(b) $\ket{\text{N00N}}^6_2$]{\includegraphics[width=0.25\textwidth]{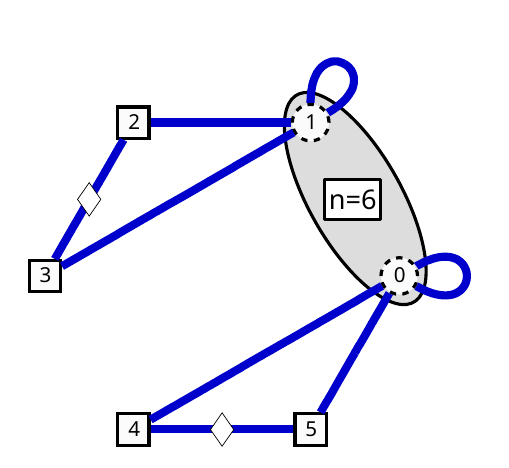}}
    \\ \vspace{-4mm}
    \subfloat[(c) $\ket{\text{N00N}}^7_2$]{\includegraphics[width=0.235\textwidth]{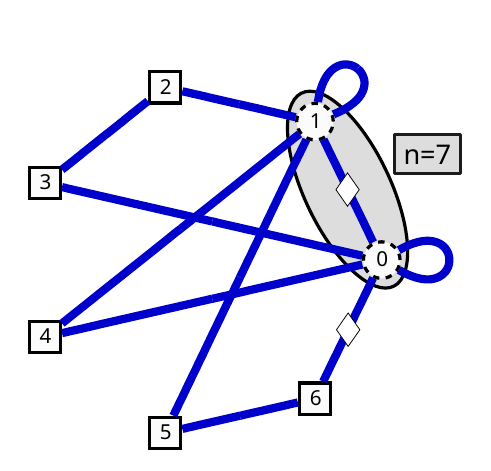}}\!\!
    \subfloat[(d) $\ket{\text{N00N}}^8_2$]{\includegraphics[width=0.24\textwidth]{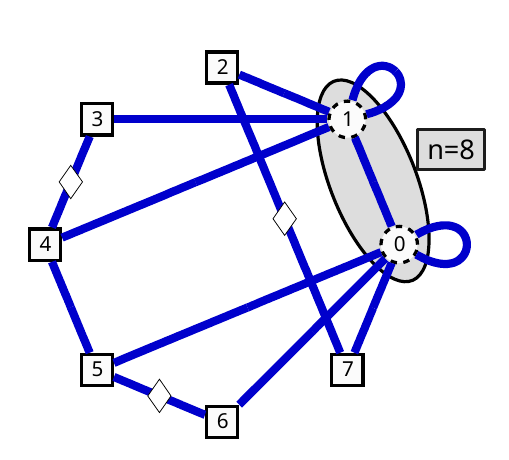}}
    \caption{Graphs for two-mode N00N states of 5, 6, 7, and 8 photons.}
    \label{fig:2mode_noons}
\end{figure}

\paragraph{Other Two-Mode States --} For the previous N00N states, the associated graphs have shown some patterns. The more obvious was the increasing need for ancillas, whether we were adding more photons or modes. Moreover, based on a few graphs, we realized how to generate N00N states with 2 and 4 photons for an arbitrary number of modes. In Fig.~\ref{fig:2mode_noons} we show how to produce the two-mode N00N state for 5, 6, 7, and 8 photons \numbering[https://github.com/artificial-scientist-lab/PyTheus/tree/main/pytheus/graphs/FockStates/noon2m5ph3anc]{experimentcounter}\numbering[https://github.com/artificial-scientist-lab/PyTheus/tree/main/pytheus/graphs/FockStates/noon2m6ph4anc]{experimentcounter}\numbering[https://github.com/artificial-scientist-lab/PyTheus/tree/main/pytheus/graphs/FockStates/noon2m7ph5anc]{experimentcounter}\numbering[https://github.com/artificial-scientist-lab/PyTheus/tree/main/pytheus/graphs/FockStates/noon2m8ph6anc]{experimentcounter}. As with the 3-photon N00N states, it seems to be an underlying pattern in the graphs solutions when increasing the number of photons.

\tocless\subsubsection{Platonic Solid States}\label{sec:platonic_state}
\begin{figure}[!t]
    \centering
    \subfloat[]{\includegraphics[width=0.21\textwidth]{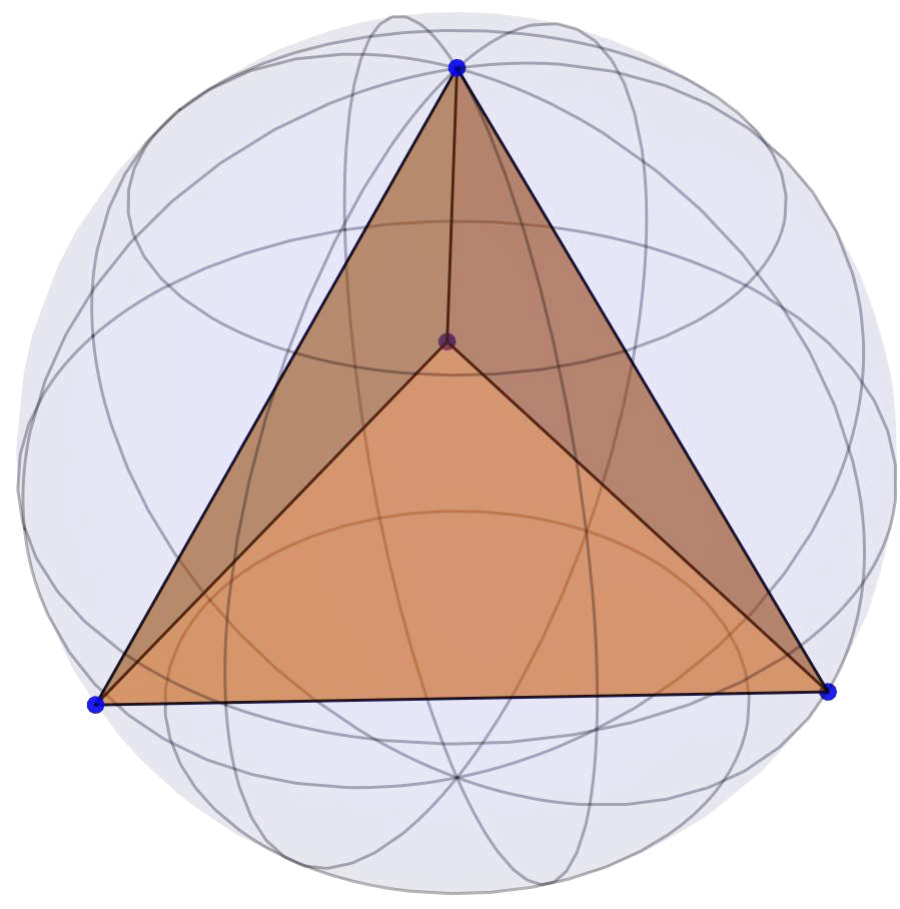}} \!\!
    \subfloat[]{\includegraphics[width=0.25\textwidth]{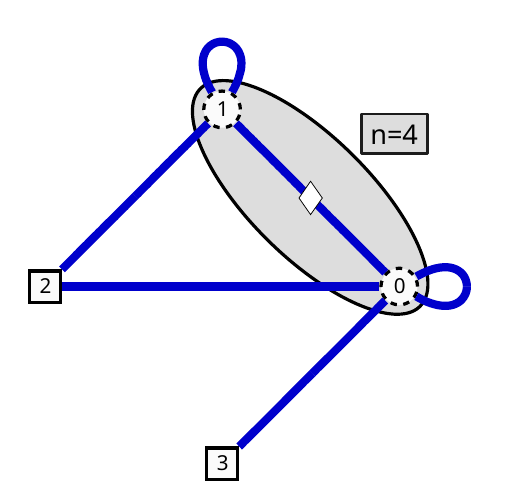}}\vspace{-1cm}
    \\
    \subfloat[]{\includegraphics[width=0.21\textwidth]{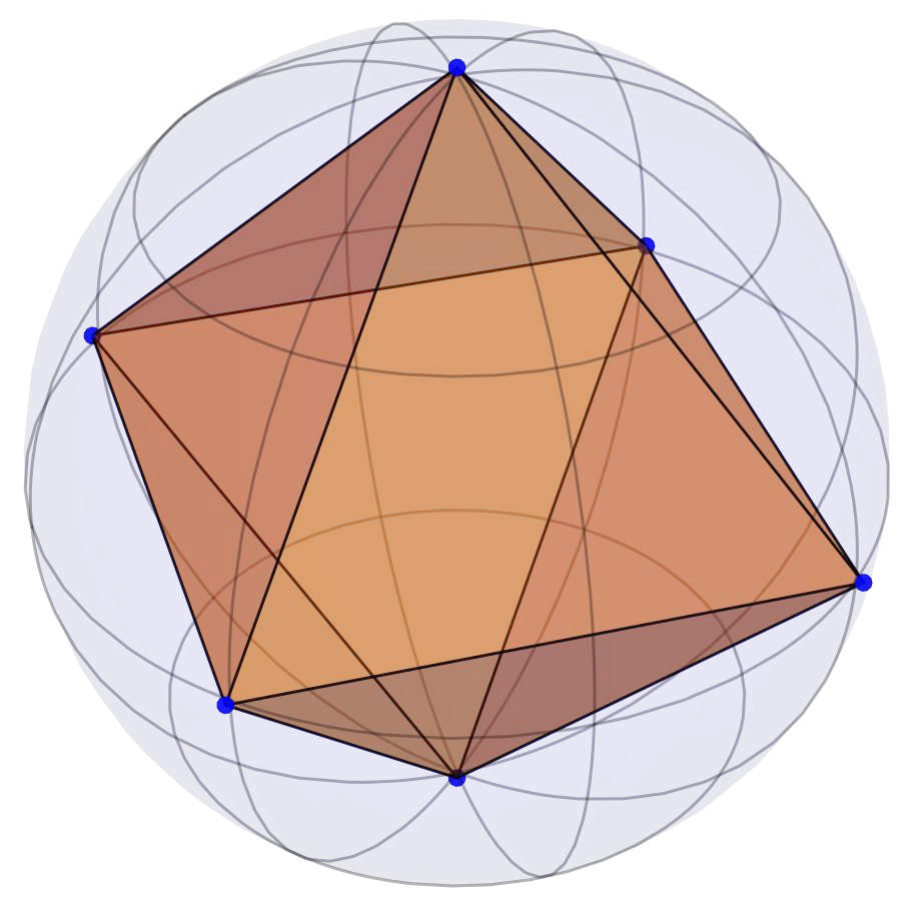}} \!\!
    \subfloat[]{\includegraphics[width=0.25\textwidth]{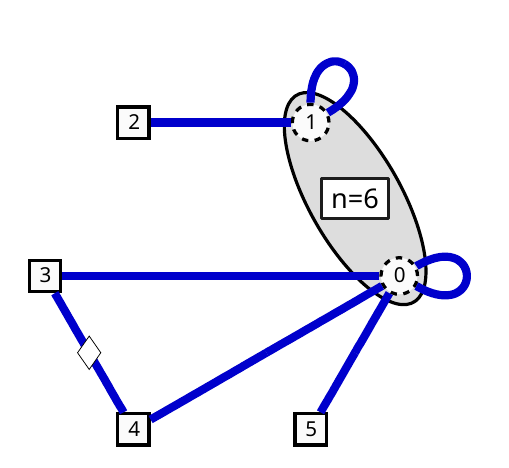}}\vspace{-1cm}
    \\ 
    \subfloat[]{\includegraphics[width=0.21\textwidth]{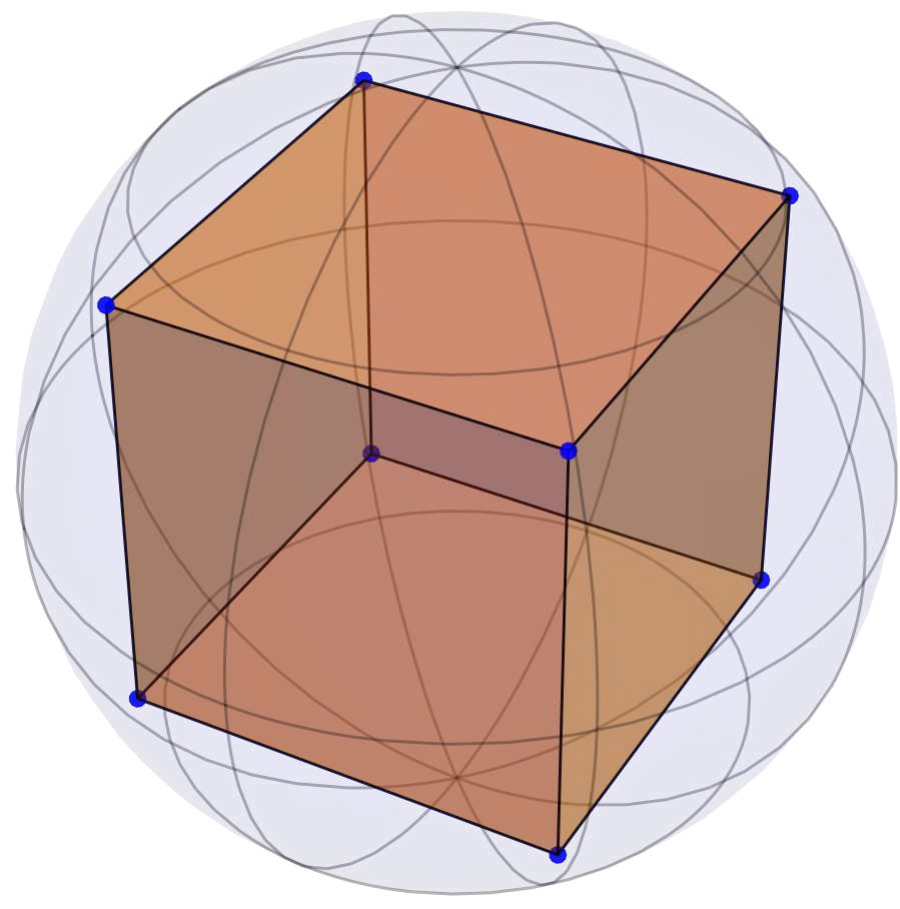}}  \!\!
    \subfloat[]{\includegraphics[width=0.25\textwidth]{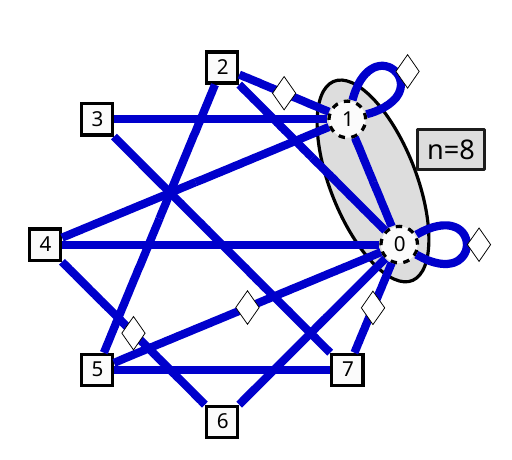}}
    
    \caption{Graphs for producing tetrahedron, octahedron, and cube quantum states.}
    \label{fig:solid_states}
\end{figure}

We have been considering states where all $N$ photons are in the same path, with the other paths empty. However, \pytheus can produce path-entangled states which do not fulfill such restrictions. Among them, a very special case leads to the highly symmetric Platonic Solid State \cite{bjork2015stars,extremalPolarization2015,Bouchard2017}, which can resolve rotations around any axis equally well. In the Platonic picture, shown in Fig.~\ref{fig:solid_states}, an $N$-photon state is mapped onto $N$ points on the Poincar\'e sphere \cite{majorana1932atomi}, offering a systematic way for visualization. This also relates to a long-lasting problem of distributing $N$ points on the Poincar\'e sphere in a highly symmetric fashion, where one can have many different solutions based on the function one tries to optimize \cite{conway1996packing,saff1997distributing}. Additionally, the concept of Platonic solids has also been used for fundamental investigations in quantum physics \cite{Tavakoli2020platonicsolids,Pal2022platonicbell}. Apart from their symmetric elegance, there is plenty of room for applications of these states, for instance, in magnetometry, polarimetry, and metrology. Researchers have started investigating how to generate these Platonic solid state. In this part, we show the examples found by \pytheus.

\paragraph{Tetrahedron --} Our first Platonic solid state is the tetrahedron \cite{Grasslwebsite}, which is the unique optimal four-photon state for characterizing polarization rotations. The state written in the Fock basis reads as 
\begin{equation}
    \ket{\Psi^{(1)}} = \frac{1}{\sqrt{3}}(\ket{4,0} + \sqrt{2}\ket{1,3}).
\end{equation}
It can be produced by the graph in Fig.~\ref{fig:solid_states} (top panels), and has already been implemented \cite{Hugothesis2022}. 

The similarity between the graphs that produce the tetrahedron and the three-photon N00N state discovered by \pytheus (see Fig.~\ref{fig:noon_multimode_all} a) shows a nice pattern: to generate states in the Fock basis, we can add an arbitrary number of photons to any path. In this way, starting with a general state $\ket{\text{M,N}} \pm \ket{\text{N,M}}$, we can obtain 
\begin{equation}\label{eq:photon_addition}
   \ket{\text{M,N}} \pm \ket{\text{N,M}} \Rightarrow \ket{\text{M+x,N+y}} \pm \ket{\text{N+x,M+y}}.
\end{equation}
We only need to connect the first and second paths to $x$ and $y$ ancillary photons, respectively. This construction can be used for an arbitrary number of modes.

\paragraph{Octahedron --}
The octahedron state reads
\begin{equation}
    \ket{\Psi^{(2)}} = \frac{1}{\sqrt{2}}(\ket{5,1} - \ket{1,5}),
\end{equation}
and was found by \pytheus using only 4 ancillary detectors (see Fig.~\ref{fig:solid_states}). Alternatively, we can apply the pattern described in Eq.~\eqref{eq:photon_addition} to the graph that produces $\ket{40} - \ket{04}$, which is shown in Fig.~\ref{fig:noon2m4ph} a).

\paragraph{Cube --}
The last platonic solid we are able to produce is the cube state, which reads \numbering[https://github.com/artificial-scientist-lab/PyTheus/tree/main/pytheus/graphs/FockStates/cube]{experimentcounter}
\begin{equation}
    \ket{\Psi^{(3)}} = \frac{\sqrt{3}}{12}(\sqrt{10}\ket{8,0} + 2\sqrt{7}\ket{4,4} + \sqrt{10}\ket{0,8}).
\end{equation}
In Fig.~\ref{fig:solid_states}, we show the graph that can be used to generate the cube state. As we can see from the expression and Fig.~\ref{fig:solid_states} (bottom), the graph cannot be obtained by modifying one of the previous graphs for a two-mode N00N state.

\subsection{Towards Quantum Simulation}\label{sec:qsim}
In this section, we present quantum state generation for states from condensed matter physics which could be interesting for quantum simulation\cite{aspuru2012photonic}. The quantum entangled states of few- or many-body systems can generally be expressed by the tensor network formulations~\cite{Schollwock2011, Cirac2021}. One of the most successful members of this family with a vast application in condensed matter physics is the 1D matrix product state~(MPS). Here one represents the quantum state of a system with periodic boundary conditions and $N$ particles as
\begin{align}
    \ket{\psi} = \sum_{s} \tr\left[ A_{1}^{(s_{1})} A_{2}^{(s_{2})} \ldots A_{N}^{(s_{N})}  \right] \ket{s_{1}\ldots s_{N}},
\end{align}
where $s_{i}=\{0,\ldots d-1\}$ denotes the local state of the $i$th particle with local physical dimension $d$. Here $A_{i}^{(s_{i})}$ is a complex matrix with dimension $\chi$, also known as the bond~(virtual) dimension. The matrix $A_{i}^{(s_{i})}$ can be viewed as a projector from a $\chi$-dimensional virtual~(correlation) vector space into the physical $d$-dimensional space~\cite{MiguelRamiro2020}. While in non-interacting systems, described by product states, the bond dimension $\chi$ is one, this quantity grows exponentially with $N$ in most strongly correlated systems. This results in employing numerical techniques to (approximately) obtain the ground states of most many-body systems. Nevertheless, one can find multiple few- or many-body ground states with $\chi>1$ but independent of the particle number. 
Here, we present a collection of these ground states with a concise description of their host Hamiltonian and their physical implication upon realization.
While our focus is on zero-temperature ground states, we emphasize that the applicability of \pytheus is not limited to these states. This is because both mixed states~\ref{sec:mixed}, which describe open quantum states, and states constitute combinations of (various) Fock states~\ref{fock_basis}, can be used to study the (grand) canonical ensemble, are treatable by \pytheus.

\tocless\subsubsection{Spin-1/2 Systems}
Due to the simplicity of spin-1/2 systems and the surge of interest in qubit quantum computation, these many-body systems received ever-growing attention.
The local physical space of these systems spanned by $\{ \ket{\uparrow}, \ket{\downarrow} \}$, or equivalently $\{ \ket{0}, \ket{1} \}$, results in the local physical dimension $d=2$. Despite such a small physical dimension, the obtained ground states can be rich and exotic. In the following, we list some of these states.

\paragraph{Spin-1/2 Wire --}
A computational quantum spin-1/2 wire governed by nearest-neighbor interactions, associated with non-zero two-point correlation functions and arbitrary local entanglement, is described by~\cite{Gross2010, MiguelRamiro2020}
\begin{align}
    \ket{\psi} =\sum_{s_i =0}^{1}
    \tr[ A (s_{n-1}) \ldots A(s_1) ] \ket{s_1 \ldots s_n},
\end{align}
where 
\begin{align}
    A(\uparrow) &=\frac{1}{\sqrt{2}} G,\quad G= \exp (\i \pi \sigma^{x}/\tau) ,\\
    A(\downarrow) &= \frac{1}{\sqrt{2}} G T (\phi), \quad T=
    \begin{pmatrix}
        e^{-\i \phi/2} & 0 \\
        0 & e^{\i \phi /2}
    \end{pmatrix},
\end{align}
where $\phi$ and $\tau$ denote the entanglement factor and period, respectively.

\begin{figure}[!h]
	\centering
	\includegraphics[width=0.26\textwidth]{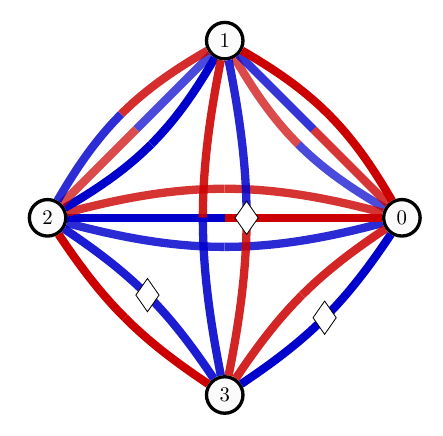}
 \caption {The associated graph for the ground state of the spin-1/2 wire with four particles.}
	\label{fig:spinhalf_wire}
\end{figure}

Fig.~\ref{fig:spinhalf_wire} displays the associated graph for the ground state of the spin-1/2 wire for four particles with $\phi= \pi/2$. The state, up to normalization, reads ~\numbering[https://github.com/artificial-scientist-lab/PyTheus/tree/main/pytheus/graphs/CondensedMatter/oneDspinhalfwire4]{experimentcounter}
\begin{align}
\ket{\psi} &= \sqrt{2}\ket{0000} + \ket{0001} + \ket{0010} + \ket{0100} \notag \\
&- \ket{0111} + \ket{1000} - \ket{1011} \notag \\
&- \ket{1101} - \ket{1110} - \sqrt{2}\ket{1111}.
\end{align}

\paragraph{Spin-1/2 States with No Adjacent Spin-ups --}
Aside from the previous spin-1/2 state, one can compute entangled states where no two neighboring spin-ups appear in the ground state. One can expect to detect these states in spin systems with nearest-neighbor interactions. In the Rydberg-atom experiments, this situation occurs due to the Rydberg blockade~\cite{Bernien2017}. The matrix product representation of these states reads 
\begin{align}
    \ket{\psi} =\sum_{s_i =0}^{1}
    \tr[ A (s_{n-1}) \ldots A(s_1) ] \ket{s_1 \ldots s_n},
\end{align}
where
\begin{align}
    A(\uparrow) &=\frac{1}{\sqrt{2}} ({\mathbb I} + \sigma^{x} ),\\
    A(\downarrow) &= 2 \sigma^{+},
\end{align}
with $\sigma^{+}=\sigma^{x}+ \i \sigma^{y}$.

\begin{figure}[t]
	\centering
	\subfloat[3-body]{\includegraphics[width=0.24\textwidth]{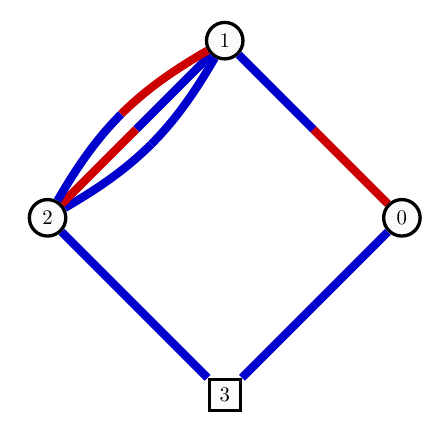}}\!\!
 \subfloat[4-body]{\includegraphics[width=0.24\textwidth]{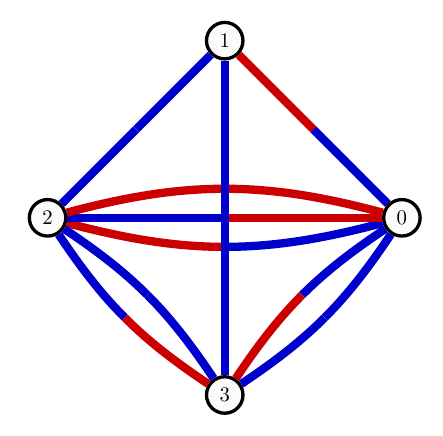}}
 \\ \vspace{-3mm}
\subfloat[5-body]{\includegraphics[width=0.24\textwidth]{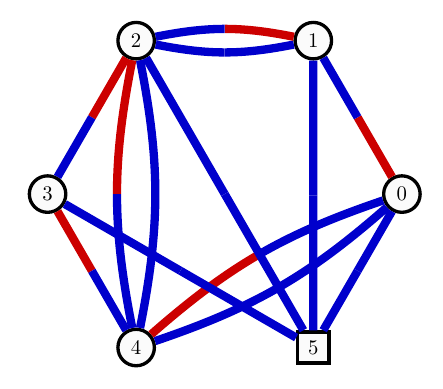}}\!\!
 \subfloat[6-body]{\includegraphics[width=0.24\textwidth]{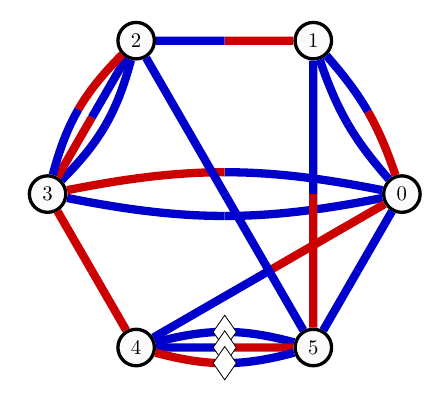}}
 \caption{Associated graphs for spin-1/2 states with no adjacent spin-ups for three, four, five, and six particles. One ancillary particle is included in states with an odd number of particles~(left column).
 }
	\label{fig:spinhalf_noadjup}
\end{figure}

Fig.~\ref{fig:spinhalf_noadjup} displays the associated graphs for this system with various numbers of particles whose ground states read

\begin{itemize}
\item  \numbering[https://github.com/artificial-scientist-lab/PyTheus/tree/main/pytheus/graphs/CondensedMatter/nbody3]{experimentcounter} Three particles with one ancillary particle:
\begin{align}
\ket{\psi_3} &= \ket{000} + \ket{001} + \ket{010} + \ket{100}.
\end{align}
\item  \numbering[https://github.com/artificial-scientist-lab/PyTheus/tree/main/pytheus/graphs/CondensedMatter/nbody4]{experimentcounter} Four particles:
\begin{align}
\ket{\psi_4} &= \ket{0000} + \ket{0001} + \ket{0010} \notag \\
&+ \ket{0100} + \ket{0101} + \ket{1000} + \ket{1010}.
\end{align}
\item  \numbering[https://github.com/artificial-scientist-lab/PyTheus/tree/main/pytheus/graphs/CondensedMatter/nbody5]{experimentcounter} Five particles with one ancillary particle:
\begin{align}
\ket{\psi_5} &= \ket{00000} + \ket{00001} + \ket{00010} \notag \\
&+ \ket{00100} + \ket{00101} + \ket{01000} \notag \\
&+ \ket{01001} + \ket{01010} + \ket{10000} \notag \\
&+ \ket{10010} + \ket{10100}.
\end{align}
\item  \numbering[https://github.com/artificial-scientist-lab/PyTheus/tree/main/pytheus/graphs/CondensedMatter/nbody6]{experimentcounter} Six particles:
\begin{align}
\ket{\psi_6} &= \ket{000000} + \ket{000001} + \ket{000010} \notag \\
&+ \ket{000100} + \ket{000101} + \ket{001000} \notag \\
&+ \ket{001001} + \ket{001010} + \ket{010000} \notag \\
&+ \ket{010001} + \ket{010010} + \ket{010100} \notag \\
&+ \ket{010101} + \ket{100000} + \ket{100010} \notag \\
&+ \ket{100100} + \ket{101000} + \ket{101010}.
\end{align}
\end{itemize}

\paragraph{Majumdar-Gosh Model --}
The one-dimensional quantum Heisenberg spin model is known
as the Majumdar-Gosh model when the value of
the next-nearest-neighbor interaction is half the value of the nearest-neighbor antiferromagnetic exchange interaction.
The Hamiltonian of this model casts~\cite{Perez-Garcia2006}
\begin{align}
    H= \sum_{i} 2 \overrightarrow{\sigma}_{i} \cdot\overrightarrow{\sigma}_{i+1} 
    +
    \overrightarrow{\sigma}_{i} \cdot\overrightarrow{\sigma}_{i+2},
\end{align}
where $\overrightarrow{\sigma}_{i}=(\sigma^{x}_{i}, \sigma^{y}_{i}, \sigma^{z}_{i})$.

 The ground states of this model Hamiltonian are dimerized states given by products of singlet configurations of spins on neighboring sites. The linear combination of these states reads~\cite{Perez-Garcia2006}
 \begin{align}
    \ket{\psi} =\sum_{s_i =0}^{1}
    \tr[ A (s_{n-1}) \ldots A(s_1) ] \ket{s_1 \ldots s_n},
\end{align}
where
\begin{align}
    A^{\uparrow} &=
    \begin{pmatrix}
        0 & 1 & 0 \\
        0 & 0 & 0 \\
        \frac{1}{\sqrt{2}} & 0 & 0
    \end{pmatrix}
    ,\quad
    A^{\downarrow}=
    \begin{pmatrix}
        0  & 0 & 1 \\
        -\frac{1}{\sqrt{2}} & 0 & 0 \\
        0  & 0 & 0
    \end{pmatrix}.
\end{align}
This ground state of the Majumdar-Gosh model is one of the simplest spin-1/2 valence-bond solids in one-dimensional systems.

\begin{figure}[!h]
    \centering
    {\includegraphics[width=0.24\textwidth]{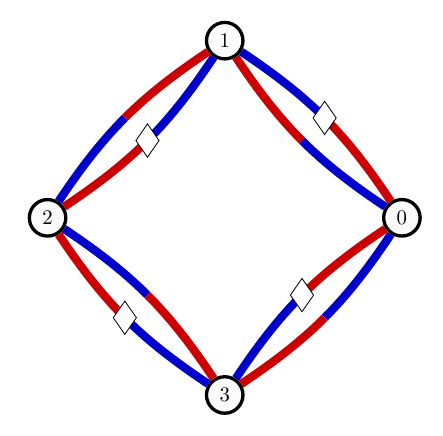}}\!\!
	{\includegraphics[width=0.24\textwidth]{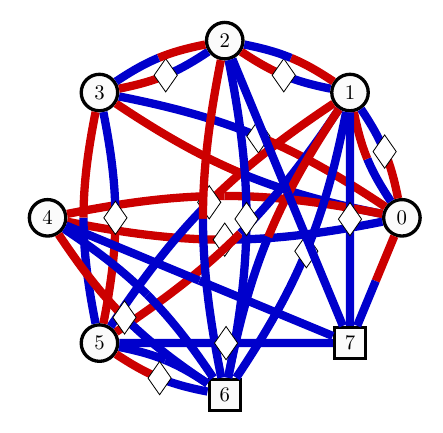}}
    \caption{Ground states of the Majumdar-Gosh model with four~(left panel) and six~(right) particles. Two ancillary particles are included in the right panel.}
    \label{fig:majumdar}
\end{figure}

 Fig.~\ref{fig:majumdar} displays the associated graphs for the four- and six-particle systems whose ground states read
\begin{itemize}
 \item \numbering[https://github.com/artificial-scientist-lab/PyTheus/tree/main/pytheus/graphs/CondensedMatter/majumdar4]{experimentcounter} Four particles:
\begin{align}
\ket{\psi^{(4)}} =&  \ket{0011} -2 \ket{0101} +\ket{0110} \nonumber \\
 +&\ket{1100}-2\ket{1010}+ \ket{1001}.
\end{align}
 \item \numbering[https://github.com/artificial-scientist-lab/PyTheus/tree/main/pytheus/graphs/CondensedMatter/majumdar6]{experimentcounter} Six particles with two ancillary particles:
 \begin{align}
 \ket{\psi^{(6)}} &=
\ket{001011}- \ket{001101}\nonumber \\
& - \ket{010011}+ \ket{010110} \nonumber \\
& + \ket{011001} - \ket{011010} \nonumber \\
&
+
\ket{100101}
-
\ket{100110}
\nonumber \\
&
-
\ket{101001}
+
\ket{101100}
\nonumber \\
&
+
\ket{110010}
-
\ket{110100}.
\end{align}
\end{itemize}

\paragraph{Dyck Words --}
The Fredkin spin-1/2 model with a three-body interaction term reads~\cite{Salberger2016, Movassagh2017, Caha2018, Adhikari2019}
\begin{align}
    {\cal H} =\sum_{i} U_{i-1} P_{i,i+1} + P_{i-1,i} D_{i+1},
\end{align}
where $U_{i} = \ket{\uparrow_{i}} \bra{\uparrow_{i}}$, $D_{i} = \ket{\downarrow_{i}}\bra{\downarrow_{i}}$ and the spin-singlet projector reads $P_{i,i+1}= \ket{S_{i,i+1}} \bra{S_{i,i+1}}$ with $\ket{S_{i,j}}= (\ket{\uparrow_{i}} \bra{\downarrow_{j}}  - \ket{\downarrow_{i}} \bra{\uparrow_{j}})$. The ground state of this model is an equally weighted superposition of spin configurations forming Dyck words. Here one may use the notation $\ket{\uparrow} = \ket{(}$ and $\ket{\downarrow} = \ket{)}$ to translate the spin states into the Dyck words. In this notation, the ground state forms balanced strings whose segments contain equal numbers of open and closed parentheses. The ground states of the Fredkin model with six and eight particles cast

\begin{figure}[!h]
	\centering
\subfloat[Dyck Word with six letters]{\includegraphics[width=0.24\textwidth]{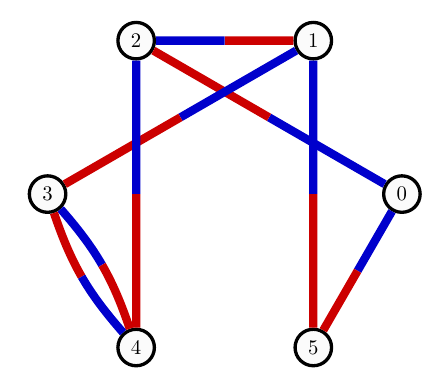}}\!\!
\subfloat[Dyck Word with eight letters]{\includegraphics[width=0.24\textwidth]{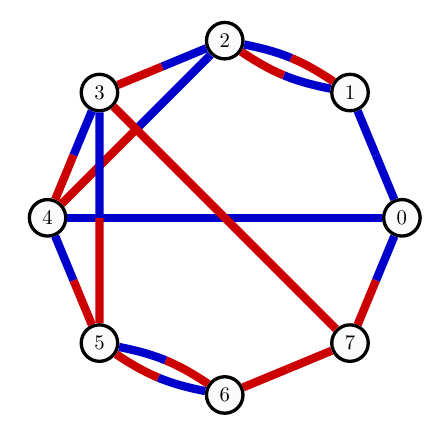}}
 \caption {The graphs corresponding to the creation of Dyck word states.}
	\label{fig:dyck}
\end{figure}

\begin{itemize}
    \item \numbering[https://github.com/artificial-scientist-lab/PyTheus/tree/main/pytheus/graphs/CondensedMatter/dyck6]{experimentcounter} Six particles:
    \begin{align}
        \ket{\psi^{(6)}}
        =\frac{1}{\sqrt{5}}
        \big[&
\ket{()()()}
+\ket{()(())}
+\ket{(())()} 
\nonumber \\
&
+\ket{(()())}
+\ket{((()))}
        \big].
    \end{align}

\item \numbering[https://github.com/artificial-scientist-lab/PyTheus/tree/main/pytheus/graphs/CondensedMatter/dyck8]{experimentcounter} Eight particles:
\begin{align}
    \ket{\psi^{(6)}}
    =\frac{1}{\sqrt{14}} \big[&
\ket{(()())()} + \ket{(()()())} \notag \\
+&\ket{(()(()))} +\ket{((()))()} \notag \\
+&\ket{((())())} +\ket{((()()))} \notag \\
+& \ket{(((())))}+\ket{()()()()} \notag \\
+&\ket{()()(())}+\ket{()(())()} \notag \\
+&\ket{()(()())} +\ket{()((()))} \notag \\
+&\ket{(())()()} +\ket{(())(())}
    \big].
\end{align}
\end{itemize}
The associated graphs for these states are shown in Fig.~\ref{fig:dyck}.

\tocless\subsubsection{Spin-1 Systems}
The quest to go beyond the two-level qubit systems in quantum computation put forward other proposals based on three-level qutrit quantum states~\cite{Williams2011, Nisbet_Jones2013}. In spin systems, this inquiry is translated into exploring the $S=1$ space spanned by $\{ \ket{-1}, \ket{0}, \ket{1} \}$ with physical dimension $d=3$. While spin-1 states can be realized experimentally in a controlled fashion~\cite{Senko2015}, often, one may encounter these states as emergent phenomena in various spin-1/2 condensed matter systems, e.g., in describing the low-energy physics of chiral threefold fermions~\cite{Bradlyn2016}. As a result, aside from technological implications, exploring the higher-spin systems may shed light on a better understanding of some emergent phenomena in other fields of physics.

\paragraph{Spin-1 Wire --}
The antiferromagnetic ground state of a spin-1 chain with nearest-neighbor interactions can be represented in the matrix product form as~\cite{Klmper1993}
\begin{align}
    \ket{\psi}
    =\tr(A_{1} A_{2} \ldots A_{N}),
\end{align}
where
\begin{align}
    A_{i} =\begin{pmatrix}
        \ket{0} & -\sqrt{a} \ket{+1} \\
        \sqrt{a} \ket{-1} & -\sigma \ket{0}
    \end{pmatrix},
\end{align}
with nonvanishing $a$ and $\sigma$.

\begin{figure}[!h]
	\centering
	\includegraphics[width=0.26\textwidth]{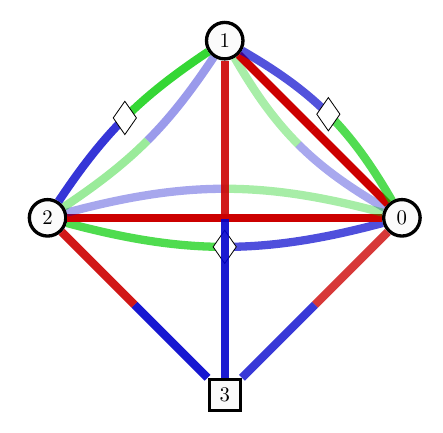}
 \caption {The associated graph for the ground states of the spin-1 chain with three particles.}
	\label{fig:spin1}
\end{figure}

Fig.~\ref{fig:spin1} displays the associated graph with one ancilla, for the ground states of the spin-1 chain with three particles, which, up to normalization, reads ~\numbering[https://github.com/artificial-scientist-lab/PyTheus/tree/main/pytheus/graphs/CondensedMatter/spin1]{experimentcounter}
\begin{align}
\ket{\psi} &= 0.3(\ket{012} + \ket{120} + \ket{201})   - 0.875\ket{111} \notag \\
&- 0.15(\ket{021} + \ket{102} + \ket{210}).
\end{align}

\paragraph{Affleck-Kennedy-Lieb-Tasaki Model --}\label{par:alktspin1}
A particular spin-1 bilinear biquadratic Hamiltonian introduced by Affleck, Kennedy, Lieb, and Tasaki~(AKLT) is given by~\cite{Affleck1987, Affleck1988}
\begin{equation}
    H=\sum_{i} \overrightarrow{S}_{i} \cdot \overrightarrow{S}_{i+1} + \frac{1}{3} (\overrightarrow{S}_{i} \cdot \overrightarrow{S}_{i+1})^2.
\end{equation}
The ground state of this solvable model, known as the AKLT state, is short-range entangled and classified in symmetry-protected-topological states~\cite{Wierschem2016}. The MPS representation of this state reads~\cite{Perez-Garcia2006, Schollwock2011}
\begin{equation}
    \ket{\psi} =\sum_{s_i =0}^{1} \tr[ A (s_{n-1}) \ldots A(s_1) ] \ket{s_1 \ldots s_n},
\end{equation}
where $s_{i}  \in  \{0,1\}$ and
\begin{equation}
    \{ A_{-1}, A_{0}, A_{+1} \} = \{ \sqrt{\frac{2}{3}} \sigma^{+}, \frac{1}{\sqrt{3}} \sigma^{z}, -\sqrt{\frac{2}{3}} \sigma^{-} \}.
\end{equation}
Here $\sigma^{\pm}=\sigma^{x} \pm \i \sigma^{y}$. The AKLT state is the spin-1 valance bond solids~\cite{Pollmann2012}.

\begin{figure}[!h]
	\centering
 \subfloat{\includegraphics[width=0.24\textwidth]{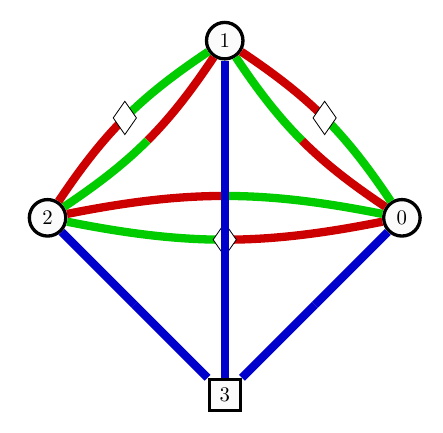}}\!\!
 \subfloat{\includegraphics[width=0.24\textwidth]{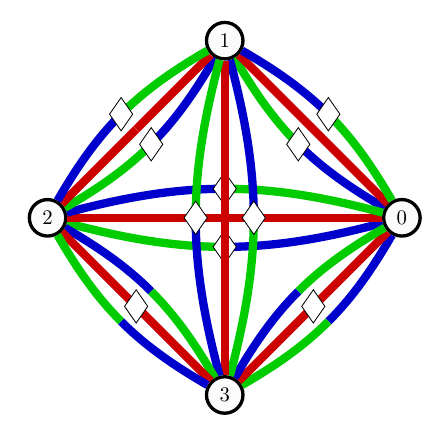}}
 \caption {The associated graphs for the ground state of the spin-1 AKLT state with three and four particles. One ancillary particle is included in the first state.}
	\label{fig:aklt3spin1}
\end{figure}

 Fig.~\ref{fig:aklt3spin1} displays the associated graphs for the three- and four-particle AKLT system whose ground states read, up to normalization ~\numbering[https://github.com/artificial-scientist-lab/PyTheus/tree/main/pytheus/graphs/CondensedMatter/aklt3spin1]{experimentcounter}\numbering[https://github.com/artificial-scientist-lab/PyTheus/tree/main/pytheus/graphs/CondensedMatter/aklt4spin1]{experimentcounter}
 \begin{align}
     \ket{\psi^{(3)}}&=  \ket{0, -1, +1} - \ket{-1, 0, +1} \notag \\
     & + \ket{-1, +1, 0}  - \ket{0, +1, -1} \notag \\
     &+ \ket{+1, 0, -1} - \ket{+1, -1, 0}, \\
     \ket{\psi^{(4)}} & = 2(\ket{-1,1,-1,1} +\ket{1,-1,1,-1}) \notag \\
     & +\ket{1,0,-1,0} -\ket{1,-1,0,0} \notag \\
     &+\ket{0,1,0,-1} -\ket{0,1,-1,0} \notag \\
     & -\ket{0,0,1,-1} +\ket{0,0,0,0} \notag \\
     & -\ket{0,0,-1,1} -\ket{0,-1,1,0} \notag \\
     &+\ket{0,-1,0,1} -\ket{-1,1,0,0}  \notag \\
     & -\ket{1,0,0,-1} + \ket{-1,0,1,0} \notag \\
     & -\ket{-1,0,0,1}.
\label{eq:AKLTspin1_3p}
 \end{align}

\begin{figure}[!h]
	\centering
\subfloat[Motzkin with three letters]{\includegraphics[width=0.24\textwidth]{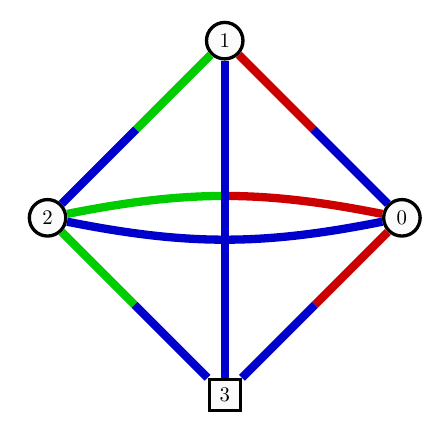}} \!\! 
\subfloat[Motzkin with four letters]{\includegraphics[width=0.24\textwidth]{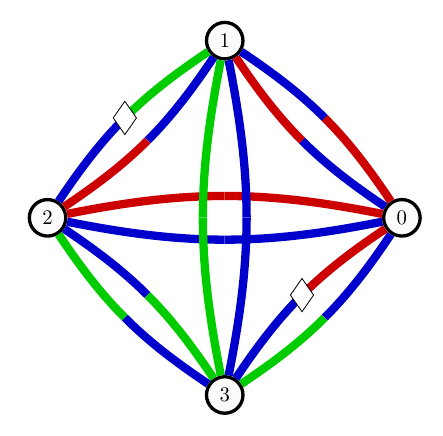}}
 \caption {The graphs corresponding to the creation of Motzkin states.}
	\label{fig:motzkin}
\end{figure}

\paragraph{Motzkin State --}
The spin-1 generalization of the Dyck words is known as the Motzkin State~\cite{Movassagh2017, Bravyi2012, Zhang2017}. The spin configurations, in this case, can be translated into strings using $\ket{-1} = \ket{)}$, $\ket{0} = \ket{-}$, and $\ket{1} = \ket{(}$. Here, similar to the Dyck words, the number of open and closed parentheses is equal. Examples of such states cast
\begin{itemize}
    \item \numbering[https://github.com/artificial-scientist-lab/PyTheus/tree/main/pytheus/graphs/CondensedMatter/motzkin3]{experimentcounter}  Three particles:
    \begin{align}
        \ket{\psi}
        =
        \frac{1}{2}
        [
\ket{---}
+\ket{()-}
+\ket{(-)}
+\ket{-()}
        ].
\end{align}
    \item \numbering[https://github.com/artificial-scientist-lab/PyTheus/tree/main/pytheus/graphs/CondensedMatter/motzkin4]{experimentcounter}  Four particles:
\begin{align}
\ket{\psi}
=
\frac{1}{\sqrt{9}}
[&
\ket{----}
+
\ket{()--}
+
\ket{-()-}
\nonumber \\
&
+
\ket{--()}
+
\ket{(--)}
+
\ket{()()}
\nonumber \\
&
+
\ket{-(-)}
+
\ket{(-)-}
+
\ket{(())}
].
\end{align}
\end{itemize}
The associated graphs for these states are presented in Fig.~\ref{fig:motzkin}.

\tocless\subsubsection{Spin-3/2 Systems}
Sharing the same motivation as spin-1 systems, the four-level spin-3/2 systems have putative implications in qudits with $d=4$ quantum computation~\cite{Nagali2010}. The local Hilbert space of these spin-3/2 systems are spanned by $\{ \ket{1}, \ket{-1} , \ket{3}, \ket{-3}  \}$ resulting in local physical dimension $d=4$.

\paragraph{Spin-3/2 Wire --}
Two exact ground states for a spin-3/2 chain with ferromagnetic character are shown to have the following matrix product representation~\cite{Niggemann1997, Alipour2008}
\begin{align}
    \ket{\psi^{\pm}}
    =\tr(A^{\pm}_{1} A^{\pm}_{2} \ldots A^{\pm}_{N}),
\end{align}
where
\begin{align}
    A^{+} = \begin{pmatrix}
        \ket{1} & -\sqrt{3} \ket{3} \\
       \ket{ -1} & - \ket{1}
    \end{pmatrix}
    , \,
    A^{-}
    =\begin{pmatrix}
        -\ket{-1} & \ket{1} \\
        -\sqrt{3} \ket{-3} &
        \ket{-1} 
    \end{pmatrix}.
\end{align}
\begin{figure}[!h]
	\centering
 \subfloat[$\ket{\psi^+}$]{\includegraphics[width=0.24\textwidth]{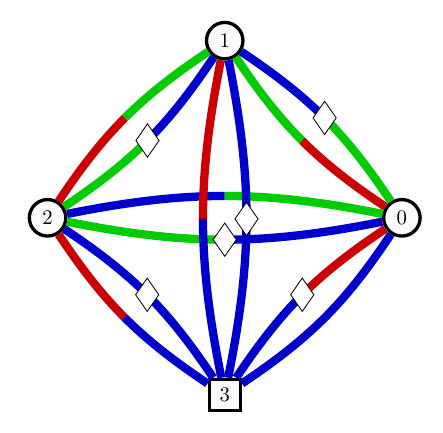}}\!\!
 \subfloat[$\ket{\psi^-}$]{\includegraphics[width=0.24\textwidth]{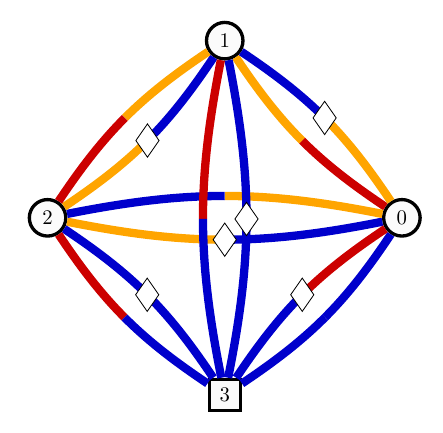}}
 \caption {The associated graphs for the ground states of the spin-3/2 chain for three particles. }
	\label{fig:spin3half}
\end{figure}

The ground states of the spin-3/2 chain, up to normalization, reads~\numbering[https://github.com/artificial-scientist-lab/PyTheus/blob/main/pytheus/graphs/CondensedMatter/spin3halfsPLUS]{experimentcounter}
\begin{align}
    \ket{\psi^{\pm}} &= \ket{-1,1,\pm 3}  - \ket{1,-1,\pm 3} + \ket{1,\pm 3,-1} \notag\\
    &- \ket{-1,\pm 3,1} + \ket{\pm 3,-1,1} - \ket{\pm 3,1,-1}.
\end{align}
As one ground state, say $\ket{\psi^{+}} $, can be achieved from the other ground state, $\ket{\psi^{-}}$, by merely replacing $\ket{3}$ with $\ket{-3}$, we only present the associated graph for $\ket{\psi^{+}} $ with three particles and one ancilla in Fig.~\ref{fig:spin3half}.

\begin{figure}[!h]
	\centering
	\subfloat[Spin-$3/2$ letter state with six letters]{\includegraphics[width=0.26\textwidth]{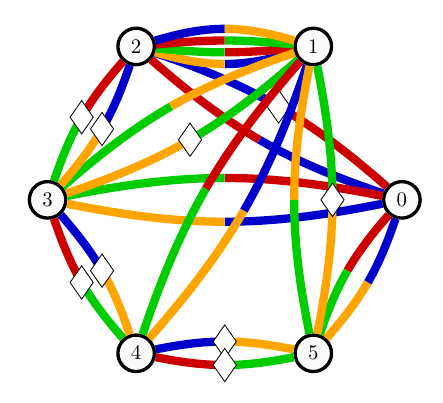}}
 \caption {The graphs corresponding to the creation of the spin-3/2 letter state.}
	\label{fig:letter}
\end{figure}

\paragraph{Spin-3/2 Letter State --}
As a straightforward generalization of the Dyck words and Motzkin letter states, discussed previously, one can introduce the letter states for the spin-3/2 states. Here one should translate the spin state into the letters as $\ket{-3}=\ket{[}$, $\ket{-1}=\ket{(}$, $\ket{1}=\ket{)}$ and $\ket{3}=\ket{]}$. Using this language, the six particle state is \numbering[https://github.com/artificial-scientist-lab/PyTheus/tree/main/pytheus/graphs/CondensedMatter/spin32letter]{experimentcounter}
\begin{align}
    \ket{\psi} =
    &\frac{1}{\sqrt{40}}
   [
     \ket{()()()} +\ket{()[]()}
    +\ket{[]()()} +\ket{()()[]}
    \nonumber \\
    &
    +\ket{[][]()} +\ket{[]()[]}
    +\ket{()[][]} +\ket{[][][]}
    \nonumber \\
    &
    +\ket{(())()} +\ket{(())[]}
    +\ket{([])()} +\ket{[()]()}
    \nonumber \\
    &
    +\ket{[()][]} +\ket{[[]]()}
    +\ket{([])[]} +\ket{[[]][]}
        \nonumber \\
    &
    +\ket{()(())} +\ket{[](())}
    +\ket{()([])} +\ket{()[()]}
        \nonumber \\
    &
    +\ket{[][()]} +\ket{[]([])}
    +\ket{()[[]]} +\ket{[][[]]}
            \nonumber \\
    &
    +\ket{(()())} +\ket{[()()]}
    +\ket{([]())} +\ket{(()[])}
            \nonumber \\
    &
    +\ket{([][])} +\ket{[[]()]}
    +\ket{[()[]]} +\ket{[[][]]}
            \nonumber \\
    &
    +\ket{((()))} +\ket{(([]))}
    +\ket{([()])} +\ket{[(())]}
            \nonumber \\
    &
    +\ket{([[]])} +\ket{[([])]}
    +\ket{[[()]]} +\ket{[[[]]]}
    ].
\end{align}
The associated graph for this state is shown in Fig.~\ref{fig:letter}.

\tocless\subsubsection{Spin-2 Systems}
Five-level computational units can further be of interest in quantum computation beyond qubits. 
The local physical space of these spin-2 units is spanned by $\{ \ket{ -2}, \ket{-1} , \ket{0}, \ket{1}, \ket{2}  \}$ with local physical dimension $d=5$. Similar to other higher-spin systems in condensed matter systems, spin-2 states can also be viewed as emergent states in some (lower-spin with $S<2$) many-body systems~\cite{Link2020}.

\paragraph{Spin-2 Wire --}

The anisotropic spin-2 chain with nearest-neighbour interactions hosts various phases depending on three parameters $(a,x,\sigma)$ with $a \in \mathbb{R}, \sigma=\pm1$~\cite{Ahrens2002}. These phases consist of three distinct antiferromagnetic Haldane phases and a weak antiferromagnetic phase.

The ground states of the weak-antiferromagnetic phase have the following matrix product representation: 
\begin{align}
    \ket{\psi^{(1)}_{0}} &= \tr [ \prod_{i=1}^{L/2} m_{2 i -1} g_{2 i }],\\    \ket{\psi^{(2)}_{0}} &= \tr [ \prod_{i=1}^{L/2} g_{2 i -1} m_{2 i }],
\end{align}
where
\begin{align}
    m =\begin{pmatrix}
        \ket{1} & x \sqrt{a} \ket{2} \\
        \sqrt{a} \ket{0} & \ket{1}
    \end{pmatrix}
    ,\,
    g =\begin{pmatrix}
        \ket{ -1} & \sqrt{a} \ket{0} \\
        x \sqrt{a} \ket{ -2 } & \ket{ -1}
    \end{pmatrix}.
\end{align}
Here we set $a=0.5$.

When the translational symmetry is not broken, i.e., $x=1$, two more states can also be identified for this phase, given by 
\begin{align}
    \ket{\psi^{(3)}_{0}} = \tr [ \prod_{i=1}^{L} m^{x=1}_{i} ], \quad     \ket{\psi^{(4)}_{0}} = \tr [ \prod_{i=1}^{L} g^{x=1}_{i} ].
\end{align}

\begin{figure}[!h]
	\centering
    \subfloat[$\ket{\psi_0^{(\text{NS})}}$]{\includegraphics[width=0.24\textwidth]{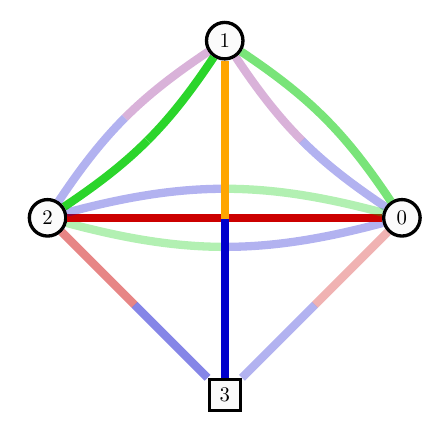}}\!\!
    \subfloat[$\ket{\psi_0^{(\text{S})}}$]{\includegraphics[width=0.24\textwidth]{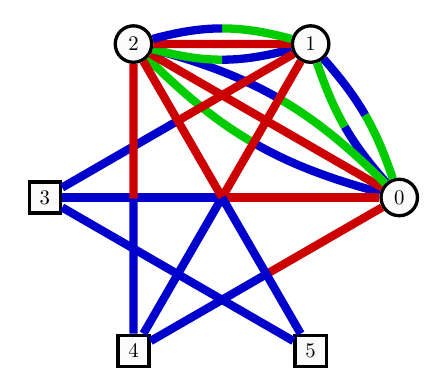}}
 \caption {The associated graphs for the three-particle ground states of the weak-antiferromagnetic phase when the translational symmetry is broken~(left) and respected~(right).}
	\label{fig:weakAF}
\end{figure}

Fig.~\ref{fig:weakAF} displays the associated graphs for the three-particle ground states of the weak-antiferromagnetic phase when the translational symmetry is broken~(top panel) and respected~(bottom). The states with broken translation symmetry, up to normalization, read \numbering[https://github.com/artificial-scientist-lab/PyTheus/tree/main/pytheus/graphs/CondensedMatter/wAF_NOsym]{experimentcounter}
\begin{align}
    \ket{\psi^{(1)}_{0}} =& 0.5(\ket{2,-1,0} + \ket{0,-1,2})  \notag\\
    +& 0.25(\ket{2,-2,1} + \ket{1,-2,2}) \notag\\
    +& \ket{1,0,0}  + \ket{0,0,1} + 4\ket{1,-1,1},\\
     \ket{\psi^{(2)}_{0}} =& \ket{-1,0,0} + \ket{0,0,-1} + 4\ket{-1,1,-1}  \notag\\
     +& 0.25(\ket{-2,2,-1} + \ket{-1,2,-2}) \notag\\
     +& 0.5(\ket{-2,1,0} + \ket{0,1,-2}) .
\end{align}
When the symmetry is respected, the states, up to normalization, are ~\numbering[https://github.com/artificial-scientist-lab/PyTheus/tree/main/pytheus/graphs/CondensedMatter/wAF_sym]{experimentcounter} 
\begin{align}
    \ket{\psi^{(3)}_{0}} &= \ket{0,1,2} + \ket{0,2,1} + \ket{1,0,2} + \ket{2,1,0}\notag\\
     &+ \ket{1,2,0} + \ket{2,0,1} + 4\ket{1,1,1},\\
     \ket{\psi^{(4)}_{0}} &= \ket{0,-1,-2} + \ket{0,-2,-1} +  \ket{-1,0,-2} \notag\\
     &+ \ket{-2,-1,0} + \ket{-1,-2,0} \notag\\
     &+ \ket{-2,0,-1} + 4\ket{-1,-1,-1}.\label{eq:weakAF_sym}
\end{align}

\begin{figure}[!h]
	\centering
	\includegraphics[width=0.26\textwidth]{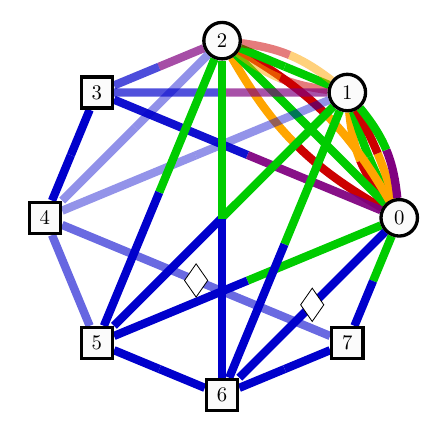}
 \caption {The associated graph for the three-particle ground state of the Haldane-antiferromagnetic-A phase.}
	\label{fig:haldaneA}
\end{figure}

The Haldane-antiferromagnetic-A is another possible phase in this spin-2 system described by the ground state 
\begin{align}
    \ket{ \psi } = \sum_{s_{i}=-2}^{2} \tr \left[ A^{s_{1}}_{1} A^{s_{2}}_{2} \ldots A^{s_{N}}_{N} \right] \ket{s_{1}, s_{2}, \ldots s_{N}}, 
\end{align}
where
\begin{align}
   \sum_{s_{m}=-2}^{2} A_{m}^{s_{m}} =
    \begin{pmatrix}
        \ket{0} & \sqrt{x} \ket{1} & a \ket{2} \\
        \sqrt{x} \ket{-1} & \gamma \ket{0} & \sqrt{x} \ket{1} \\
        a \ket{ -2} & \sqrt{x} \ket{ -1} & \ket{0}
    \end{pmatrix}.
\end{align}
Fig.~\ref{fig:haldaneA} displays the graph associated with the ground state, with three particles and five ancillas, for the Haldane-antiferromagnetic-A phase. Up to normalization, the state reads \numbering[https://github.com/artificial-scientist-lab/PyTheus/tree/main/pytheus/graphs/CondensedMatter/haldaneA_3]{experimentcounter}
\begin{align}
    \ket{\psi_A} =& \ket{-1,0,1} + \ket{-1,1,0} + \ket{0,-1,1}  \notag\\
                 +& \ket{0,1,-1} + \ket{1,-1,0} + \ket{1,0,-1}  \notag\\
                 +& 0.5(\ket{-2,0,2} + \ket{0,-2,2} + \ket{-2,2,0}) \notag\\
                 +& 0.5(\ket{2,0,-2}  + \ket{0,2,-2}  + \ket{2,-2,0}) \notag\\
                 +& 0.25(\ket{1,-2,1}  + \ket{-1,-1,2} + \ket{-1,2,-1}) \notag\\
                 +& 0.25(\ket{1,1,-2} + \ket{-2,1,1} + \ket{2,-1,-1})\notag\\
                 +&6\ket{0,0,0} . 
\end{align}

The ground states of the second antiferromagnetic Haldane phase, referred to as Haldane-antiferromagnetic-B, reads
\begin{align}
    \ket{ \psi_B } = \sum_{s_{i}=-1}^{1} \tr \left[ A^{s_{1}}_{1} A^{s_{2}}_{2} \ldots A^{s_{N}}_{N} \right] \ket{s_{1}, s_{2}, \ldots s_{N}}, 
\end{align}
where
\begin{align}
      \sum_{s_{m}=-1}^{1} A_{m}^{s_{m}} =
    \begin{pmatrix}
        \ket{0} & \sqrt{a} \ket{1}  \\
        \sqrt{a} \ket{-1} & \sigma \ket{0} 
    \end{pmatrix}.
\end{align}

Similarly, the ground state of the third Haldane-antiferromagnetic phase, known as Haldane-antiferromagnetic-C, is
\begin{align}
    \ket{ \psi_C } = \sum_{s_{i}\in \{ 0,\pm2\}}\tr \left[ A^{s_{1}}_{1} A^{s_{2}}_{2} \ldots A^{s_{N}}_{N} \right] \ket{s_{1}, s_{2}, \ldots s_{N}}, 
\end{align}
where
\begin{align}
     \sum_{s_{i} \in \{ 0,\pm2\}} A_{m}^{s_{m}} =
    \begin{pmatrix}
        \ket{0} & \sqrt{a} \ket{2}  \\
        \sqrt{a} \ket{-2} & \sigma \ket{0} 
    \end{pmatrix}.
\end{align}

Notice that the last Haldane phases, B and C, have essentially the same Hamiltonian as the weak-antiferromagnetic phase when the translational symmetry is preserved (see Eq.~\eqref{eq:weakAF_sym}). One can obtain one from the other by performing $\ket{\pm1} \leftrightarrow \ket{\pm2}$.

\tocless\subsubsection{Quantum Many-Body Scars}
The phenomena in which weakly entangled nonthermal quantum eigenstates are embedded in the eigensystem of non-integrable (thermal) systems is dubbed `quantum many-body scars'~\cite{Serbyn2021}. The well-known examples of these scar states are shown to be present in the AKLT spin chain models~\cite{Moudgalya2018, Moudgalya2018b}. The experimental realizations of such states are also reported in Rydberg-atom quantum simulators~\cite{Bernien2017, Choi2019}. In the following, we suggest that one may also detect these states using quantum optics experiments.

\paragraph{Onsager’s Scars in Disordered Spin Chains --}

A non-integrable quantum spin chain that exhibits quantum many-body scars can be described by the coherent state with parameter $\beta$ as~\cite{Shibata2020}
\begin{align}
    \ket{\psi_{\beta}} &=\tr[ A_{p_1} B_{p_1} \ldots A_{p_n} B_{p_n} ] \ket{p_1,\ldots ,p_n}
    ,
    \end{align}
    where $A,B$ are $n \times n$ matrices with matrix elements
    \begin{align}
    (A_{p})_{ij}
    &=
    \beta^{p} \delta_{ip} \delta_{j0}
    + \frac{(-1)^{j+1} \beta^{p} }{\sin (\pi (n-j) /n)} \delta_{n-p,j-i},\\
    (B_p)_{ij} &= \beta^{p} \delta_{ip} \delta_{j0} +
    \frac{(-1)^{n-j} \beta^{p}}{ \sin(\pi (n-j) /n)} \delta_{n-p,j-1}.
\end{align}
Here $0 \leq i,j \leq n-1$ with $n \geq 2$. This representation ensures that no certain spin configurations occur over three consecutive sites in $\ket{\psi_{\beta}}$.

\begin{figure}
	\centering
	\includegraphics[width=0.25\textwidth]{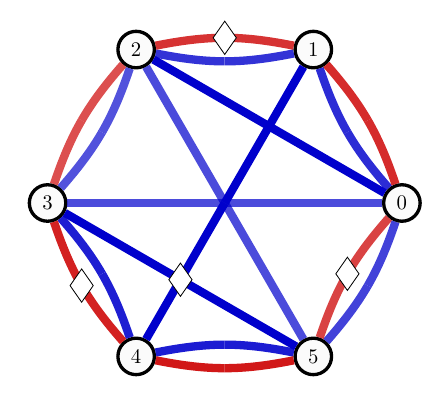} \hspace{-4mm}
	\includegraphics[width=0.23\textwidth]{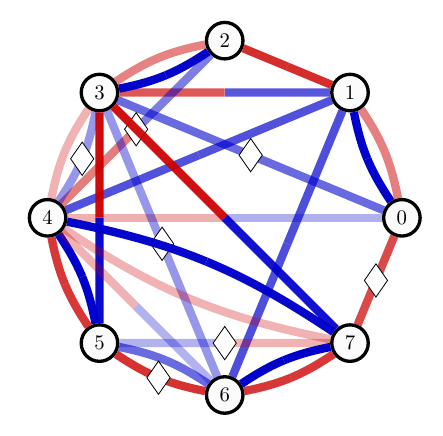}
 \caption {The associated graphs for the Onsager's scar in a system with six~(left panel) and eight~(right) qubits. Here we set $\beta=1/\sqrt{2}$.}
	\label{fig:onsager}
\end{figure}

Fig.~\ref{fig:onsager} displays the graph associated with the Onsager's scar in a system with six and eight spins, with $\beta=1/\sqrt{2}$. Up to normalization, the states read \numbering[https://github.com/artificial-scientist-lab/PyTheus/tree/main/pytheus/graphs/CondensedMatter/onsager6]{experimentcounter}\numbering[https://github.com/artificial-scientist-lab/PyTheus/tree/main/pytheus/graphs/CondensedMatter/onsager8]{experimentcounter}
\begin{align}
    \ket{\psi_{\beta}^{(6)}} &= +2(\ket{000011} -\ket{000110} +\ket{001100}   \notag \\
        &+ \ket{110000} -\ket{011000} -\ket{100001})  \notag \\
        &+\ket{011110} +\ket{001111} +\ket{100111}    \notag \\
        &-\ket{101101} -\ket{011011} +\ket{110011}   \notag \\
        & - \ket{110110} +\ket{111001} +\ket{111100}  \notag \\
        &+ 4\ket{000000} ,
\end{align}
\begin{align}
    \ket{\psi_{\beta}^{(8)}} =& 
    8\ket{00000000}+4(\ket{00000011}-\ket{00000110}\nonumber \\
    &+\ket{00001100}-\ket{00011000}+\ket{00110000}\nonumber \\
    &-\ket{01100000}+\ket{11000000}-\ket{10000001})\nonumber \\
    &+2(\ket{00001111}-\ket{00011011}+\ket{00011110}\nonumber \\
    &+\ket{00110011}-\ket{00110110}+\ket{00111100}\nonumber \\
    &-\ket{01100011}+\ket{01100110}-\ket{01101100}\nonumber \\
    &+\ket{01111000}+\ket{10000111}-\ket{10001101}\nonumber \\
    &+\ket{10011001}-\ket{10110001}+\ket{11000011}\nonumber \\
    &-\ket{11000110}+\ket{11001100}-\ket{11011000}\nonumber \\
    &+\ket{11100001}+\ket{11110000})+\ket{00111111}\nonumber \\
    &+\ket{10110111}-\ket{10111101}-\ket{01101111}\nonumber \\
    &-\ket{10011111}+\ket{11001111}-\ket{11011011}\nonumber \\
    &+\ket{11011110}+\ket{01111011}-\ket{01111110}\nonumber \\
    &-\ket{11100111}+\ket{11101101}+\ket{11110011}\nonumber \\
    &-\ket{11110110}-\ket{11111001}+\ket{11111100}\nonumber \\
    &+\ket{11111111}.   
\end{align}

\paragraph{Scars in the PXP Model --}
\begin{figure}[!h]
	\centering
	\includegraphics[width=0.24\textwidth]{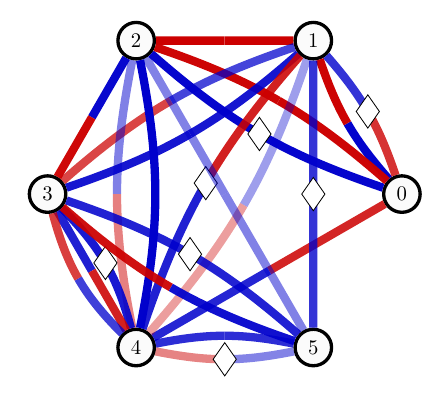}\hspace{-3mm}
	\includegraphics[width=0.24\textwidth]{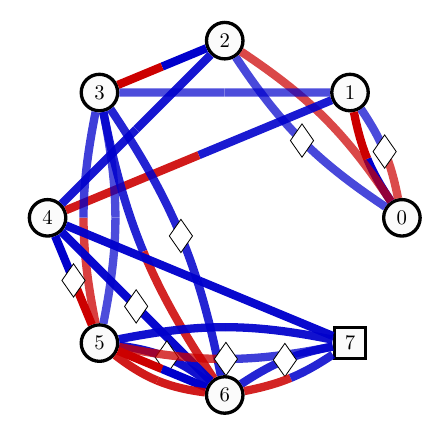}
 \caption {The graphs associated with the ground state of the PXP model with six~(left panel) and eight~(right) particles.}
	\label{fig:PXP}
\end{figure}
Another platform where quantum scars emerge is in the $PXP$ model. This model on a chain with $L$ sites and with the periodic boundary condition is~\cite{Lin2019}
\begin{align}
    {\cal H} = \sum_{i} P_{i-1} X_{i} P_{i+1} + P_{L} X_{1} P_{2} + P_{L-1} X_{L} P_{1},
\end{align}
where $P=\ket{0}\bra{0}$ and $X=\ket{0}\bra{1}+\ket{1}\bra{0}$. One of the ground states of this model in the matrix product representation yields
\begin{align}
    \ket{\text{PXP}} = \sum_{\{ \sigma \}}
    \tr [
B^{\sigma_{1}} C^{\sigma_{2}}
\ldots
B^{\sigma_{L-1}} C^{\sigma_{L-1}}
    ]
    \ket{\sigma_{1} \sigma_{2}  \ldots \sigma_{L}},
\end{align}
where
\begin{align}
    B^{0} &= 
    \begin{pmatrix}
         1 & 0 & 0 \\
         0 & 1 & 0
    \end{pmatrix}
    ,\quad 
   B^{1} = \sqrt{2} 
   \begin{pmatrix}
       0 & 0 & 0 \\
       1 & 0 & 1
   \end{pmatrix}
   ,\\
   C^{0} &=
   \begin{pmatrix}
       0 & -1 \\
       1 & 0 \\
       0 & 0
   \end{pmatrix}
   ,\quad 
   C^{1} = \sqrt{2} 
   \begin{pmatrix}
       1 & 0 \\
       0 & 0 \\
       -1 & 0
   \end{pmatrix}.
\end{align}
Fig.~\ref{fig:PXP} displays the graph associated with the ground state of the PXP model with six and eight particles. Up to normalization, these states read \numbering[https://github.com/artificial-scientist-lab/PyTheus/tree/main/pytheus/graphs/CondensedMatter/pxp6]{experimentcounter} \numbering[https://github.com/artificial-scientist-lab/PyTheus/tree/main/pytheus/graphs/CondensedMatter/pxp8]{experimentcounter}
\begin{align}
    \ket{\text{PXP6}} &= \ket{000000} - \ket{000010} \notag\\
     &- \ket{101000} - \ket{100000} \notag\\
    &+ \ket{101010}  - 2\ket{010100}  \notag\\
    &+ \sqrt{2}(\ket{100100} + \ket{010000}) ,
\end{align}
\begin{align}
\ket{\text{PXP8}} &= \ket{00000000} + \ket{00001000} \notag \\
 & + \ket{10000000} + \ket{10101010} \notag \\
 & -\ket{00001010} -\ket{10000010} \notag \\
 & -\ket{10100000} -\ket{10101000} \notag \\
 & -2( \ket{01010000}+\ket{10010100}) \notag \\
 & + \sqrt{2}(\ket{01000010} - \ket{00000100} \notag \\
 & + \ket{10010000} - \ket{01000000}\notag \\
 & + \ket{10100100} + 2\ket{01010100} ).
\end{align}

\subsection{Quantum Communication}\label{experiments:qcommunication}
Quantum communication refers to communication protocols that involve the transfer of quantum states. Machine learning has been previously employed for the artificial discovery of quantum communication schemes \cite{wallnofer2020machine}.

The first experimental realization of entanglement swapping was presented as a way of entangling two photons that never interacted \cite{pan1998experimental}. In this section we show graphs corresponding to experiments that create entanglement between two parties  (each party with single or multiple photons) without interaction between the parties. Entanglement swapping has a strong connection to other quantum communication tasks, e.g.\ teleportation. This opens the door to use \pytheus for discovering related quantum information protocols.

\begin{figure}[!t]
\centering
\includegraphics[width=0.48\textwidth]{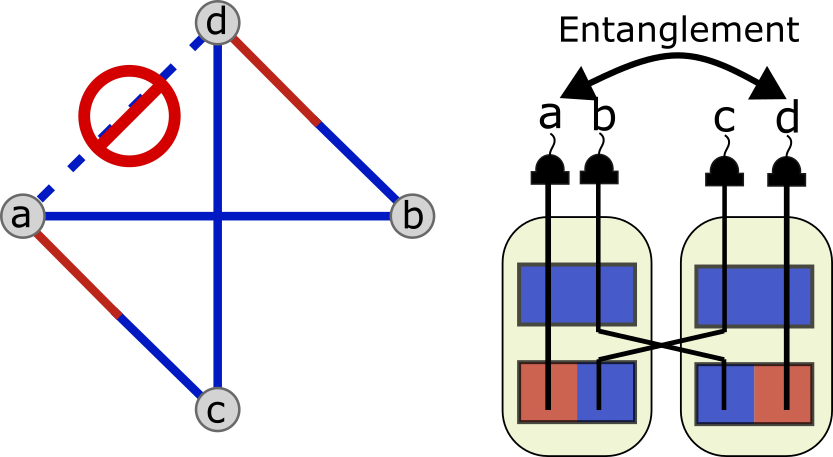}
\caption {Entanglement between two qubits without a common source. Because the vertices a and d have no connection in the graph, each of the corresponding photons is created in one of two disjoint subsystems (green boxes).}
\label{fig:entswap}
\end{figure}

\paragraph{Entangling Two Photons that Never Interacted without Bell Pairs --}
Respecting the constraints that are placed on two particles that do not interact (presented in section \ref{sec:qcommunication}) we show a graph that entangles two photons from independent sources in Fig.~\ref{fig:entswap} . The corresponding setup constructed by path identity does not require the initial creation of entangled Bell pairs \numbering{experimentcounter}. 

\begin{figure}[t]
\centering
\subfloat[Three qubit ES]{\includegraphics[width=0.22\textwidth]{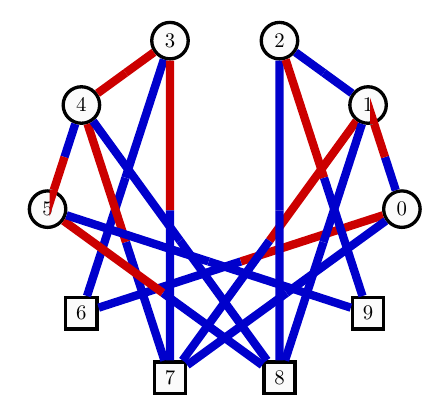}}\!\!
\subfloat[Outline of Experiment]{\includegraphics[width=0.26\textwidth]{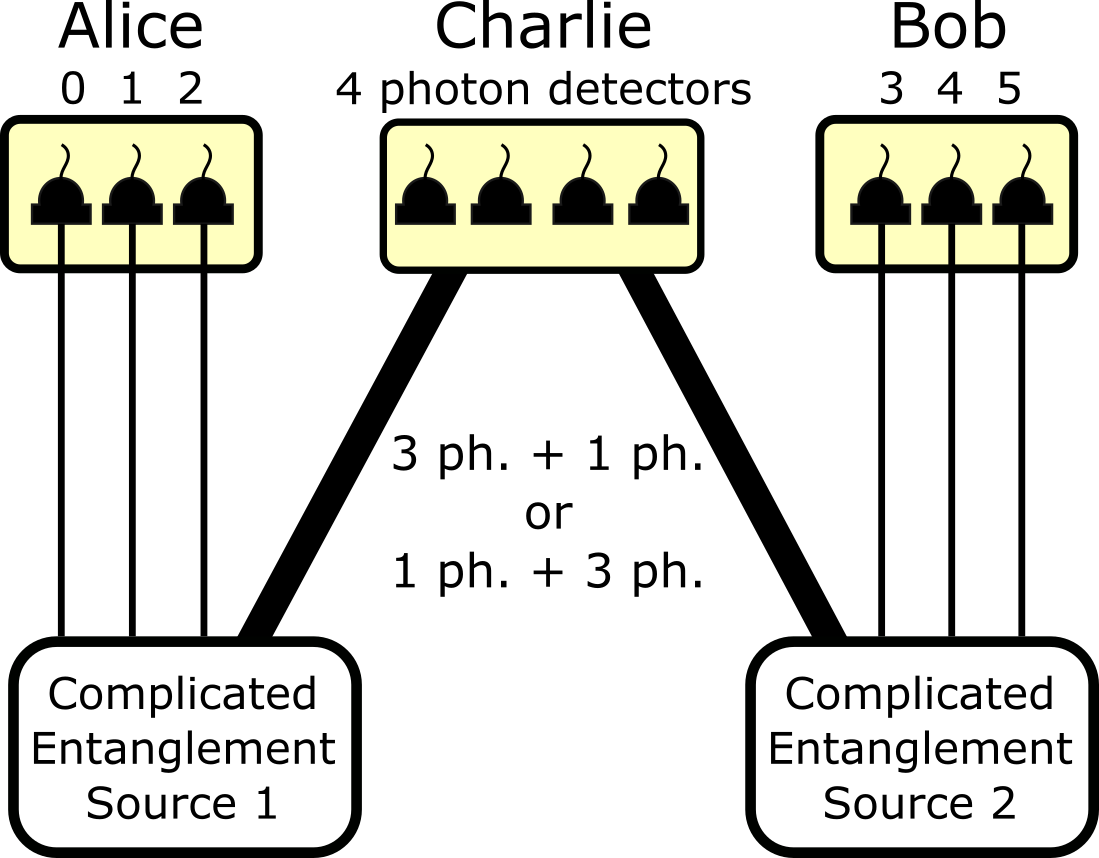}}
\caption{\textbf{Entanglement swapping of 3 Bell pairs measuring only 4 photons.}  Left: abstract graph for multi-particle entanglement swapping. Right: outline of a multi-particle Entanglement swapping experiment. Alice and Bob each receive three particles from the two independent sources 1 and 2 (each a collection of SPDC crystals). When Charlie measures a coincidence in all four detectors Alice and Bob share three Bell pairs (0\&3, 1\&4, 2\&5) without them having interacted. This is possible due to a superposition of events where Charlie either receives three photons from Alice and one photon from Bob or one photon from Alice and three photons from Bob.}
\label{fig:mmultiswap}
\end{figure}

\paragraph{Multiparticle Entanglement Swapping --}
Performing simultaneous entanglement swapping on multiple pairs of particles is one of the key players in achieving resource-efficient quantum communication. Entanglement swapping between two qubits requires two additional particles for a Bell state measurement. Performing this swapping experiment in parallel for $n$ qubit pairs would require $2n$ additional photons. We prompted \pytheus to find experiments that beat this naive baseline. With this, we found a three-qubit entanglement swapping experiment (Fig.~\ref{fig:mmultiswap}), which produces the state \numbering[https://github.com/artificial-scientist-lab/PyTheus/tree/main/pytheus/graphs/Communication/3pES]{experimentcounter}
\begin{align}
    \ket{\Phi^+}_{03}\otimes\ket{\Phi^+}_{14}\otimes\ket{\Phi^+}_{25},
\end{align}
where the photons $0,1,2$ are separated from $3,4,5$. This experiment requires four additional particles instead of the six particles necessary for the parallel case described above.

Similarly, entanglement swapping between a pair of qutrits requires four ancillary particles. We found that "performing entanglement swapping for two pairs of qutrits could also beat the naive baseline by only requiring six ancillas instead of eight, in total, to produce \numbering[https://github.com/artificial-scientist-lab/PyTheus/tree/main/pytheus/graphs/Communication/2pES_3d]{experimentcounter}
\begin{align}
    \ket{\Phi^+}^{3d}_{02}\otimes\ket{\Phi^+}^{3d}_{13},
\end{align}
where the photons $0,1$ are separated from $2,3$, and $\ket{\Phi^+}^{3d}$ is the first three-dimensional Bell state.

\begin{figure}[t]
\centering
\subfloat[2 qutrit ES]{\includegraphics[width=0.24\textwidth]{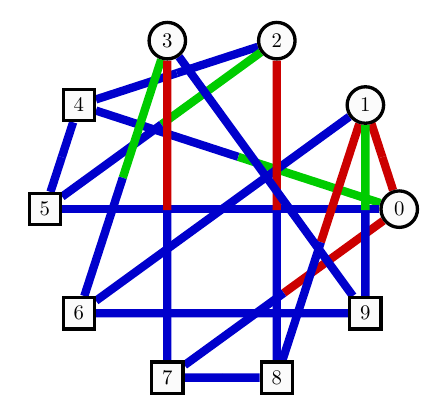}}\!\!
 \subfloat[2 qutrit ES (s.p.s)]{\includegraphics[width=0.24\textwidth]{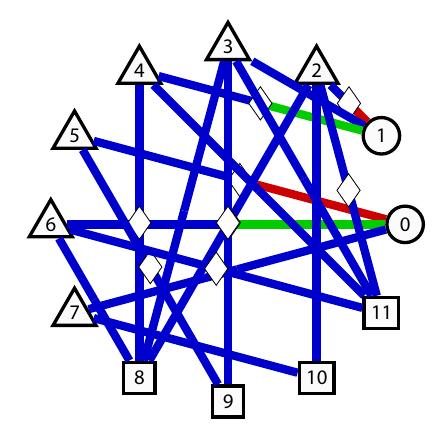}}
\caption {Graphs for entanglement swapping experiments.}
\label{fig:entanglementswapping}
\end{figure}

\paragraph{Entanglement Swapping with Single-Photon Sources}
It is also possible to use single-photon sources for entanglement swapping \cite{basset2019entanglement,troiani2014entanglement,zopf2019entanglement}. In Fig.~\ref{fig:entanglementswapping}, we show a graph \numbering[https://github.com/artificial-scientist-lab/PyTheus/tree/main/pytheus/graphs/Communication/ES3d_sp]{experimentcounter} for a higher-dimensional case, performing three-dimensional entanglement swapping with single photon sources. In total, six single photon sources are necessary. 

\subsection{Quantum Measurements}\label{exp:quantum_measurements}
An important tool in quantum information is multi-particle measurements such as the Bell state measurement \cite{weinfurter1994experimental} or the GHZ analyzer \cite{ghzanalyzer}. Here we present a wider range of measurements that \pytheus has discovered. Many others, such as POVM (positive operator-valued measure) or their symmetric, informationally complete special case (SIC-POVMs) could be additional targets for future research, given their exceptional importance for quantum information tasks \cite{bergou2010discrimination,bent2015experimental} as well as their connection to exciting number theoretical questions \cite{caves2002unknown}.

\tocless\subsubsection{Analyzers}
\begin{figure}[t]
	\centering
	\subfloat[GHZ analyzer 3d]{\includegraphics[width=0.24\textwidth]{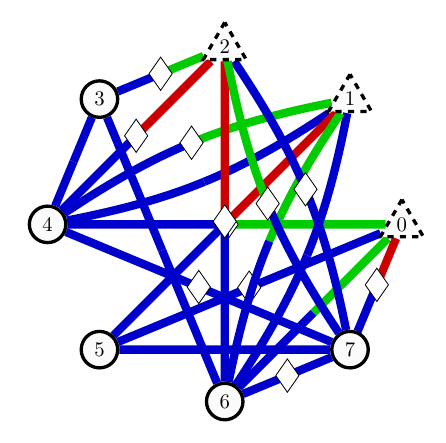}}\!\!
    \subfloat[GHZ analyzer 4d]{\includegraphics[width=0.24\textwidth]{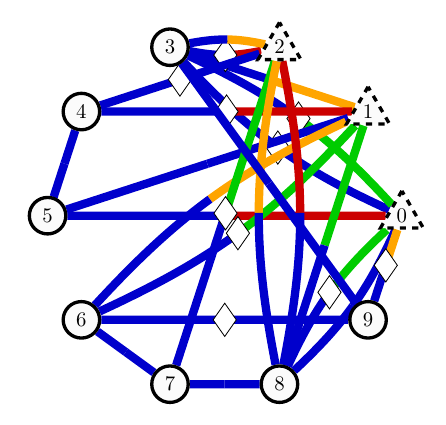}}\\
     \subfloat[Higuchi-Sudbery analyzer]{\includegraphics[width=0.24\textwidth]{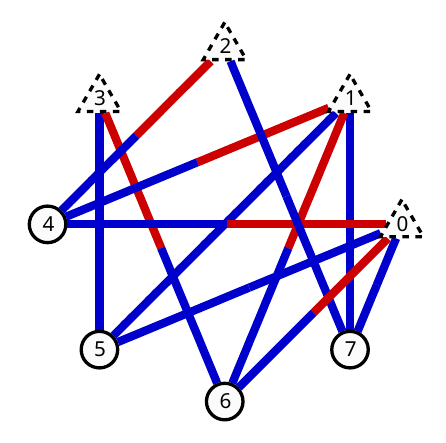}}\!\!
     \subfloat[W analyzer]{\includegraphics[width=0.24\textwidth]{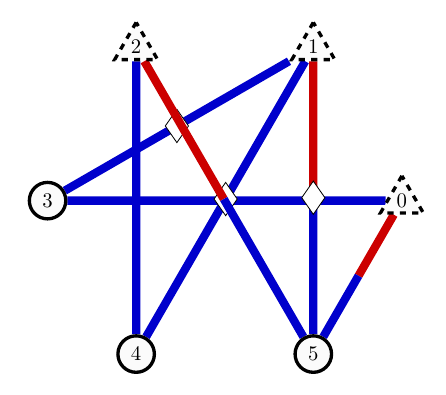}}\\
     \subfloat[Yeo-Chua analyzer]{\includegraphics[width=0.24\textwidth]{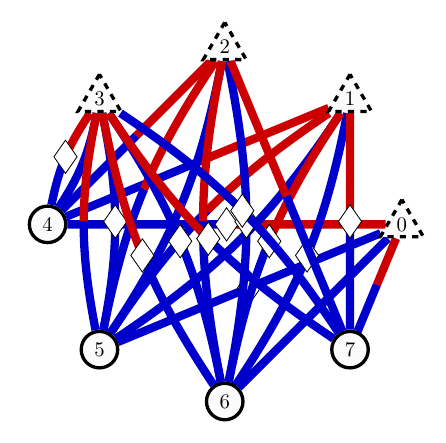}}
	\caption {Graphs for analyzers. When all detectors click, the incoming photons were in the corresponding state.}
	\label{fig:analyzers}
\end{figure}

Here, we use the word \textit{analyzer} to refer to experimental setups that confirm a collection of photons to be in a particular state. Their formulation in terms of graphs is described in \ref{sec:measurements}. The two-dimensional GHZ analyzer realized in Ref.~\cite{ghzanalyzer} can be extended towards higher dimensions, giving the 3d GHZ analyzer \numbering[https://github.com/artificial-scientist-lab/PyTheus/tree/main/pytheus/graphs/Measurements/ghz_analyzer_3d]{experimentcounter} and the 4d GHZ analyzer \numbering[https://github.com/artificial-scientist-lab/PyTheus/blob/main/pytheus/graphs/Measurements/ghz_analyzer_4d]{experimentcounter}. Further, we show analyzers for the W state \numbering[https://github.com/artificial-scientist-lab/PyTheus/tree/main/pytheus/graphs/Measurements/W_measurement]{experimentcounter}, the Higuchi-Sudbery state (shown in Eq.~\eqref{eq:HSstate}) \numbering[https://github.com/artificial-scientist-lab/PyTheus/tree/main/pytheus/graphs/Measurements/HS_measurement]{experimentcounter} and the Yeo-Chua state (shown in Eq.~\eqref{eq:YC}) \numbering[https://github.com/artificial-scientist-lab/PyTheus/tree/main/pytheus/graphs/Measurements/YC_measurement]{experimentcounter} in Fig.~\ref{fig:analyzers}.

\tocless\subsubsection{Mean King's Problem}\label{subsubsec:meanking}

In 1987, L. Vaidman, Y. Akaronov, and D. Z. Albert \cite{vaidman1987} devised an interesting quantum communication task that can be solved only with quantum resources. Later referred to as the Mean King's Problem (MKP), it involves two parties. Alice, who sends a quantum state which she created to another party -- the Mean King. The Mean King then performs a projective measurement on the state in a basis of his choice out of a collection of mutually unbiased bases (MUBs). Alice is then allowed to perform one more measurement, after which the King declares his measurement basis and Alice must correctly guess his result. Should Alice guess incorrectly even once, she will suffer a cruel fate, for the Mean King is exceptionally intolerant of poor guesses. This task has applications in  quantum key distribution, wherein even the slightest discrepancy between the shared secret key of Alice and Bob implicates (in principle) the presence of an eavesdropper.  

Over the years, various generalizations to the initial solution, proposed in the original paper \cite{vaidman1987}, have been introduced. Here, we employ a generalization proposed by Hayashi, Horibe, and Hashimoto \cite{hayashi2005} for quantum states of dimension $D$ which is a prime power. This solution proposes that Alice first prepares a maximally entangled, two-photon state $\ket{\psi_o}$ and sends one of her photons to the Mean King.  After the King makes a projective measurement in one of the $(D+1)$ MUBs, Alice retrieves her photon and performs a measurement in the basis of states credited to Vaidman, Akaronov, and Albert (VAA states). There are $D^2$ VAA states in all. For $D=2$, the first VAA state can be written as 
 \begin{align}
        \ket{\phi_1} &= \frac{1}{2} (\sqrt{2}\ket{00} + e^{-i\frac{\pi}{4}}\ket{01} + e^{i\frac{\pi} {4}}\ket{10})
\end{align}
For $D=3$, the third VAA state can be written as 
 \begin{align}
        \ket{\phi_3} &= \frac{1}{\sqrt{3}} (\ket{00} + \alpha(\ket{02} + \ket{10} + \ket{01} + \ket{20}) \\ \nonumber
        &+ \beta(\ket{12}) + \gamma(\ket{21})) 
      \end{align}
where $\omega=e^{-i 2\pi/3}$, $\alpha=(\omega^2 + 2\omega)/3$, $\beta=(\omega^2 + 2)/3$, and $\gamma=(\omega + 2)/3$, 


Experimentally realizing Alice's measurement in this basis is non-trivial. Setups that recreate the two-dimensional VAA measurement using two qubits encoded in a single photon  \cite{oliver2003,Englert2001} and that realize extensions to the Mean King's problem \cite{chengQiuHu2020} have been proposed; but the original solution to the Mean King's Problem involves two photons, to which no experimental realization has been proposed for any prime-power $D$. Here we provide examples discovered by \pytheus for select prime-dimensional cases.  
\begin{figure}[t]
	\centering
	\subfloat[MKP (2d)]{\includegraphics[width=0.24\textwidth]{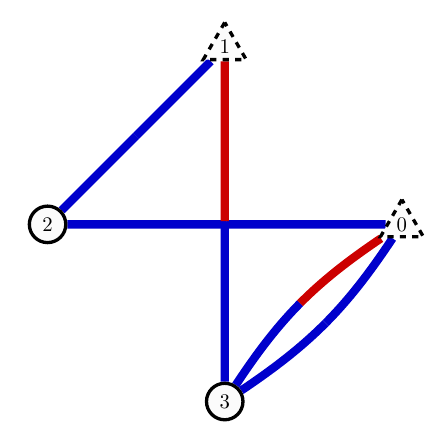}}\!\!
    \subfloat[MKP (3d)]{\includegraphics[width=0.24\textwidth]{graph_MKP_3d.pdf}}
	\caption {Graphs realizing the VAA state measurement for the MKP in the 2-dimensional case and 3-dimensional case.}
	\label{fig:mkp}
\end{figure}

\paragraph{Two-dimensional Case --} The left side of Fig.~\ref{fig:mkp} shows the graph of Alice's measurement in the VAA basis \numbering[https://github.com/artificial-scientist-lab/PyTheus/tree/main/pytheus/graphs/Measurements/MKP_2d]{experimentcounter}. This graph was found with \pytheus by optimizing for the VAA state projection into $\ket{\phi_{1}}$. Surprisingly however, during the translation of the graph to an optical setup, we found that several other VAA states can be projected with the same setup, by adding detectors at unused beam splitter ports. Several of the remaining VAA states can then be distinguished by the set of possible simultaneous two-detector click events that they each trigger. Since any result of the Mean King's measurement can be expressed in terms of a superposition of VAA states, Alice is able to guess the result by performing the measurement of her input state in the VAA basis and guessing the state that can be expressed in terms of Alice's measurement result. This procedure works regardless of the Mean King's choice of MUB in his measurement.   

\paragraph{Three-Dimensional Case --}
The right-hand side of Fig. ~\ref{fig:mkp} shows the graph for the projection of the state into the one of the three-dimensional VAA states $\ket{\phi_{3}}$ \numbering[https://github.com/artificial-scientist-lab/PyTheus/tree/main/pytheus/graphs/Measurements/MKP_3d]{experimentcounter}. As with the two-dimensional case, we add again detectors at the empty port of beam splitters and find that we can distinguish between more three of the nine states without any additional modifications. After post-selection, Alice has a guaranteed chance to correctly guess the Mean King's measurement result so long as the King does not choose the second MUB for his measurement. Were the King to pick that MUB, two of the three results of the measurement are expressed in terms of VAA states that Alice cannot distinguish, giving her a 50\% chance of escaping the Mean King's cruelty.


\subsection{Quantum Gates}\label{experiments:gates}
Universal quantum gates, which rely on the interaction between two or more photons, can be realized with non-linearities induced by measurements \cite{knill2001scheme}. Quantum gates based on this approach, such as a CNOT between two qubits, have been realized experimentally for almost twenty years \cite{gasparoni2004realization}. However, since then, the attempt to experimentally discover a wide range of quantum gates has continued due to the advancement of experimental resources which can deal with higher dimensional systems. 

\begin{figure}[t]
	\centering
    \subfloat[Heralded CNOT(2,3)]{\includegraphics[width=0.24\textwidth]{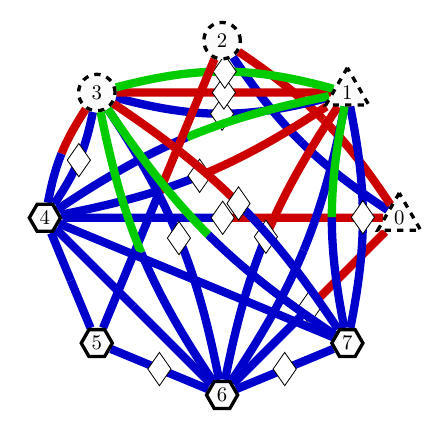}}\!\!
    \subfloat[Heralded Toffoli]{\includegraphics[width=0.24\textwidth]{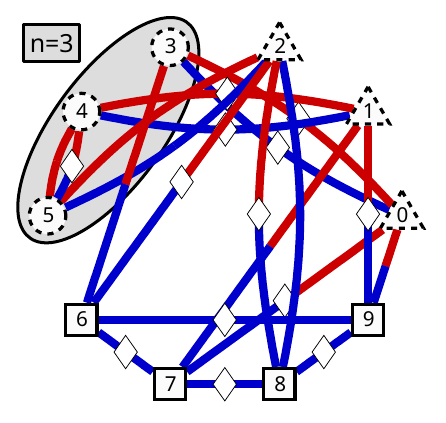}}
	\caption {Graphs for heralded quantum gates.}
	\label{fig:heraldedgates}
\end{figure}

\paragraph{Heralded --}
We call a quantum gate heralded when only ancillary particles are detected, meaning that the outgoing particles are not measured. The photons exiting such a gate can continue into further components of a longer circuit. At the end of the circuit it is necessary to confirm that two photons have exited the gate, but not which path they followed.
We find that a higher dimensional version of the CNOT gate with a control qubit and a target qutrit ($\text{CNOT}(2,3)$ \numbering[https://github.com/artificial-scientist-lab/PyTheus/tree/main/pytheus/graphs/Gates/cnot23]{experimentcounter}) can also be realized with two ancilla photons.
Further, we find a heralded Toffoli gate with four ancillas \numbering[https://github.com/artificial-scientist-lab/PyTheus/tree/main/pytheus/graphs/Gates/toffoli]{experimentcounter}; see Fig.~\ref{fig:heraldedgates}.

\paragraph{Reduced Input Space --}
For some tasks in which quantum gates are applied, there is some prior knowledge about the input state. This knowledge can be exploited to reduce the experimental resources required to perform the transformation \cite{q_mem_advantage,NonDetLogicOperations}. A wide range of constraints on the input state can be encoded by our framework. Here we show examples of heralded quantum gates where a target qubit is prepared in the zero state before entering the gate. 

\begin{figure}[t]
	\centering
	\subfloat[heralded CNOT(3,3)]{\includegraphics[width=0.24\textwidth]{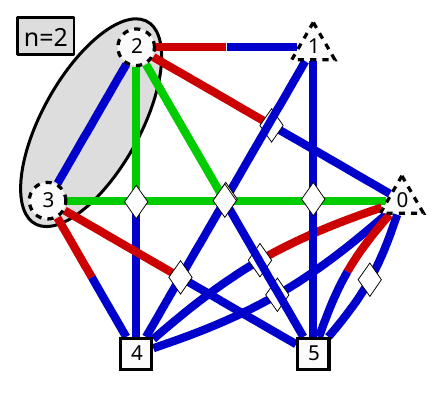}}
	\subfloat[heralded Fredkin]{\includegraphics[width=0.24\textwidth]{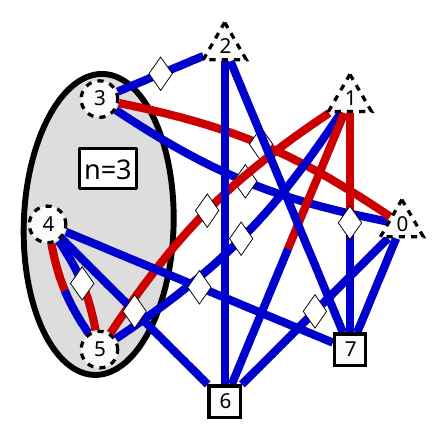}}\!\!
    \subfloat[heralded CNOT(4,4)]{\includegraphics[width=0.24\textwidth]{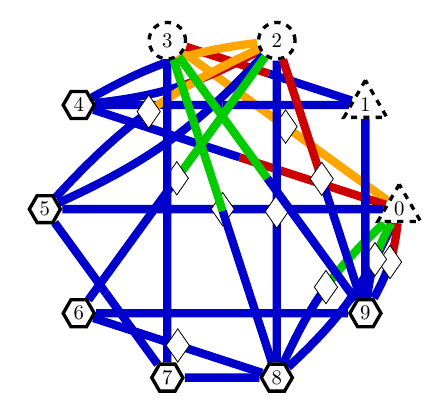}}
	\caption {Graphs for heralded quantum gates with reduced input space.}
	\label{fig:onzero}
\end{figure}

\pytheus found gates $\text{CNOT}(3,3)$ \numbering[https://github.com/artificial-scientist-lab/PyTheus/tree/main/pytheus/graphs/Gates/cnot33_0]{experimentcounter}, $\text{CNOT}(4,4)$ \numbering[https://github.com/artificial-scientist-lab/PyTheus/blob/main/pytheus/graphs/Gates/cnot44_0]{experimentcounter} and the Fredkin gate \numbering[https://github.com/artificial-scientist-lab/PyTheus/tree/main/pytheus/graphs/Gates/fredkin_0]{experimentcounter} acting on a target input photon in the computational zero state; see Fig.~\ref{fig:onzero}. A post-selected Fredkin gate on the full input space has been experimentally realized \cite{fredkin16}.

\begin{figure}[t]
	\centering
	\subfloat[CNOT(2,3)]{\includegraphics[width=0.24\textwidth]{graph_cnot23_post.pdf}}\!\!
    \subfloat[CNOT(2,4)]{\includegraphics[width=0.24\textwidth]{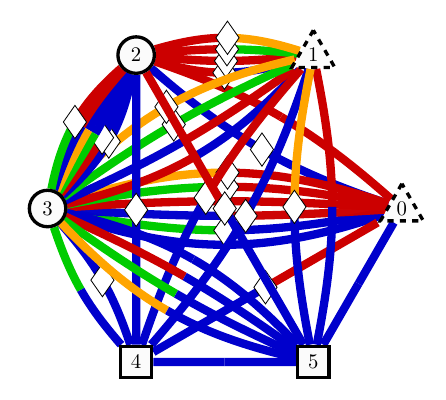}}\\
    \subfloat[CNOT(3,3)]{\includegraphics[width=0.24\textwidth]{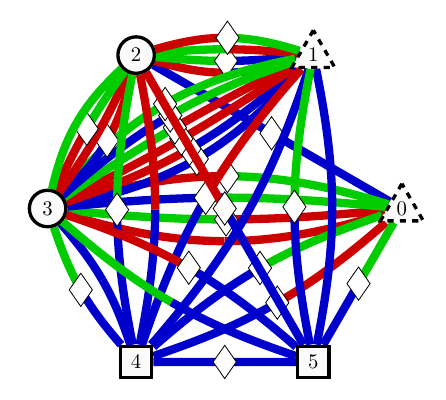}}\!\!
    \subfloat[3d Toffoli with two ancilla photons]{\includegraphics[width=0.24\textwidth]{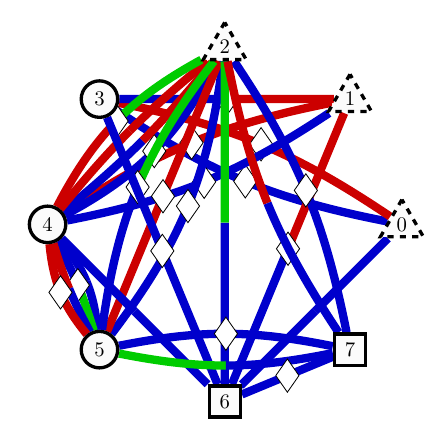}} 
	\caption {Graphs for quantum gates with post-selection.}
	\label{fig:postselectedgates}
\end{figure}

\paragraph{Post-Selected --}
When post-selecting a gate, all outgoing photons are detected and thus destroyed. Such a gate does not require as many resources as a heralded gate, but it imposes restrictions, such as not mixing the two output paths, on the remainder of the circuit, to ensure that the presence of a photon in each output of the gate can be ascertained.
    \pytheus found different post-selected high-dimensional CNOT gates, such as $\text{CNOT}(2,3)$ \numbering[https://github.com/artificial-scientist-lab/PyTheus/blob/main/pytheus/graphs/Gates/cnot23_post]{experimentcounter}, $\text{CNOT}(2,4)$ \numbering[https://github.com/artificial-scientist-lab/PyTheus/tree/main/pytheus/graphs/Gates/cnot24_post]{experimentcounter}, $\text{CNOT}(3,3)$ \numbering[https://github.com/artificial-scientist-lab/PyTheus/tree/main/pytheus/graphs/Gates/cnot33_post]{experimentcounter} and the Toffoli gate on a qutrit \numbering[https://github.com/artificial-scientist-lab/PyTheus/tree/main/pytheus/graphs/Gates/toffoli_3d_post]{experimentcounter}; see Fig.~\ref{fig:postselectedgates}.

\tocless\subsubsection{Single Photon Sources for Quantum Gates}
\begin{figure}[t]
	\centering
	\subfloat[heralded CNOT(2,2) s.p.s\centering]{\includegraphics[width=0.24\textwidth]{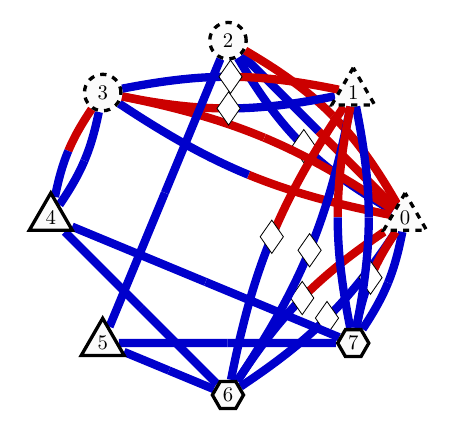}}\!\!
    \subfloat[heralded CNOT(2,3) s.p.s\centering]{\includegraphics[width=0.24\textwidth]{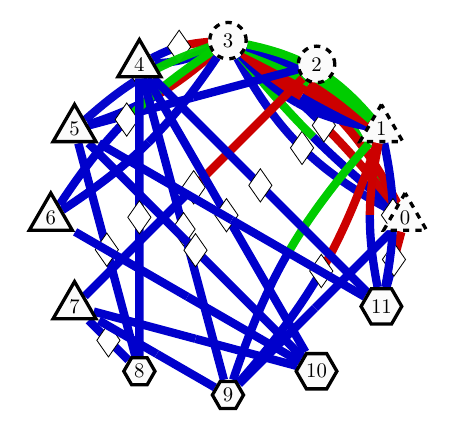}}\\
    \subfloat[post-selected CNOT(2,3) s.p.s\centering]{\includegraphics[width=0.24\textwidth]{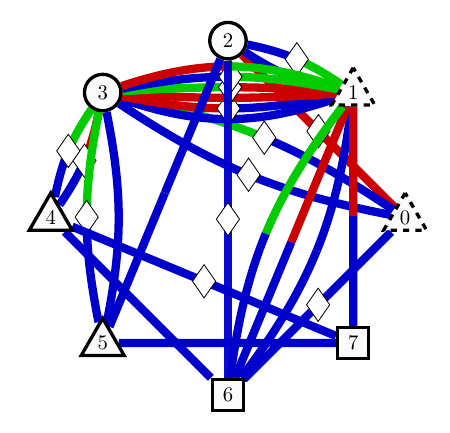}}
	\caption {Graphs for quantum gates with single-photon sources as an additional resource.}
	\label{fig:spsgates}
\end{figure}

    Experiments exploiting single-photon sources as an additional resource for the construction of quantum gates have been previously demonstrated \cite{spscnot}. Fig.~\ref{fig:spsgates} displays graphs for a heralded $\text{CNOT}(2,2)$ with two additional input photons \numbering[https://github.com/artificial-scientist-lab/PyTheus/blob/main/pytheus/graphs/Gates/cnot22_sp]{experimentcounter}, a heralded $\text{CNOT}(2,3)$ with three additional input photons \numbering[https://github.com/artificial-scientist-lab/PyTheus/tree/main/pytheus/graphs/Gates/cnot23_sp]{experimentcounter}, and a post-selected $\text{CNOT}(2,3)$ with two additional input photons \numbering[https://github.com/artificial-scientist-lab/PyTheus/tree/main/pytheus/graphs/Gates/cnot23_sp_post]{experimentcounter}.\newline

\subsection{Combinatorial Measures}
In this section, we demonstrate how \pytheus can discover exceptional structures without computing the quantum state. This ability might be useful for finding experimental configurations with particular symmetries or properties that are interesting, independent of the resulting quantum state. Such metrics can also be combined with other objectives mentioned in the previous sections.
\tocless\subsubsection{Assembly Index}
As an example for structural property discovery, we consider the \textit{assembly index}, which has been invented in chemistry for the search of extraterrestrial life \cite{marshall2017probabilistic, marshall2021identifying}. Specifically, the assembly index counts the complexity of building up a combinatorial structure, for instance, molecules or --- in our case --- graphs. It counts the number of independent processes that are necessary to create a structure. The hypothesis is that a structure with a large assembly index cannot be formed by natural processes and would require complex (living) systems for its generation. Strong indications in favor of this hypothesis have been found in the study of millions of molecules on earth \cite{marshall2021identifying}.

Our goal here is to discover graphs with very high assembly indices. The assembly index is a discrete structural metric that cannot be optimized directly using gradients. However, we can use the weights of edges to transform the discrete metric into a continuously differentiable metric. A general procedure is to compute the average combinatorial value of the weighted graph from sampled discrete graphs. Here, the weights $|\omega_i|^2$ are used as the probability for sampling discrete graphs. While this procedure is differential, it is computationally expensive as many sampled graphs need to be evaluated to obtain an average assembly index of the whole graph.

\begin{figure}
	\centering
	\subfloat[4 qubits, 8 edges,\\assembly index 8\centering]{\includegraphics[width=0.24\textwidth]{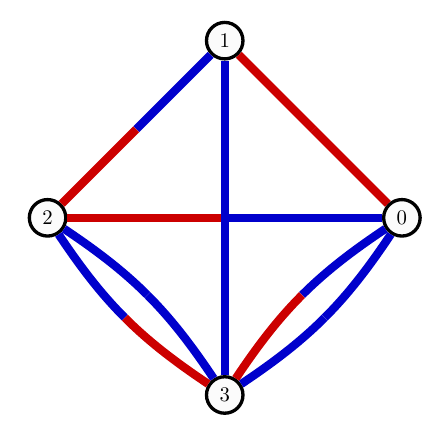}}\hspace*{\fill}
	\subfloat[6 qubits, 10 edges,\\assembly index 10\centering]{\includegraphics[width=0.24\textwidth]{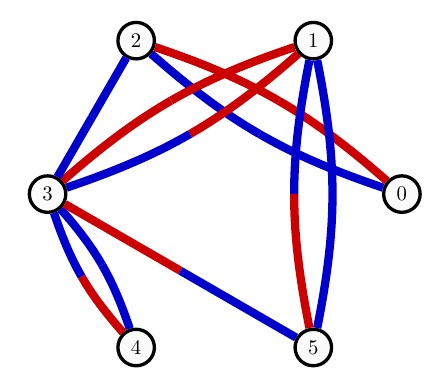}}
	\caption {Graphs for assembly index.}
	\label{fig:assemblyindex}
\end{figure}

Alternatively, here we restrict ourselves to graphs with exactly eight edges (with 4 \numbering[https://github.com/artificial-scientist-lab/PyTheus/tree/main/pytheus/graphs/AssemblyIndex/assembly4]{experimentcounter} and 6 \numbering[https://github.com/artificial-scientist-lab/PyTheus/tree/main/pytheus/graphs/AssemblyIndex/assembly6]{experimentcounter} vertices). Here, the eight highest-weighted edges are used to generate the graph, and in addition, weights are used to change the order of edges. The graphs in Fig.~\ref{fig:assemblyindex} present graphs with very high assembly indices with 4 vertices and 8 edges as well as 6 vertices and 10 edges.

\section{Outlook}\label{sec:outlook}
This article uses a list of examples to showcase the broad range of possibilities of digital discovery in quantum optics. The goal to design one hundred novel and intriguing experiments has led us to collect a diverse set of tasks for which experimental setups could be designed. To make the discoveries possible, we expanded the theoretical framework underlying the algorithm and produced faster, more versatile open-source software. We hope that this work can be an inspiration for experimental physicists in two ways: (1) to explore some of the experiments as they are proposed in this article (2) to see that \pytheus can be used to design experiments for a wide range of targets and for individually tailored specifications of experimental restrictions and resources. To explore particular examples that have not been covered in this article, the open-source library \pytheus can be used. Many more avenues leave room for significant further exploration, particularly within the optimization for targets other than fidelity, as well as more specific experimental constraints. For this, the \pytheus framework can readily be expanded in the future.

A number of extensions that would be interesting include the discovery of experiments that maximize success probability in terms of heralding efficiency or Bell state measurement efficiency, especially as experimentally available technologies became increasingly powerful \cite{bayerbach2022bell}. The potential of \pytheus on suggesting experimental setups can pave the way to find experiments to simulate quantum states, which are nowadays merely realizable in extreme (thermodynamics) conditions, such as high pressure and low temperature~\cite{Blume2012}. An exciting extension would be the analysation whether we can extend state generation to encoding the dymanics of a quantum state into the weights of the graphs. Along with this line of research, it is theoretically tantalizing to explore the putative correlation between the complexity of time-evolved states~\cite{Parker2019} and the complexity of associated graphs as outputs of \pytheus.

As a final thought, we want to view our results through an additional lens. We have created a dataset of 100 hand-selected quantum experiments that are some aspects exceptional (which \pytheus has discovered). This dataset might be large enough for highly efficient artificial intelligence algorithms to bootstrap an intuition of what properties make an experiment interesting, and allow them to produce proposals for new, hopefully equally interesting quantum experiments. It would sure be exciting to investigate the physical properties of the proposals that the machine believes are \textit{interesting}. In the best case, it could act as an inspiration for new ideas for the human scientist \cite{krenn2022scientific, rudolph2023terry}.

\tocless\section{Acknowledgment}
The authors thank Jian-Wei Wang and Jake Bulmer for their useful comments. The authors thank Matthias Bär for his support in software development and Jhon Alejandro Montañez for help in an early version of the software.
S.S. thanks the Galileo Galilei Institute for Theoretical Physics for hospitality during the completion of this work. T.J. and E.K. acknowledge the support of the Canada Research Chairs (CRC) and Max Planck-University of Ottawa Centre for Extreme and Quantum Photonics.
N.T. is a recipient of an Australian Research Council Discovery Early Career Researcher Award (DE220101082). This work was partially supported by the Australian Research Council Centre of Excellence for Quantum Computation and Communication Technology (Grant No. CE170100012). 

\bibliographystyle{unsrtnat}
\bibliography{main.bib}
\end{document}